\documentclass[10pt,aps,pre,groupedaddress,showpacs]{revtex4-1}

\pdfoutput=1
\usepackage{bm}
\usepackage{amssymb}
\usepackage{amsmath}
\usepackage{amsfonts}
\usepackage{graphicx}
\usepackage{epstopdf}
\usepackage{color}
\usepackage{xcolor}

\newcommand{\dbar}[1]{\overline{\overline{\bm{#1}}}}

\newcommand{\gap}{\dot{\gamma}}

\graphicspath{{./}{./figs/}}

\begin{document}

\title{Shear-banding and Taylor-Couette instability in thixotropic yield stress fluids}


\author{Mathieu Jenny}
\email[]{mathieu.jenny@univ-lorraine.fr}
\author{S\'ebastien Kiesgen de Richter}
\email[]{sebastien.kiesgen@univ-lorraine.fr}
\author{Nicolas Louvet}
\author{Salahedine Skali-Lami}
\affiliation{Universit\'e de Lorraine, LEMTA, UMR 7563, Vandoeuvre-l\`es-Nancy, F-54500, France \\
CNRS, LEMTA, UMR 7563, Vandoeuvre-l\`es-Nancy, F-54500, France}


\date{\today}

\begin{abstract}
In the present work, we study the flow of thixotropic yield stress fluids between two concentric cylinders. In order to take into account the thixotropy, the constitutive relation uses a structural parameter which is driven by a kinetic equation. Here, the Houska's model is considered. Depending on the breakdown rate of the structural parameter, localization or shear-banding are observed. We show that for fragile structures, a shear-banding flow may be observed although for stronger structures, only localisation of the flow is observed such as in Bingham fluids. Physical explanations of the shear-banding discussed by several authors in the literature highlight that the shear-banding may be associated with a discontinuity into the structure of the material and a non-monotonic evolution of the stress according to the constitutive relation with the strain rate.  Solving numerically the flow, we show that such a rheological model based on the existence of a structural parameter is able to predict shear-banding. Moreover, the consequences of the thixotropy on the linear stability of the azimuthal flow is studied in a large range of parameters. Although the thixotropy allows shear banding for the base flow, it does not modify fundamentaly the stability of the Couette flow compared to a simple yield stress fluid. The apparent shear-thinning behaviour depends on the thixotropic parameters of the fluid and the results about the onset of the Taylor vortices in shear-thinning fluids are retrieved. Nevertheless, the shear-banding modifies the stratification of the viscosity in the flowing zone such that the critical conditions are mainly driven by the width of the flowing region.
\end{abstract}


\maketitle

\section{Introduction}

\begin{figure*}
\centering
\includegraphics[width=0.49\textwidth]{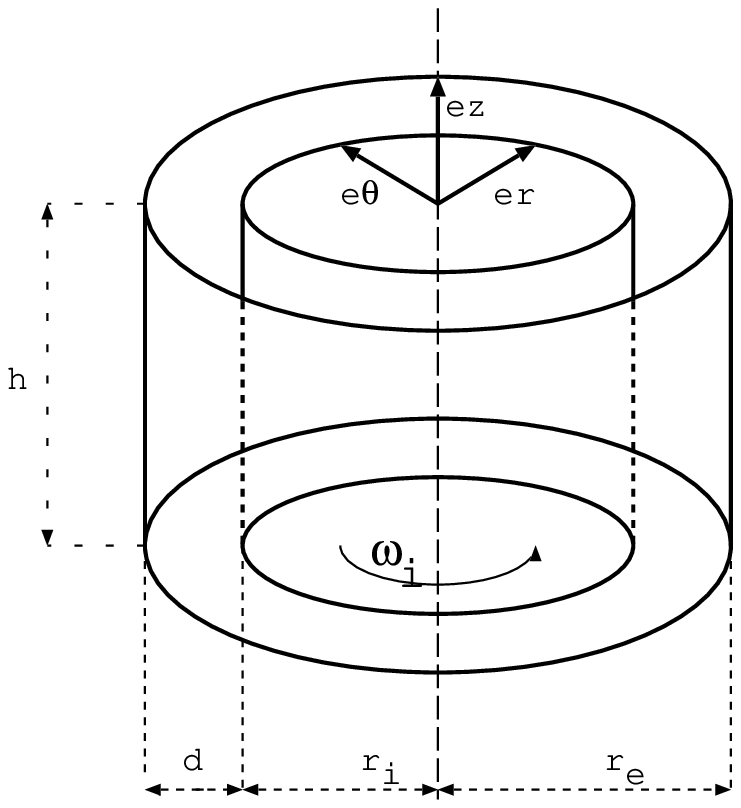}
\hspace{0.1\textwidth}
\includegraphics[width=0.3\textwidth]{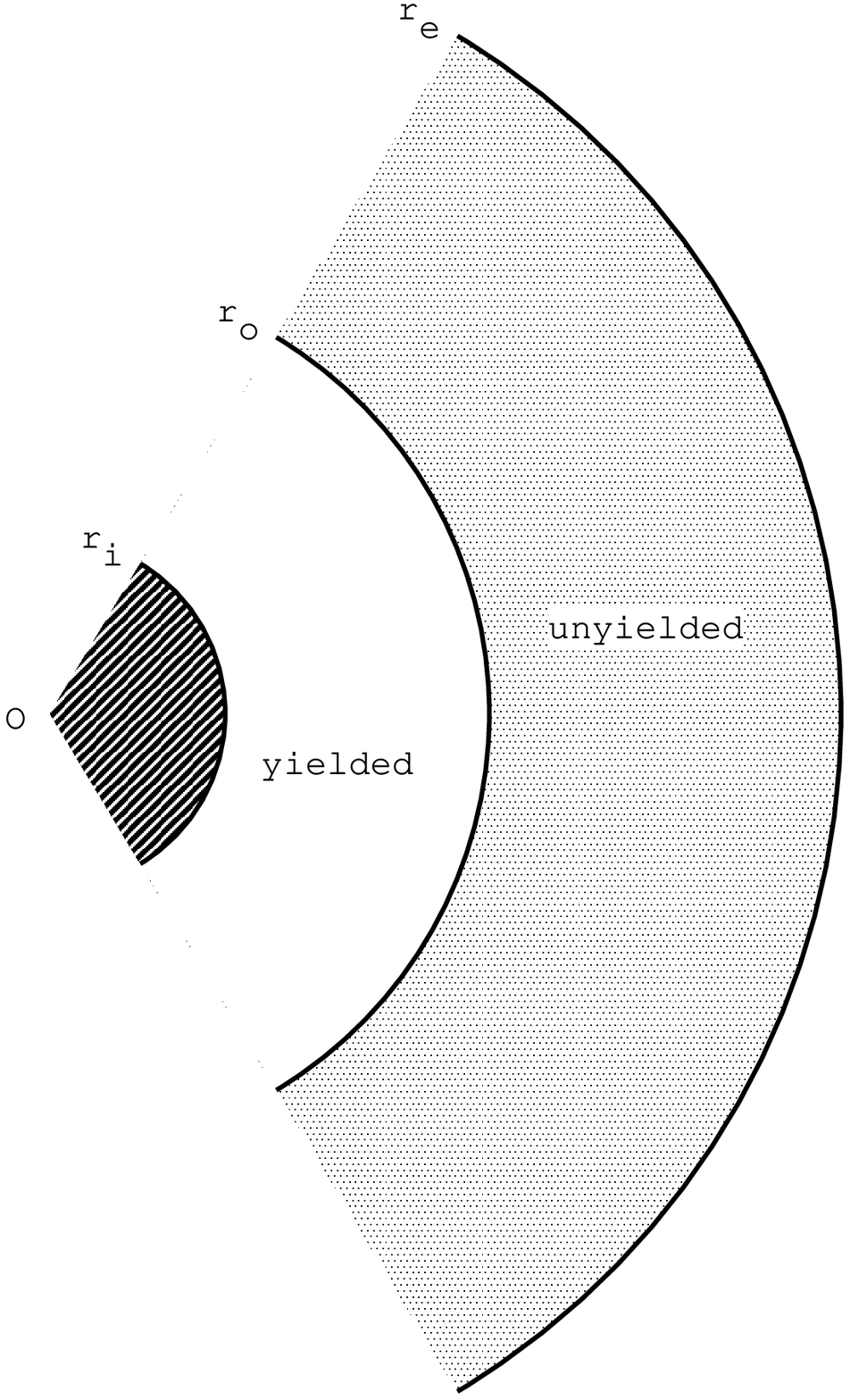}
\caption{Taylor-Couette geometry. $\mathbf{e}_r$, $\mathbf{e}_{\theta}$ and $\mathbf{e}_z$ are the unit vectors of the cylindrical coordinates system $(r,\theta,z)$. \label{fig:tc}}
\end{figure*}

Shear banding occurs in many complex materials which manifest non-linear rheological behavior such as micellar solutions, granular paste or colloidal gels \citep{divoux2015}.  It has been shown that the coexistence of static and flowing regions can be associated with the existence of a yield stress which is often related to the existence of a rigid network between the elements of the fluid which has to be broken for the system starts to flow. Shear banding occurs, in many cases, in systems which exhibit a competition between at least two mechanisms: a breakdown process due to an external applied shear and a spontaneous restructuring of the fluid due to the non-linear interaction between its elements \citep{olmsted2008}. This feature suggests that the onset of shear banding is related to a coupling between the constitutive law of the material and the evolution of its internal structure. However questions remain on the quantitative effects of such a structure parameter on the behavior of the flow and its stability. The structure parameter represents some physical properties of the instantaneous structure state of the fluid. The constitutive law depends on this structural parameter whose evolution is driven by a kinetic equation. In this picture, thixotropy models    \citep{coussot1993,cheng2003,roussel2004} such as Houska's model are relevant to predict flow instabilities and heterogeneities in complex structured fluids. Ovarlez \textit{et al.} 2009   \citep{ovarlez2009} and Coussot \& Ovarlez 2010 \citep{coussot2010} recently discuss the physical origin of shear localization and shear banding in complex fluids. Authors  highlight the existence of a discontinuity of the shear rate profile during shear banding contrary to shear localization where the shear rates goes to zero continuously as one approaches the static region. In the last case, the shear rate is zero on both side of the static and flowing regions. "Shear localization" refers to a stress heterogeneity inherent to the geometry for a yield stress fluid whereas "shear banding" is directly related to the non-linear rheology of the material. The onset of shear banding or shear localization highly depends on the stress distribution. Analysing in depth these phenomenon needs a control of that distribution. In that context, Couette flows are particularly relevant to study these phenomena since the shear stress distribution is heterogeneous but well controlled contrary to cone or plate geometry.
Taylor-Couette flow is often used in rheology as a reference shear flow. Moreover, since the historical work of Taylor \citep{taylor1923}, it is a paradigm for studies of stabilities and transition to turbulence. With Newtonian fluids, the transitional regimes observed in Taylor-Couette flows have been widely studied \citep{andereck1986} when the inner or the outer cylinders rotates. In the present work, we only consider the case where the inner cylinder rotates at a given angular velocity $\omega _i$ such as the velocity at the inner radius $r_i$ is $v_i=\omega _i r_i$ (fig. \ref{fig:tc}). The outer cylinder is static. In this configuration, when the velocity of the inner cylinder is sufficiently low, the purely azimuthal steady flow is stable for viscous fluids \citep{taylor1923,andereck1986,escudier1995,alibenyahia2012}. One can notice that an elastic instability of the flow may occurs even at very low velocity \citep{larson1990} for non-Newtonanian fluids. According to the studies of the hydrodynamic stability of shear-thinning \citep{li2004,alibenyahia2012} and Bingham fluids \citep{landry2006,alibenyahia2012,chen2015} in Couette flow,  it is observed that when the viscosity is scaled with the inner wall shear-viscosity, shear-thinning has a stabilizing effect, \textit{i. e.} the appearance of the Taylor-vortices is delayed. For a Bingham fluid, the dependence of the wavelength of the Taylor vortices to the yield stress changes when a non-yielded zone appears close to the outer cylinder. Chen \textit{et al.} \cite{chen2015} have shown that the effects of yield stress on the energy transient growth and flow structure of the optimal perturbations are quite different for the wide or narrow-gap cases. Considering axisymmetric optimal modes for corotating cylinders, the peak of the amplitude of optimal perturbation is found to be shifted toward the inner cylinder with increasing yield stress for the wide-gap case whereas the peak is shifted toward the outer cylinder for the narrow-gap case. Three dimensional perturbations were also investigated in \citep{alibenyahia2012} but it was found that the most unstable modes are axisymmetric, even for very strong shear-thinning fluids.

Certainly, Taylor vortices do occur in yield stress fluids. This is evidenced by the numerical experiments of Lockett \textit{et al.} (1992) \cite{lockett1992}, Coronado-Matutti, Souza Mendes \& Carvalho (2004) \cite{coronado2004} and Jeng  \& Zhu (2010) \cite{jeng2010}, by phenomenological evidence from applications such as oil well drilling (\textit{i.e.} observed changes in frictional pressure), and also by only few direct experimental studies. Nouar, Devienne \& Lebouche (1987) \cite{Nouar1987} and Naimi, Devienne \& Lebouche (1990) \cite{Naimi1990} have studied axial flow through an annulus with rotating inner cylinder, using CMC and Carbopol solutions, respectively. The former behaves as a power-law fluid and the latter has a yield stress, as well as shear-thinning behaviour. In both cases only axisymmetric Taylor vortices are reported, with their appearance retarded by the presence of an axial flow. Naimi \textit{et al.} (1990) \cite{Naimi1990} reports that the yield stress appears to stabilize the flow. Fardin \textit{et al.} \cite{fardin2009} show experimental evidences of Taylor-like vortices in shear-banding flow of giant micelles. This observation suggests that the Taylor-vortices can commonly occur in a very large range of thixotropic or non-thixotropic non-Newtonian fluids.

In this article, we propose a numerical study of the stationary flow of a thixotropic yield stress fluid in a Couette configuration. For modelling steady state flows, we use the Houska's model where the constitutive law depends on a structural parameter driven by a kinetic equation.It is commonly admitted in the litterature \cite{olmsted2000} that a diffusion term is recquired to model shear banding. We show, in this article, that the stress selection at the interface by the structural diffusion term admits a limit value equals to the yield stress of the fully structured fluid when the structural diffusion coefficient tends to zero. We show that the inherent thixotropy of the model controls the transition from a solution where a static and a flowing regions coexists in which the shear rate profile is continuous (shear localization) to a solution where the two regions coexist but in which the shear rate profile is discontinuous (shear banding). We show that the onset of shear banding is controled by the competition between restructuring effects and breakdown effects due to the flow. Although the thixotropy initiates shear banding in the base flow,  a linear analysis  of the stability of flow solutions shows that the nature of the linear unstable mode which is steady and axisymmetric in a large range of explored parameters does not depend on the thixotropic character of the flow.

\section{CONSTITUTIVE EQUATIONS AND NUMERICAL APPROACH}

\subsection{Modified Houska's model}

The Houska's model \cite{houska1981,houska1983} is built from the Hershel-Bulkley model, commonly used for non-elastic yield stress fluids, considering that the consistency $K$ and the yield stress $\tau_0$ depend linearly on the structural parameter $\lambda$. Thus, for a non-zero strain rate, \textit{i. e.} $\hat{\gap} \neq 0$, the stress tensor $\dbar{\hat{\tau}}$ is given by :

\begin{equation}
\dbar{\hat{\tau}} = \left( (K+ \Delta K \lambda) \hat{\gap} ^{n_c} + \tau_0 + \tau_1 \lambda \right) \, \frac{\dbar{\hat{\gap}}}{\hat{\gap}} \label{eq:taud}
\end{equation}

\noindent The parameters $\Delta K$ and $\tau_1$ determine the sensitivity of the consistency and the yield stress respectively with the structural parameter $\lambda$. $n_c$ is the shear thinning index. The second invariant of the strain rate tensor $\dbar{\hat{\gap}}$ is defined by

\begin{equation}
\hat{\gap} = \left( \frac{1}{2} \hat{\dot{\gamma}}_{ij} \hat{\dot{\gamma}}_{ij} \right)^{1/2}
\end{equation}

\noindent using the Einstein summation convention where the elements of the strain tensor $\hat{\dot{\gamma}}_{ij} = \partial \hat{v}_i / \partial \hat{x}_j + \partial \hat{v}_j / \partial \hat{x}_i$ are defined with the components of the fluid velocity $\hat{\mathbf{v}}$. The indices $i$ and $j$ stand for the cylindrical coordinates $r$, $\theta$ and $z$ (fig. \ref{fig:tc}) and the hat $\hat{.}$ denotes a dimensional variable.
According Olmsted \textit{et al.} \cite{olmsted2000}, a stress diffusion has to be added to keep the unicity of the steady solution in case of shear-banding by selecting the stress at the interface between the bands. Fardin \textit{et al.} \cite{fardin2015} show that the length based on this stress diffusion is at the order of magnitude of the molecular size in wormlike micelles. Thus, the structural parameter $\lambda$ is determined by the kinetic equation :

\begin{equation}
\frac{\partial \lambda}{\partial \hat{t}} + \hat{\mathbf{v}} . \bm{\nabla} \lambda = a (1-\lambda) - b \lambda \hat{\gap} ^m + \mathcal{D} \Delta \lambda \label{eq:lambdad}
\end{equation}

\noindent $\mathcal{D}$ is the structural diffusion coefficient. The diffusion term in eq. (\ref{eq:lambdad}) has been added to the original version of Houska's model to take into account the stress diffusion, \textit{via} the structural diffusion, in our model. $a$ and $b$ are respectively the building and the breakdown parameters. The thixotropic breakdown index $m$ is taken equal to $1$ in the following. The values of the structural parameter are within the range $0 \leq \lambda \leq 1$. The value $\lambda=1$ means that the fluid is fully structured and, at the opposite, $\lambda=0$ means that it is fully unstructured. The evolving dynamics of the stress is governed by the equation (\ref{eq:lambdad}).

\subsection{Nondimensional equations}

To non-dimensionalize the constitutive equations of the flow in a cylindrical Couette geometry, we choose the following references for the density, the velocity and the length, respectively:

\begin{equation}
\rho_{ref}=\rho , \quad
v_{ref} = v_i , \quad
l_{ref} = d,
\end{equation} 

\noindent where $\rho$ is the density of the fluid. For the non-Newtonian fluids, several choices may be done for the reference viscosity. The choice of the reference viscosity will be discussed in the section \ref{sec:muref}. The reference viscosity is the plastic viscosity of the fluid at the given reference strain rate $v_i/d$. When the strain rate is $v_i/d$, the corresponding structural parameter $\lambda_{ref}$ is given by eq. (\ref{eq:lambdad}) at equilibrium:
\begin{equation}
\lambda_{ref}=\frac{a}{a+b(v_i/d)^m}=\frac{1}{1+b^{\star}/a^{\star}} \label{eq:lambdar}
\end{equation}
\noindent where the nondimensional building and breakdown parameters are 
\begin{equation}
a^{\star} =\frac{a d}{v_i} \quad \text{and} \quad b^{\star}=b \left( \frac{v_i}{d}\right)^{m-1} \, , \label{eq:ab}
\end{equation}
\noindent respectively. Thus, the reference viscosity built using the reference strain rate  reads :
\begin{equation}
\mu_{ref} = \mu_0 (1+\Delta K^{\star} \lambda_{ref})
\label{eq:mur}
\end{equation}
\noindent where $\mu_0= K  (v_i/d)^{n_c-1}$ can be recognized as the standard reference viscosity of a power law fluid. $\Delta K^{\star}=\Delta K / K$ is the reduced thixotropic consistency factor. This parameter characterizes the dependence of the plastic viscosity with the inner structure of the fluid in comparison with the intrinsic consistency $K$ which depends on the solvent. The reference viscosity $\mu_{ref}$ depends on the ratio of the breakdown parameter $b^{\star}$ over the building parameter $a^{\star}$. The reference viscosity decreases when the ratio $b^{\star}/a^{\star}$ increases, \textit{i. e.} when the inner structure of the fluid becomes more and more fragile. 

Using the previous reference dimensions, the Navier-Stokes and mass conservation equations for incompressible fluids are :

\begin{eqnarray}
\frac{\partial \mathbf{v}}{\partial t} + (\mathbf{v} . \bm{\nabla})  \mathbf{v} &=& - \bm{\nabla} p + \frac{1}{Re} \bm{\nabla} . \dbar{\tau} \label{eq:ns} \\
\bm{\nabla} . \mathbf{v} &=& 0  \label{eq:div}
\end{eqnarray}

\noindent where $\mathbf{v}= \hat{\mathbf{v}}/v_i$ stands for the reduced velocity and $p=\hat{p}/(\rho v_i^2)$ for the reduced pressure. One can notice that $\hat{p}$ is the modified pressure including the hydrostatic pressure. The Reynolds number is defined using the reference viscosity (\ref{eq:mur}) by:
\begin{equation}
Re = \frac{Re_0}{1+\Delta K^{\star} \lambda_{ref}} \quad \text{where} \quad Re_0=\frac{\rho v_i d}{\mu_0}\label{eq:Re}
\end{equation}
\noindent Thus, the reduced stress tensor reads:

\begin{equation}
\dbar{\tau} = \left[ \left( \frac{1 + \Delta K^{\star} \lambda}{1 + \Delta K^{\star} \lambda_{ref}} \right) \gap ^{n_c} + Bn \left( \frac{1 + \tau_1^{\star} \lambda}{1 + \tau_1^{\star} \lambda_{ref}} \right)  \right] \, \frac{\dbar{\gap}}{\gap} \, , \label{eq:tau}
\end{equation}

\noindent where $\gap$ and $\dbar{\gap}$ are the nondimensional strain rate and strain tensor, $\tau_1^{\star}=\tau_1 / \tau_0$ is the reduced thixotropic yield stress. The equation (\ref{eq:tau}) involves the Bingham number which is the ratio between the yield stress and the plastic viscous stress:
\begin{equation}
Bn=Bn_0 \frac{1+\tau_1^{\star} \lambda_{ref}}{1+\Delta K^{\star} \lambda_{ref}} \quad \text{where} \quad Bn_0=\frac{\tau_0}{K(v_i/d)^{n_c}} \label{eq:Bn}
\end{equation}

\noindent is the standard Bingham number of a Hershel-Bulkley fluid. For a strong yield stress behavior, the Bingham number is high and localized flows are expected. According the equation (\ref{eq:tau}), the reduced yield stress is:
\begin{equation}
\tau_y = Bn \left( \frac{1 + \tau_1^{\star} \lambda}{1 + \tau_1^{\star} \lambda_{ref}} \right)\, . \label{eq:tauy}
\end{equation}

For the structural parameter, the non-dimensional version of the equation (\ref{eq:lambdad}) is: 

\begin{equation}
\frac{\partial \lambda}{\partial t} + \mathbf{v} . \bm{\nabla} \lambda = a^{\star} (1-\lambda) - b^{\star} \lambda \gap ^m  + \mathcal{D}^{\star} \Delta \lambda \, . \label{eq:lambda}
\end{equation}

\noindent with the non-dimensional structural diffusion coefficient

\begin{equation}
\mathcal{D}^{\star}=\frac{\mathcal{D}}{v_i d} \, .
\end{equation}

\subsection{Boundary conditions for the  flow \label{sec:bc}}

The inner and outer radii of the Couette setup are now defined by
\begin{eqnarray}
r_i&=&\frac{\eta}{1-\eta} \, ,\\
r_e&=&\frac{1}{1-\eta} \, ,
\end{eqnarray} 

\noindent with $\eta=r_i/r_e$  the radii ratio.

\begin{figure}
\centering
\includegraphics[width=0.65\columnwidth]{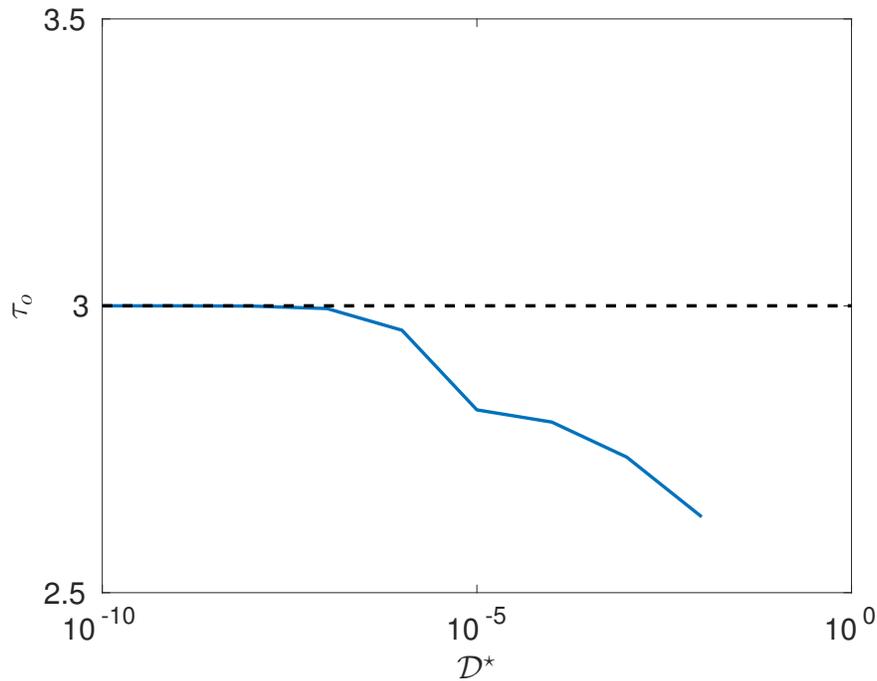}
\caption{Stress at the interface $\tau_o=\tau(r_o)$ between the fluid and solid-like region when the diffusive coefficient $\mathcal{D}^{\star}$ tends to zero. The dashed-line stands for the value of the yield stress of the fully structured material $\tau_{ys}$. $Bn=2$, $\Delta K^{\star}=1$, $\tau_1^{\star}=1$ and $b^{\star}=2$.\label{fig:diffuse}}
\end{figure}

The velocity vector $\mathbf{v}$ is written in the cylindrical basis as $\mathbf{v}=v_r \mathbf{e}_r + v_{\theta} \mathbf{e}_{\theta} + v_z \mathbf{e}_z$. We only consider the case where the inner cylinder rotates and the outer cylinder is fixed. Thus, the boundary conditions are:

\begin{itemize}
\item At the inner radius $r=r_i$, the velocity components are $v_{\theta}=1$ and $v_r=v_z=0$.
\item At the outer radius of the flowing zone $r=r_o$, the velocity components are $v_r=v_{\theta}=v_z=0$.
\item In our case, there is a material limit at $r=r_e$. Thus, the outer radius $r_o$ is given by the following criterion:\\
 If $\tau (r_e) \geq \tau_y$, $r_o=r_e$, else,  $\tau(r_o)=\tau_y$.
 \item If $\mathcal{D}^{\star} \neq 0$, $\partial \lambda / \partial r = 0 $ at $r=r_i$ and $r=r_e$.
\end{itemize}
At  $r=r_o$, the yield stress $\tau_y$ given by eq. (\ref{eq:tauy}) becomes
\begin{equation}
\tau_{yo}=Bn \left( \frac{1 + \tau_1^{\star} \lambda_o}{1 + \tau_1^{\star} \lambda_{ref}} \right), \label{eq:tauyo}
\end{equation}
 
\noindent with $\lambda_o$ the structural parameter at the interface. Thus, the last boundary condition becomes $\tau(r_o)=\tau_{yo}$ and the stress $\tau_{yo}$ at the interface is defined by eq. (\ref{eq:tauyo}). Nevertheless, the stress condition at the interface between the fluid and solid-like zone is well defined only if the structural parameter $\lambda$ is continuous across the interface. When the shear banding appears, the structural parameter $\lambda$ is sharply discontinuous across the interface when no diffusive gradient term is added ($\mathcal{D}^{\star}=0$) in the governing equation of $\lambda$ and thus the yield stress can not be well defined (Olmsted \textit{et al.} \citep{olmsted2000} and Lu \textit{et al.} \citep{lu2000} in 2000). Authors showed that a spatially local model, \textit{i. e.} without any diffusive gradient of the stress (diffusive term for the structural parameter in our case), will not correctly predict a shear banded state. The steady state depends then on the flow/numerical noise history because it will select arbitrary a stress value at the interface. When adding a diffusive term, the continuity of the yield stress or the structural parameter across the interface is ensured. This kind of diffusive term has been recently interpret as a non-local effect at the molecular scale in flow of micellar suspensions in \cite{fardin2015}. As the value of the stress diffusion coefficient, similar to the structural diffusion coefficient, is found to be very small \cite{radulescu2003,fardin2015}, we will focus on the cases where $\mathcal{D}^{\star}$ tends to zero. In this case, we found a limit value for the stress at the interface in case of shear-banding and steady flows. The limit is the yield stress of the fully structured material $\tau_{ys}$ (Fig. \ref{fig:diffuse}). Thus, we can solve the steady equations with $\mathcal{D}^{\star}=0$, setting the stress at the interface :
\begin{equation}
\tau(r_o)= \tau_{ys} = Bn \left( \frac{1 + \tau_1^{\star}}{1 + \tau_1^{\star} \lambda_{ref}} \right), \label{eq:tauymax}
\end{equation}

\subsection{NUMERICAL METHOD FOR TRANSIENT FLOWS}

\subsubsection{Arbitrary Lagrangian Eulerian method}

To perform the numerical solution, we use a finite difference method for the spatial discretisation. The mesh is split in two regions: the first one is the fluid zone for radii $r_i \leq r \leq r_o$ and the second one is the solid zone for radii $r_o \leq r \leq r_e$. The mesh points are regularly spread between the inner radius $r_i$ and the outer radius $r_o$ in the first region and between $r_o$ and the external radius $r_e$ in the second region. The number of points are the same in the two regions. In order to solve the transient flow, we use an Arbitrary Lagrangian-Eulerian (ALE) method \cite{hirt1974,donea1982} for the interface tracking. The method consists in solving the governing equations in the gap considering a mesh where one node (1D problem) sticks to the interface. To keep a constant radial length between two successive nodes of the mesh at each time step, the arbitrary velocity of nodes is given by:

\begin{equation}
\left\lbrace
\begin{array}{l}
\mathbf{u} = u_o \left( \frac{r-ri}{r_o-r_i} \right) \mathbf{e}_r \text{ in the flowing side, } r \leq r_o ,\\
\mathbf{u} = u_o \left( \frac{r_e-r}{r_e-r_o} \right) \mathbf{e}_r \text{ in the solid side, } r \geq r_o ,
\end{array} \right.
\label{eq:meshvel}
\end{equation}

\noindent where $\mathbf{e}_r$ is the radial unit vector. $u_o$ is the radial velocity of the interface. The set of equations (\ref{eq:ns}, \ref{eq:div} and \ref{eq:lambda}) has to be written in the moving domain \cite[see][]{hirt1974,donea1982}:

\begin{eqnarray}
\frac{\partial \mathbf{v}}{\partial t} + ((\mathbf{v} - \mathbf{u}) . \bm{\nabla})  \mathbf{v} &=& - \bm{\nabla} p + \frac{1}{Re} \bm{\nabla} . \dbar{\tau} \, , \label{eq:nsale} \\
\bm{\nabla} . \mathbf{v} &=& 0 \, , \label{eq:divale} \\
\frac{\partial \lambda}{\partial t} + (\mathbf{v} - \mathbf{u} ). \bm{\nabla} \lambda &=& a^{\star} (1-\lambda) - b^{\star} \lambda \gap ^{m} + \mathcal{D}^{\star} \Delta \lambda \, . \label{eq:lambdaale}
\end{eqnarray}

\begin{figure*}
\centering
\begin{tabular}{cc}
(a) & (b) \\
\includegraphics[width=0.48\textwidth]{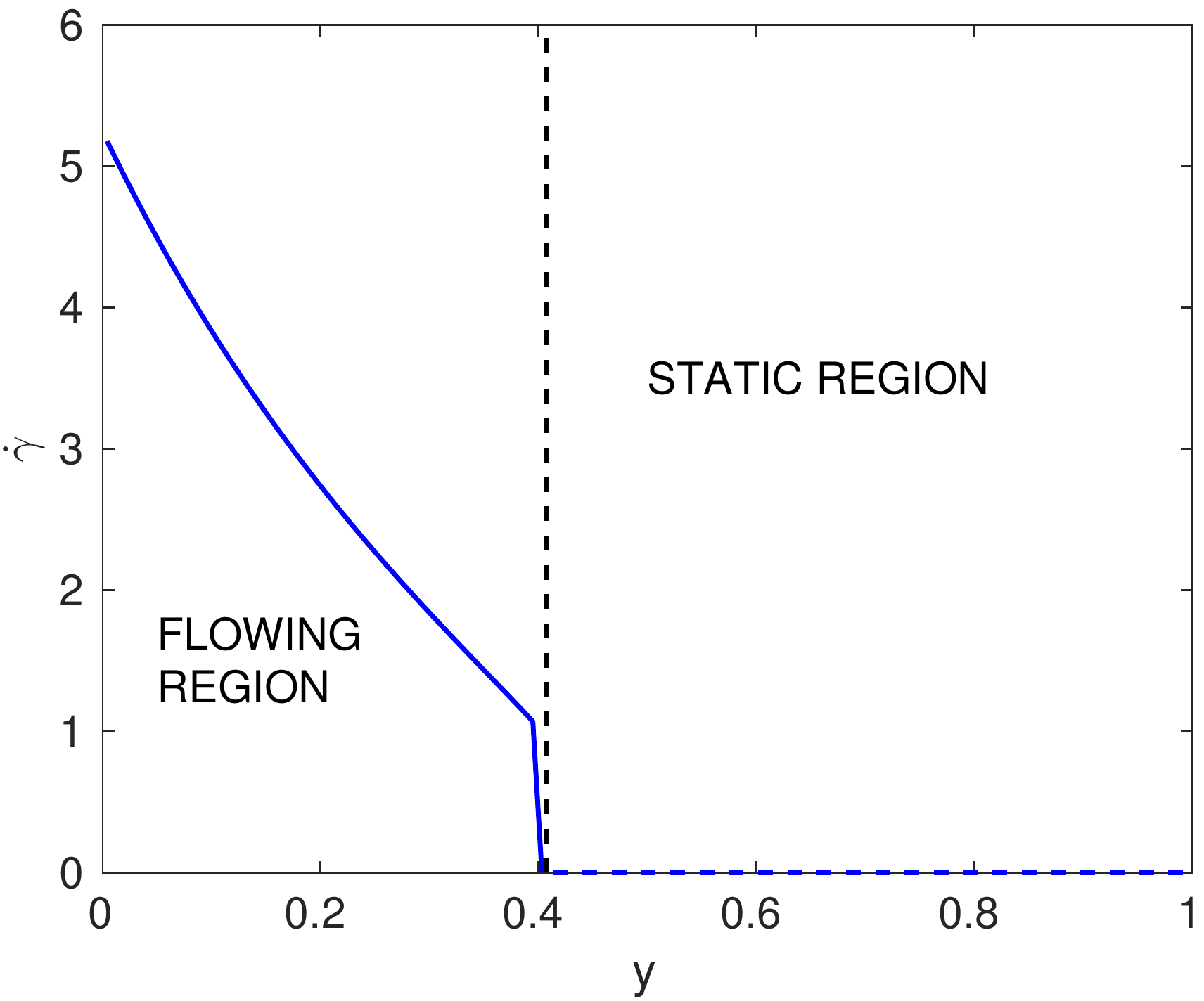} &
\includegraphics[width=0.48\textwidth]{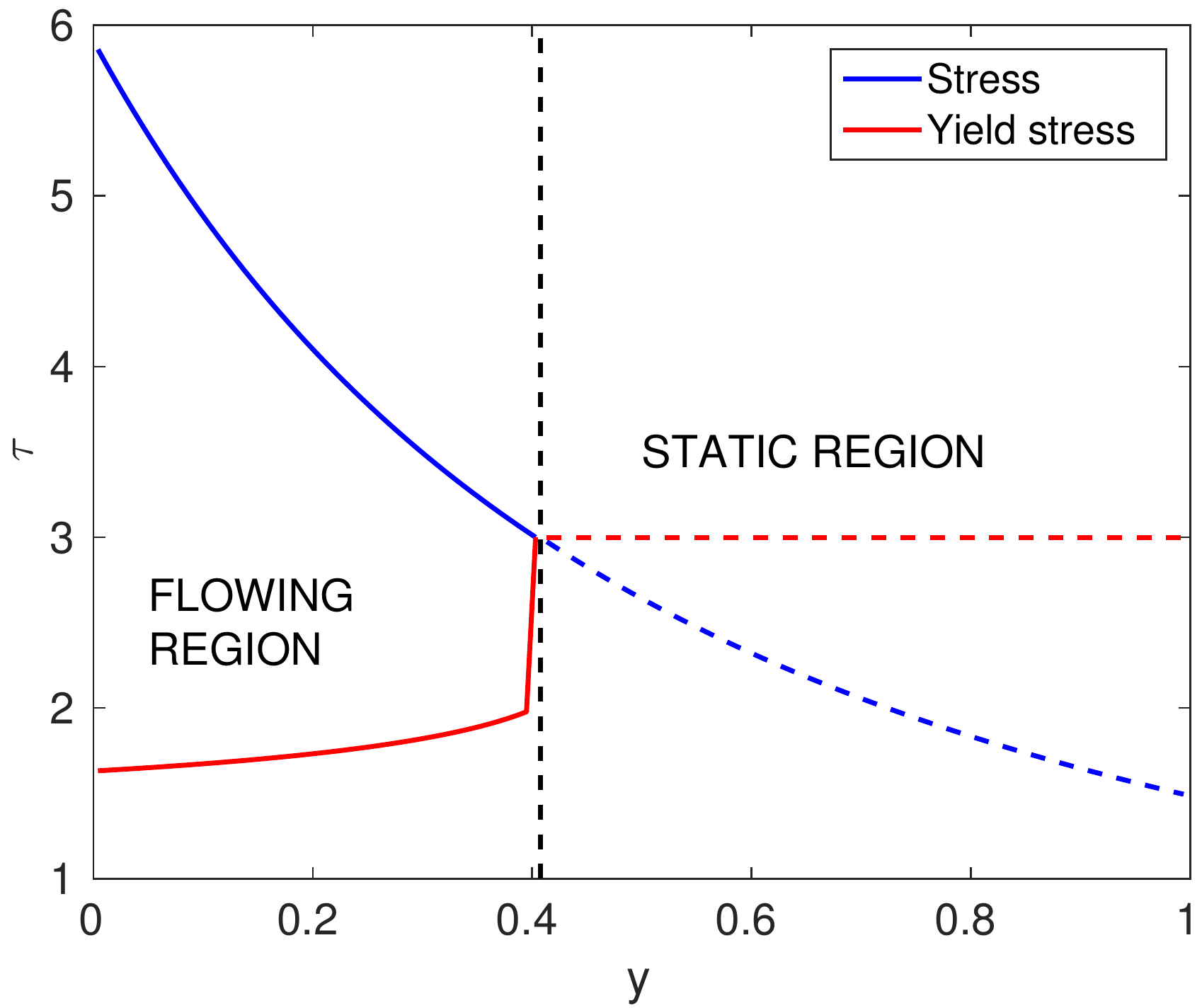}
\end{tabular}
\caption{(a) : Strain rate $\dot{\gamma}$ in the gap. (b) : Stress and yield stress in the gap. Asymptotic shear banding at $t=100$, starting at rest, interface velocity $u_o=-4.85 \times 10^{-14}$ ($|u_o| \ll 1$). $Re=10$, $Bn=2$, $n_c=1$, $\Delta K^{\star}=1$, $\tau_1^{\star}=1$, $a^{\star}=1$, $b^{\star}=2$ and $\mathcal{D}^{\star}=0$. $50$ nodes per region. \label{fig:tb2}}
\end{figure*}

\noindent In the last equation (\ref{eq:lambdaale}), the convective gradient term, $\mathbf{u} \bm{\nabla} \lambda$ ensures the continuity of the structural parameter and the yield stress across the interface as long as the interface velocity is not zero, even if there is no diffusive term in the initial equation (\ref{eq:lambda}), \textit{i.e.} when $\mathcal{D}^{\star}$ tends to zero, as shown in the figure \ref{fig:tb2}. Thus, in the case $Re \neq 0$ and $u \neq 0$ (finite time), the stress at the interface given by eq. (\ref{eq:tauyo}) is defined even if $\mathcal{D}^{\star}$ tends to zero in shear banded flows because of the smoothing due to the motion of the interface. The steady state of the shear banded flow, \textit{i. e.} with a sharp discontinuity, can only be considered as an asymptotic state when the velocity tends to zero at $t \rightarrow + \infty$. One can remark that the figure \ref{fig:tb2}-b shows that the stress value at the interface tends to the yield stress of the fully structured material $\tau_{ys}$ as found previously with $\mathcal{D}^{\star}$ tending to zero in the steady flows.

To compute the flow in the fluid region $r_i \leq r \leq r_o$, one solves the set of equations (\ref{eq:nsale}, \ref{eq:divale} and \ref{eq:lambdaale}) with the boundary conditions given in the section \ref{sec:bc}. The stress at the interface is equal to the local yield stress. This latter condition  provides the interface velocity $u_o$. The structural parameter is also computed in the solid-like region considering $\dot{\gamma}=0$. One can notice that when $r_o=r_e$, there is no interface between a fluid and a solid-like region. Thus, the stress has just to be above the local yield stress, \textit{i. e.}  $\dot{\gamma} \geq 0$ everywhere.

\subsubsection{Time discretisation}

The time scheme used for the velocity is a semi-implicit time scheme at the order one. Here, we focus on the 1D equations where considering $\mathbf{v}=V(t,r) \mathbf{e}_{\theta}$ and $\mathbf{u}=U(t,r) \mathbf{e}_{r}$, $U$ being defined by eq. (\ref{eq:meshvel}):

\begin{equation}
Re V_{n+1} - Re \Delta t \left( U_{n+1} \left. \frac{\partial V}{\partial r} \right|_n + U_n \left. \frac{\partial V}{\partial r} \right|_{n+1} \right) + \frac{\Delta t}{r^2} \frac{\partial (r^2 \tau_{n+1})}{\partial r} = Re V_n - Re \Delta t U_n \left. \frac{\partial V}{\partial r} \right|_{n} \label{eq:nsnum}
\end{equation}

\noindent The index $n$ denotes the time step $t_n$ and $V_n$, for instance, stands for $V(t_n,r_n)$. The radii $r_n$ and $r_{n+1}$ are related by $r_{n+1} = r_n + U_n \Delta t$ with the time step defined as $\Delta t= t_{n+1} - t_n$. The stress $\tau = -\tau_{r \theta}$ is given by the equation (\ref{eq:tau}). To implicit the time scheme, one writes the developpment for the stress at the order one in time:

\begin{equation}
\tau_{n+1} = \tau_{n} + \left. \frac{\partial \tau}{\partial \dot{\gamma}} \right|_n (\dot{\gamma}_{n+1} - \dot{\gamma}_n) +  \left. \frac{\partial \tau}{\partial \lambda} \right|_n (\lambda_{n+1} -\lambda_n) \label{eq:taunp1}
\end{equation}

\noindent where $\dot{\gamma}= - \dot{\gamma}_{r \theta} = V/r - \partial V / \partial r$. For numerical stability reasons, the time discretization scheme used

\begin{equation}
\lambda_n + \Delta t U_{n+1} \left. \frac{\partial \lambda}{\partial r} \right|_{n+1} \simeq \lambda_n (r_{n+1}) \label{eq:timesch}
\end{equation}

\noindent $\lambda_n(r_{n+1})$ is obtained by the interpolated value of the former field $\lambda_n$ at the current node position $r_{n+1}$. Thus, the equation (\ref{eq:lambdaale}) is rewritten as:

\begin{equation}
(1 + \Delta t ( a^{\star} + b^{\star} \dot{\gamma}_n ^m - \mathcal{D}^{\star} \Delta. )) \lambda_{n+1} + m \Delta t b^{\star} \lambda_n \dot{\gamma}_n^{m-1} \dot{\gamma}_{n+1}= \lambda_n(r_{n+1}) + \Delta t ( a^{\star} + m b^{\star} \lambda_n \dot{\gamma}_n^{m}) \label{eq:lambdanum}
\end{equation}

\begin{figure*}
\centering
\begin{tabular}{cc}
(a) & (b) \\
\includegraphics[width=0.48\textwidth]{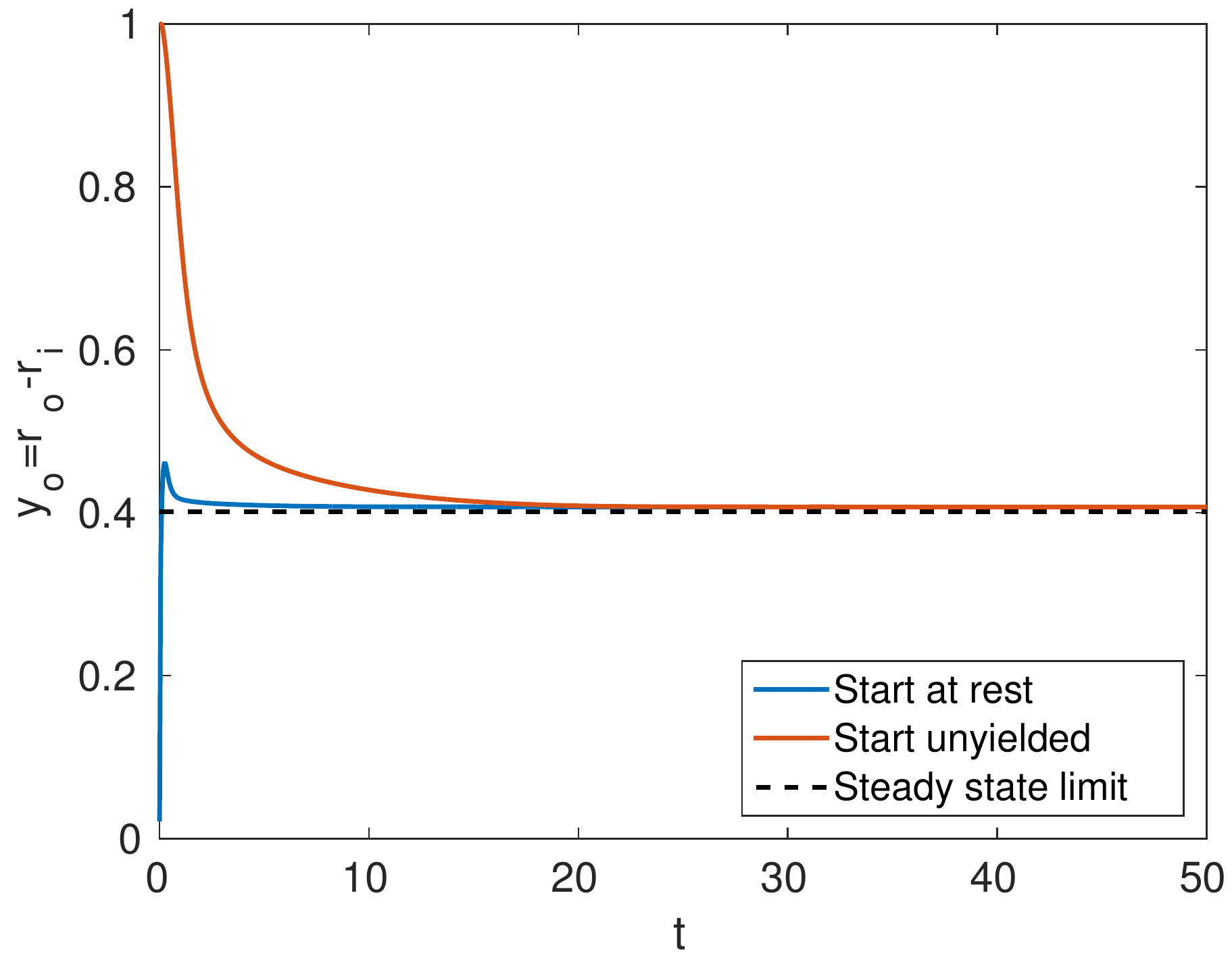} &
\includegraphics[width=0.48\textwidth]{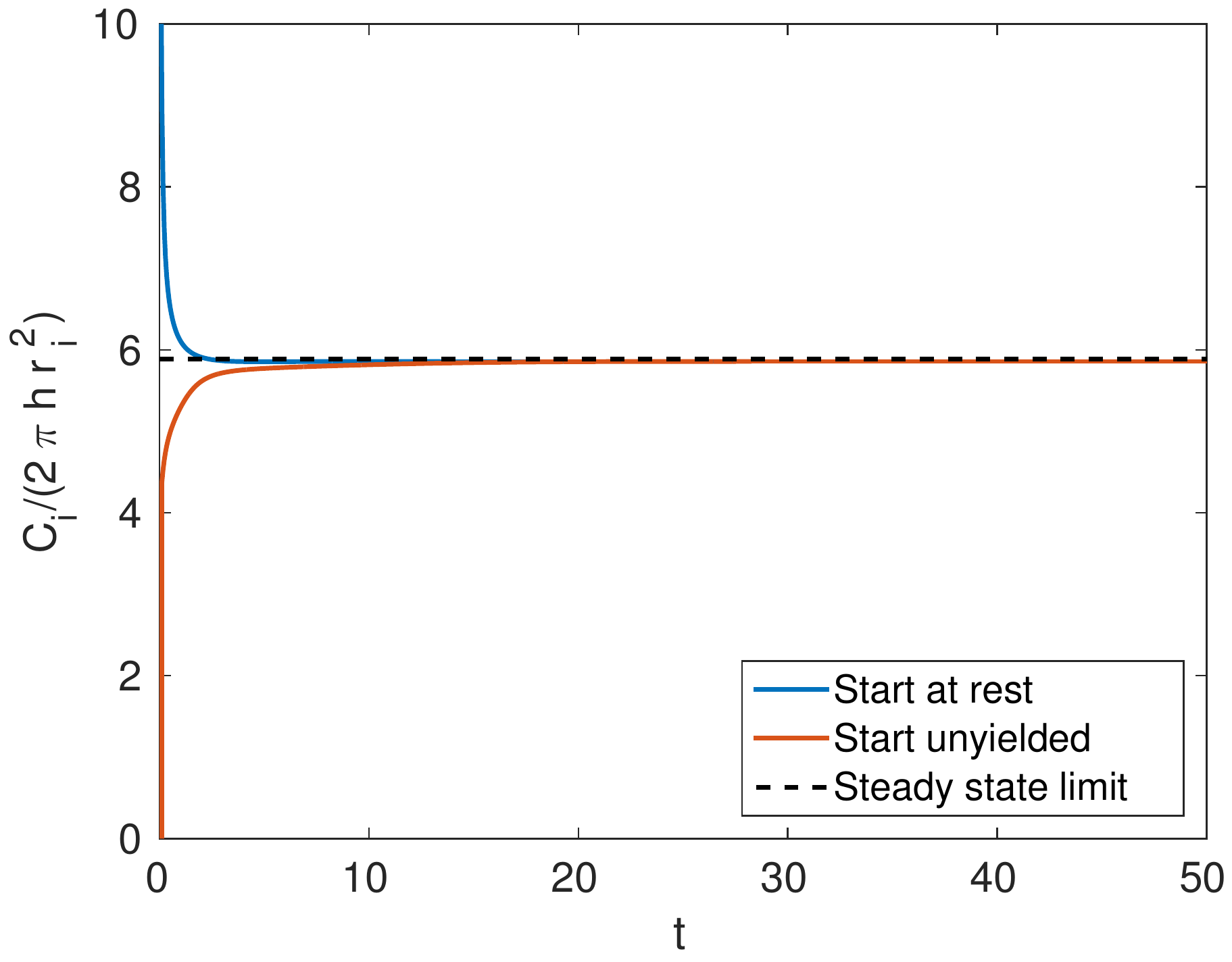}
\end{tabular}
\caption{(a) : Position of the interface $y_o=r_o-r_i$ \textit{vs} time. (b) : Torque on the inner cylinder \textit{vs} time. The steady state limit is compute using the steady version of the code described below. $Re=10$, $Bn=2$, $n_c=1$, $\Delta K^{\star}=1$, $\tau_1^{\star}=1$, $a^{\star}=1$, $b^{\star}=2$ and $\mathcal{D}^{\star}=0$. $50$ nodes per region. \label{fig:rotb2}}
\end{figure*}

To test that the final steady state does not depend on the starting condition, as it would be expect when $\mathcal{D}^{\star}=0$ \cite{olmsted2000}, we define two cases. The first one is a start at rest with the material fully structured, \textit{i. e.} $V_0=0$ and $\lambda_0=1$ everywhere. As the thickness of the fluid region collapses at the initial time, we take an arbitrary profile for the first non-zero velocity at $t_1=\Delta t$. The velocity $V_1$ is defined with a second order polynomial function of the radius. The velocity field $V_1$ verifies all the boundary conditions of the velocity and the strain rate is zero at the interface $r_{o,1}$. The stress $\tau_1$ in the initial fluid region is calculated in order to balance the pulse momentum provided to accelerate the flow from $0$ to the arbitrary velocity $V_1$ (eq. \ref{eq:nsnum}). Thus, the stress depends on $U_0$, \textit{i. e.} on $u_{o,1} = ( r_{o,1} - r_i ) / \Delta t$. The position $r_{o,1}$ of the interface is calculated to equalize the stress at the interface and the yield stress. From this time, the velocity profile is computed solving numerically the equations (\ref{eq:nsnum}), (\ref{eq:taunp1}) and (\ref{eq:lambdanum}). In the second scenario, the calculation starts from a fully unstructured and flowing material in the whole gap. The figures \ref{fig:rotb2} show that both transient stage tends to the same steady state even if the asymptotic steady state is a shear banded flow as it will be discussed in the following. The asymptotic case corresponds to the steady state where the interfacial stress is the yield stress of the fully structured material. Setting $\mathcal{D}^{\star} \neq 0$, in the right-hand-side of the equation (\ref{eq:lambda}) would select an intermediary value for the stress at the interface in a shear banded flow. This stress would be smaller than the yield stress of the fully structured material and depend on the scale length defined by the diffusive coefficient.  Indeed, it would smooth the variation of $\lambda$ between its values in the flowing region near the interface (\textit{i. e.} $<1$) and $1$ in the far distance, at the scale of the diffusive length, in the solid-like region.

In the following, we will focus on the case where the diffusive coefficient $\mathcal{D}^{\star}$ is $0$. Knowing that the asymptotic steady state of the transient stage corresponds to the steady state where the stress at the interface $\tau_o=\tau(r_o)=\tau_{ys}$ defined by the equation (\ref{eq:tauymax}), \textit{i. e.} eq. (\ref{eq:tauyo}) with $\lambda_o = 1$, whatever shear banding or not, one can solve directly the steady equations.


\subsection{NUMERICAL METHODS FOR STEADY FLOWS}

 Only the fluid region needs to be considered to solve the steady flow. One uses the same finite difference method for the spatial discretisation as for the unsteady problem. The stress points are taken between two successive velocity points to insure the numerical accuracy of the scheme for the velocity. For the derivative operations, the standard second order centred scheme is used. The numerical method used for the spatial discretisation is quite well established and is similar to the ones used, for instance, in \cite{jenny2007,jenny2012b,pourjafar2015}. Moreover, a validation and a convergence test are performed in the subsection \ref{ssec:valid}.

To calculate the base flow, we consider the steady axisymetric solution of the equations (\ref{eq:ns}), (\ref{eq:div}) and (\ref{eq:lambda}), \textit{i. e.} $\mathbf{v}_b = V_b(r) \mathbf{e}_{\theta}$ and $\lambda = \lambda_b (r)$.  In the fluid domain, \textit{i. e.} $r_i \leq r \leq r_o$, the only non-zero element of the strain rate tensor is $\gap_{r \theta}$. The strain rate does not reach zero in the flowing region and its sign is always negative. Thus, one can write:
\begin{equation}
\gap_b = - \gap_{r \theta,b} = \frac{V_b}{r} - \frac{\partial V_b}{\partial r} \label{eq:grth}
\end{equation}

The only non-zero element of the stress tensor $\dbar{\bm{\tau}}$ is then $\tau_{r \theta}$. Considering the previous assumptions for the flow, the azimuthal component of the equation (\ref{eq:ns}) becomes:

\begin{equation}
\frac{\partial \tau_{r \theta,b}}{\partial r} + \frac{2 \tau_{r \theta,b}}{r} =0, \label{eq:taurth}
\end{equation}

\noindent leading to the well known result for the steady Couette flow:
\begin{equation}
\tau_{r \theta,b} = - \frac{C}{r^2}\, . \label{eq:taub}
\end{equation}
\noindent where the positive constant $C$ is related to the torque imposed by the inner rotating cylinder. The radius $r_o$ can be obtained from the stress condition on the interface between the yielded and unyielded regions:
\begin{equation}
r_o=\sqrt{\frac{C}{\tau_{yo}}}\, . \label{eq:ro}
\end{equation}

If $r_o \geq r_e$ according the equation (\ref{eq:ro}), all the material in the gap flows and $r_o=r_e$. In the next section, the flow curves show that the minimal value $\tau_{min}$ of the stress may be below $\tau_{yo}$. Thus, for $\tau_{min} re^2 \leq C < \tau_{yo} r_e^2$, an alternative to the equation (\ref{eq:ro}) is to set $r_o=r_e$. In practice, it means that if there is no interface at the initial state, the fluid region fits the whole gap for $\tau_{min} re^2 \leq C < \tau_{yo} r_e^2$ and if there is a solid-like region in the initial state, the flowing region is confined between $r_i$ and $r_o<r_e$ according Eq. (\ref{eq:ro}).

To compute the flow velocity in the yielded region, we calculate the strain rate in the yielded region by solving the regular setup of equations at each point of the mesh:
\begin{eqnarray}
\lambda_b &=& \frac{1}{1+(b^{\star}/a^{\star})\gap_b^m} \label{eq:lsteady}\\
\gap_b &=& \left( \frac{(C/r^2-\tau_{yb})(1+\Delta K^{\star} \lambda_{ref})}{1+\Delta K^{\star} \lambda_b} \right)^{1/n_c} \, , \label{eq:gsteady}
\end{eqnarray}
\noindent with $\tau_{yb}$ the yield stress given by eq. (\ref{eq:tauy}) replacing $\lambda$ by $\lambda_b$. Once we obtain the strain rate $\gap_b$ and the structural parameter $\lambda_b$ for a given constant $C$ by solving the setup of equations (\ref{eq:lsteady}--\ref{eq:gsteady}), the linear equation (\ref{eq:grth}) is solved to calculate the fluid velocity with $V_b(r_o)=0$ as boundary condition. Finally, one has to find the value of $C$ such as $V_b(r_i)=1$ using the algorithm available in Matlab to calculate the zero of a real non-linear function. If needed, the pressure $P_b$ of the base flow can be obtained by integrating the equation:

\begin{equation}
\frac{\partial P_b}{\partial r} = \frac{V_b^2}{r}
\end{equation}

\noindent and setting the inner pressure $P_b(r_i)=0$ for instance.




\section{Effect of the thixotropy on the Base flow}


In the next sections, we set the thixotropic index breakdown to $m=1$. It seems reasonable to argue that the structural parameter $\lambda$ modifies the viscous term and the yield stress with the same order of magnitude. Thus, we fix $\Delta K^{\star} = \tau_1^{\star}$ and so $Bn=Bn_0$. As we focus our study on the cases where both a flowing and a solid-like regions exist, \textit{i. e.} $r_o<r_e$, the Bingham number is fixed to $Bn=2$. Finally, for steady state flow only the ratio $b^{\star}/a^{\star}$ appears and we choose to set $a^{\star}=1$ without loss of generality.

\subsection{Steady state flow curves}

The base flow is computed using 50 nodes in the flowing  region of the gap to insure a good accuracy for the linear stability analysis as shown in the validation subsection \ref{ssec:valid}.  

\begin{figure}
\includegraphics[width=0.65\columnwidth]{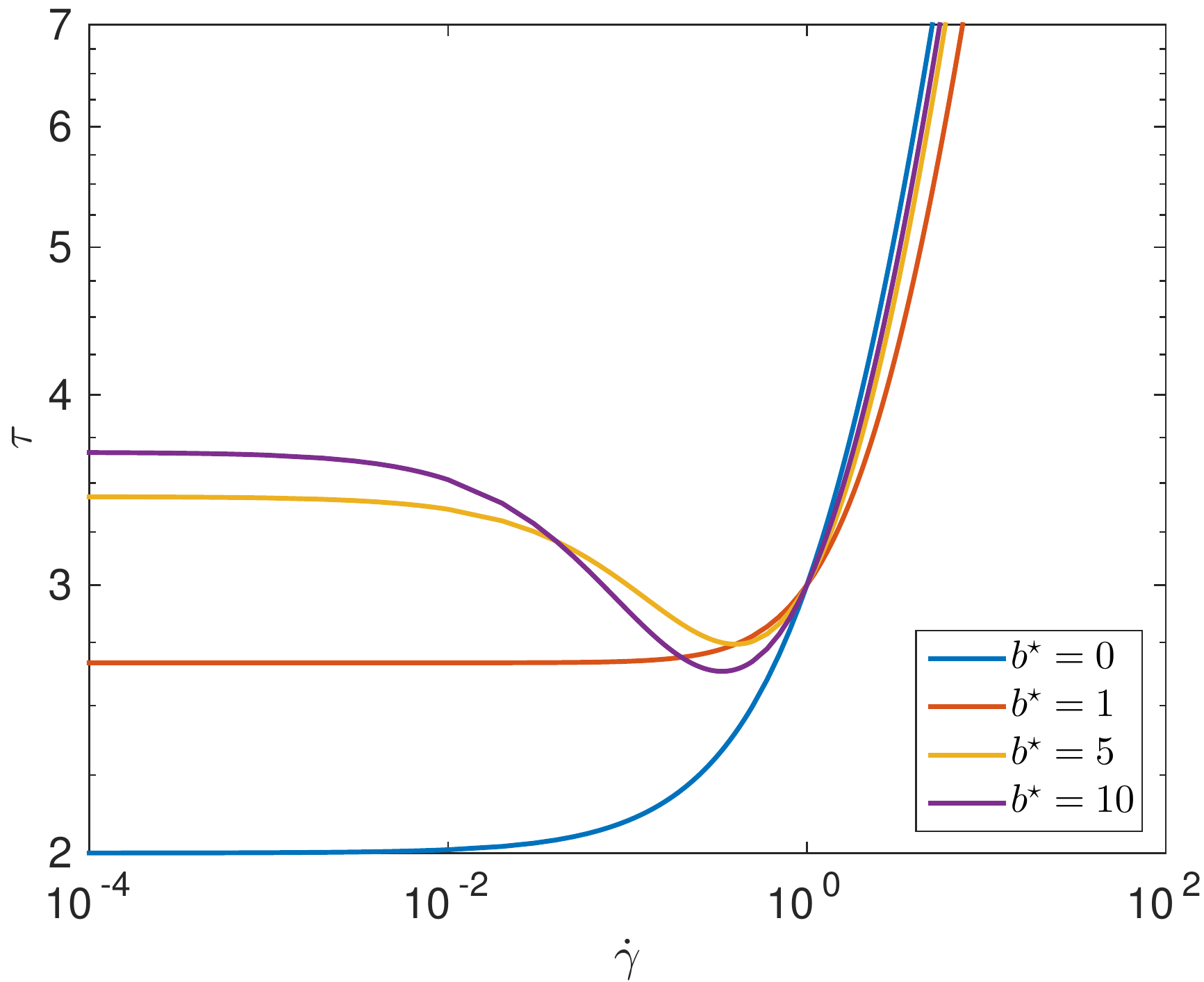}
\caption{Composite flow curve, stress $\tau$ \textit{vs} the strain rate $\gap$ with $Bn=2$, $n_c=1$, $\Delta K^{\star}=1$, $\tau_1^{\star}=1$ and $a^{\star}=1$. \label{fig:taudegamma}}
\end{figure}

We report in Fig.\ref{fig:taudegamma} the evolution of the composite flow curves under controlled shear rate for different values of the breakdown parameter $b^{\star}$. The composite curves are obtained straightfully by replacing the structural parameter $\lambda$ by its relation to $\dot{\gamma}$ (\ref{eq:lsteady}) in the constitutive law (\ref{eq:tau}). As $b^{\star}$ increases, the composite curve drops from a monotonic to a non-monotonic behavior which presents an unstable branch leading to shear banding \citep{moller2008}. This result indicates that, in the range of parameters studied here, Houska's model is able to predict shear banding.

\begin{figure*}
\centering
\begin{tabular}{cc}
(a) & (b) \\
\includegraphics[width=0.48\textwidth]{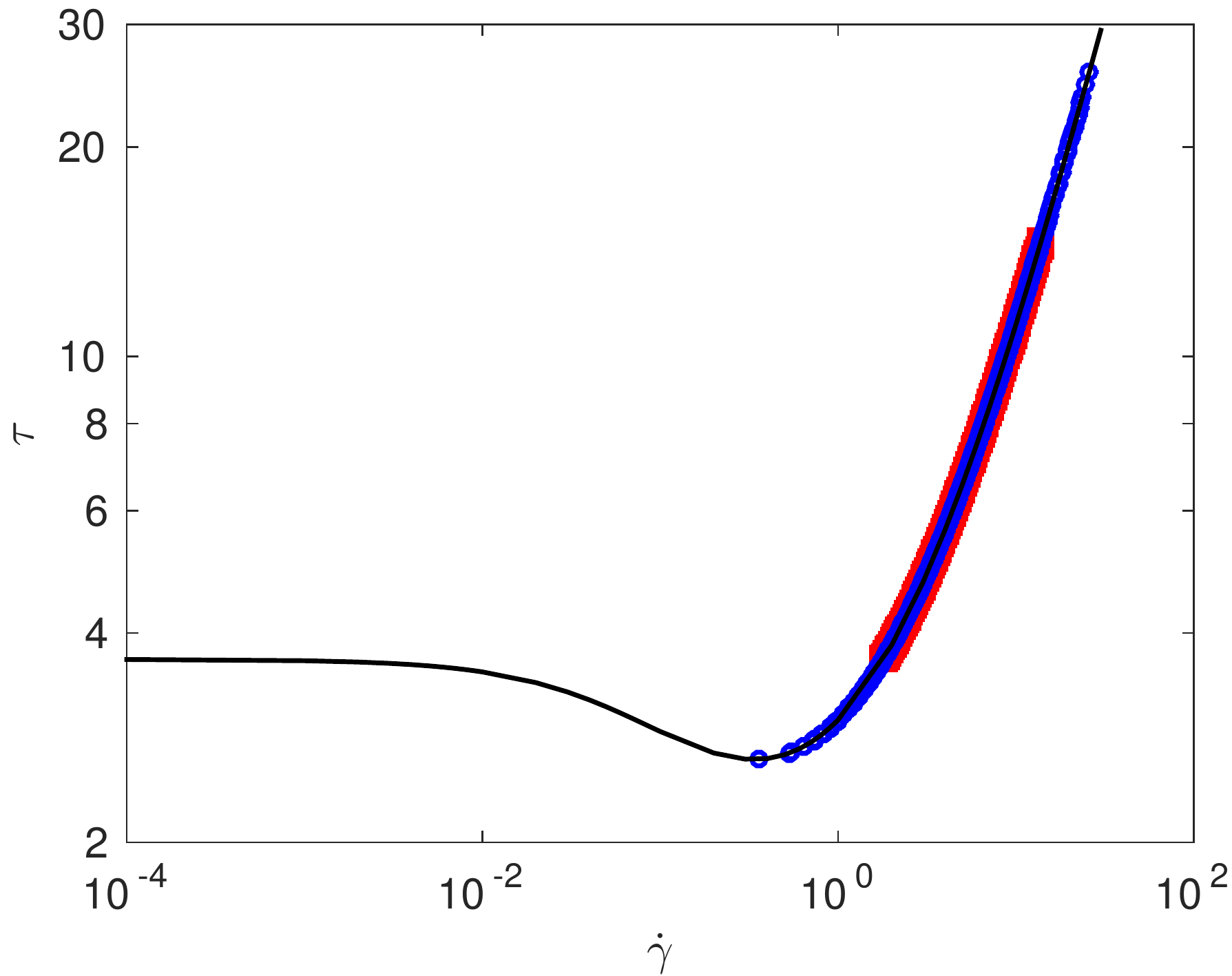} &
\includegraphics[width=0.48\textwidth]{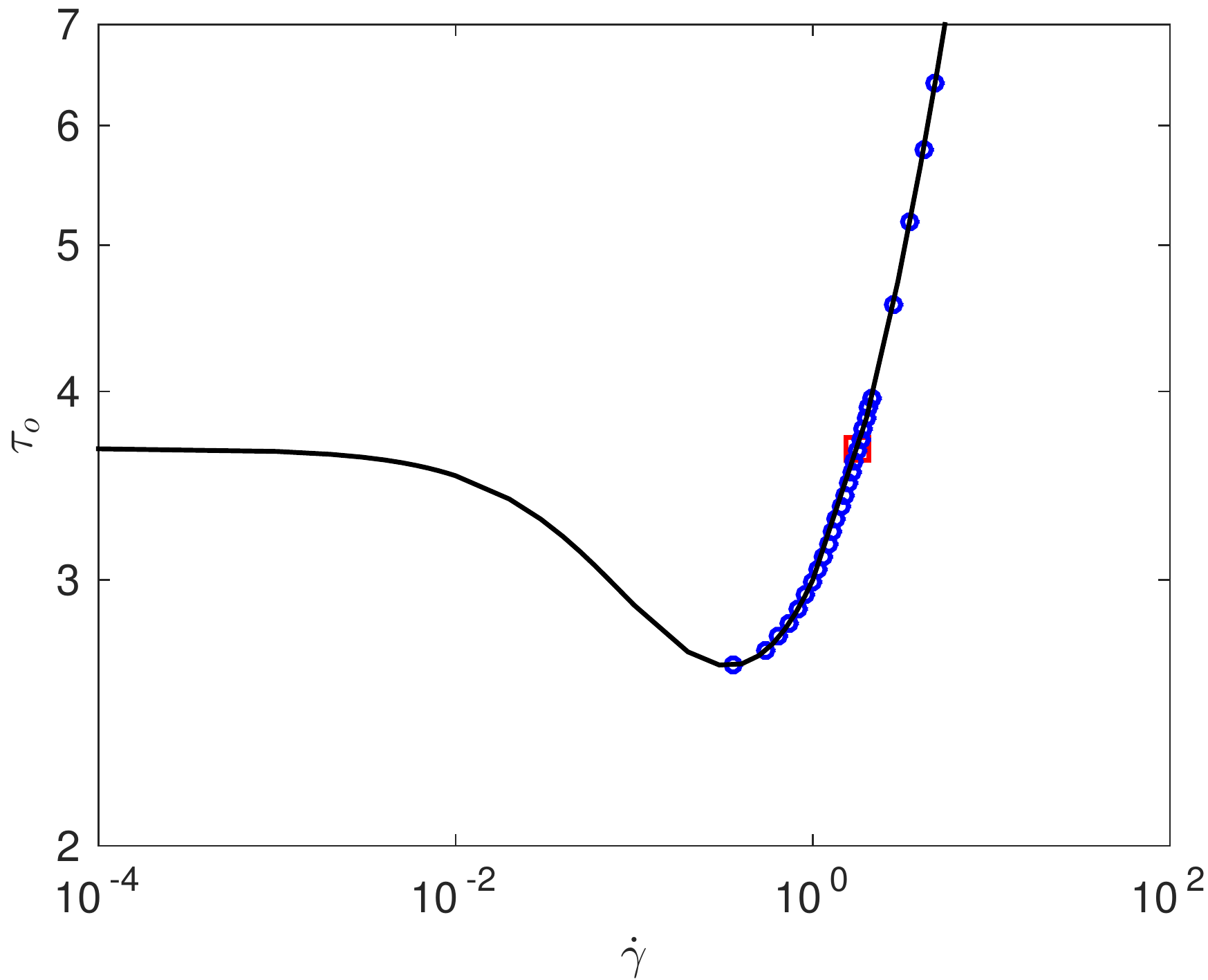}
\end{tabular}
\caption{(a): Local stress $\tau$ \textit{vs} strain rate $\gap$ in the whole fluid region. (b): Local stress $\tau_o$ at the outer radius $r_o$ \textit{vs} the strain rate. The global strain rate $\dot{\gamma}_i = v_i /d$ is imposed whithin the range $0.1 \leq \dot{\gamma}_i \leq 9$. Markers $\circ$ stand for cases where all the gap is flowing, $r_o=r_e$. Markers $\square$ stand for the cases with shear banding, $r_o<r_e$. The full black line is the composite curve. $Bn=2$, $\Delta K^{\star}=1$, $\tau_1^{\star}=1$, $a^{\star}=1$ and $b^{\star}=10$.\label{fig:localstress}}
\end{figure*}

\begin{figure*}
\centering
\begin{tabular}{cc}
(a) & (b) \\
\includegraphics[width=0.48\textwidth]{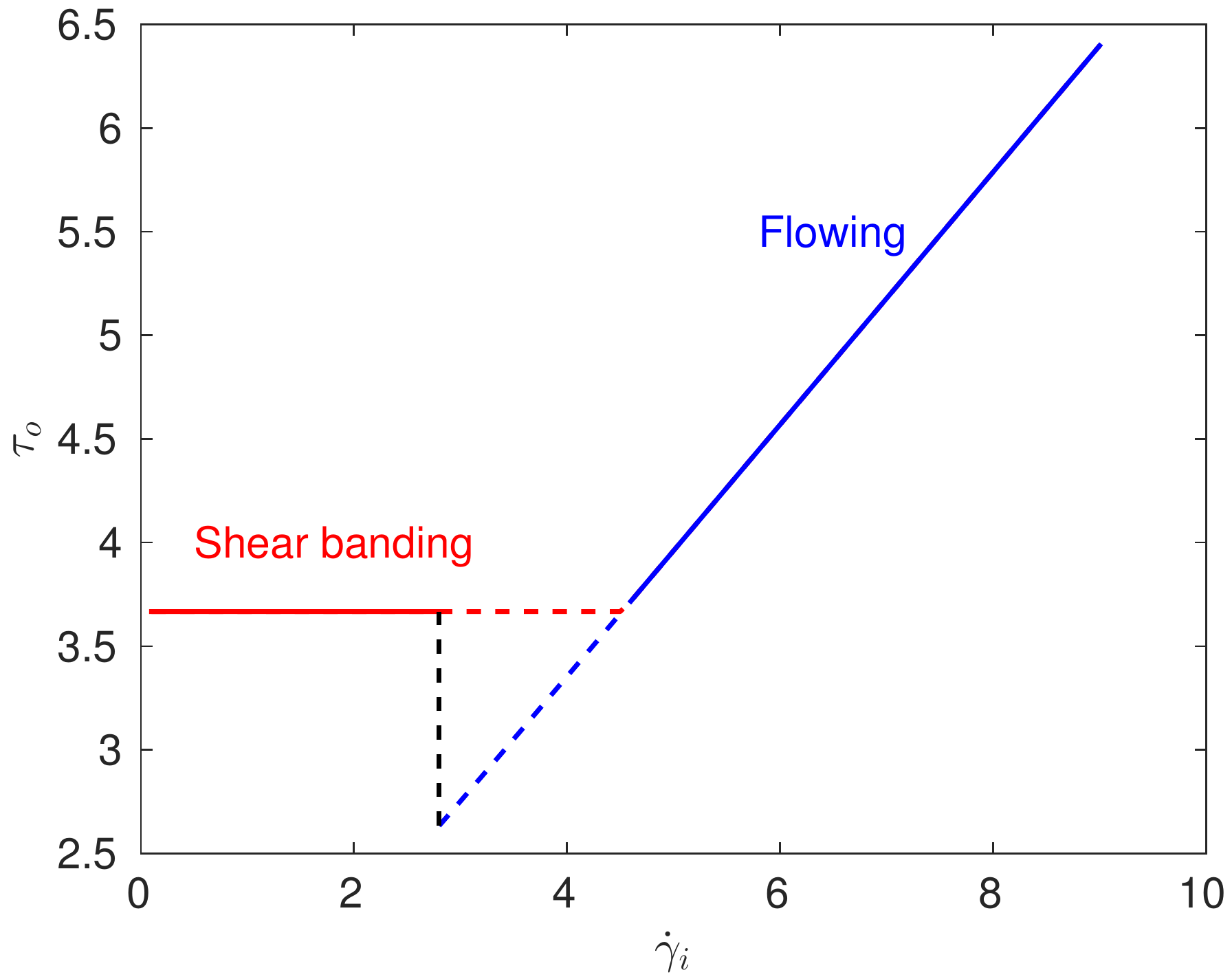} &
\includegraphics[width=0.48\textwidth]{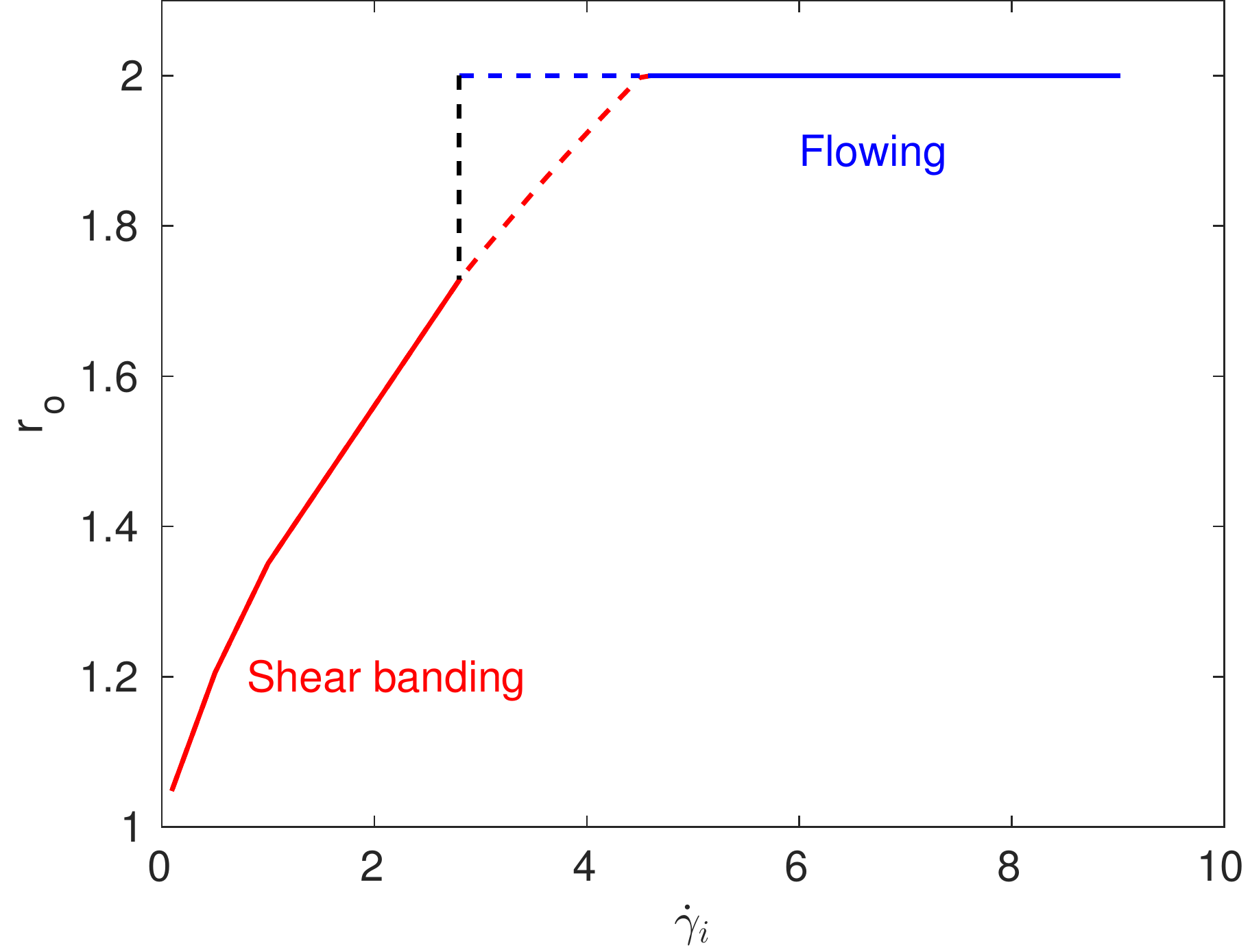}
\end{tabular}
\caption{(a): Local stress $\tau_o$ at the outer radius $r_o$ \textit{vs} the controled strain rate $\dot{\gamma}_i$. (b): Outer radial limit $r_o$ of the fluid region \textit{vs} the global strain rate $\dot{\gamma}_i$. Color dashed-lines stand in the range of $\gap _i$ where either the shear banding (red) or the fully flowing regime (blue) is possible depending on the history of the flow. $Bn=2$, $\Delta K^{\star}=1$, $\tau_1^{\star}=1$, $a^{\star}=1$ and $b^{\star}=10$.\label{fig:rheo}}
\end{figure*}

The existence of a critical shear rate $\gap_c$ in  thixotropic yield stress fluids is often explained in terms of an underlying decreasing branch of the flow curve at low shear rates \citep{olmsted2008}. In such a scenario, the constitutive relation of the material is a decreasing function of the strain rate between $0$ and $\gap_0$ \cite[see][for instance]{divoux2015}. In the case of the shear localization, the constitutive  law is a growing function of $\gap$ and its derivative is strictly positive. Thus, the shear-banding may appear if the sign of the derivative changes at a critical strain rate $\gap_0>0$. In other words, the necessary condition to allow the shear-banded flow is:

\begin{equation}
\exists \gap_0 \geq 0, \text{ such as } \frac{\partial \tau}{\partial \gap} = 0 \label{eq:sbcond} 
\end{equation}

\noindent  Thus, the range of strain rate $[0,\gap_0]$, where $\partial \tau / \partial \gap < 0$, is unstable and the shear-banding allows to avoid the forbidden values of $\gap$. In the steady-state, the derivative of the stress given by the Houska's model is:

\begin{equation}
\frac{\partial \tau}{\partial \gap} = \gap ^{n_c-1} \left( \begin{array}{l}
\frac{n_c (b^{\star} / a^{\star})^2 \gap ^{2 m} + (b^{\star} / a^{\star}) \left( 2 n_c + (n_c-m) \Delta K^{\star} \right) \gap ^{m} + n_c (1+ \Delta K^{\star})}{\left( 1 + \Delta K^{\star} \lambda_{ref} \right) \left( 1+(b^{\star} / a^{\star}) \gap^m \right) ^2} \\
- \frac{m (b^{\star} / a^{\star}) Bn \tau_1^{\star} \gap ^{m-n_c}}{\left( 1 + \tau_1^{\star} \lambda_{ref} \right) \left( 1+(b^{\star} / a^{\star}) \gap^m \right) ^2}
\end{array} \right) \label{eq:dtaudg}
\end{equation}

\noindent When eq. (\ref{eq:sbcond}) admits at least one solution with the function (\ref{eq:dtaudg}), the shear-banding appears if the stress values reach levels that correspond to multiple possible strain rates according the steady constitutive function of the material. For Bingham-like behavior where $m=n_c=1$, eq. (\ref{eq:sbcond}) using the derivative of $\tau$ given by eq. (\ref{eq:dtaudg}) admits only one positive root $\gap_0$:

\begin{equation}
\gap_0 = \frac{a^{\star}}{b^{\star}}\left( \sqrt{Bn \tau_1^{\star} \frac{b^{\star}}{a^{\star}} \left( \frac{1 + \Delta K^{\star} \lambda_{ref}}{1+\tau_1^{\star} \lambda_{ref}} \right) - \Delta K^{\star}} - 1 \right)\, , \label{eq:gapc}
\end{equation}

\noindent if
  
\begin{equation}
1+\Delta K^{\star} - Bn \tau_1^{\star} \frac{b^{\star}}{a^{\star}} \left( \frac{1+\Delta K^{\star} \lambda_{ref}}{1+\tau_1^{\star} \lambda_{ref}} \right) < 0 \label{eq:bc}
\end{equation}

\noindent The derivative of $\tau$ (\ref{eq:dtaudg}) is negative for $\gap \in [0,\gap_0[$. In case of the localisation of the flow in $r \in [r_i,r_o]$, $r_o<r_e$, the boundary condition at $r=r_o$ for the stress is: 

\begin{equation}
\tau(\gap)=\tau_{yo} \, . \label{eq:tauatro}
\end{equation}

\noindent The latter equation (\ref{eq:tauatro}) admits alway $\gap=0$ as zero but if the condition (\ref{eq:bc}) is verified, this zero $\gap=0$ is within the forbidden range for $\gap$, \textit{i. e.} $[0,\gap_0[$, and thus, we observe the shear-banding. The second root of eq. (\ref{eq:tauatro}) is 

\begin{equation}
\gap_c=Bn \tau_1^{\star} \left( \frac{1 + \Delta K^{\star} \lambda_{ref}}{1+\tau_1^{\star} \lambda_{ref}} \right) - \frac{1+\Delta K^{\star}}{b^{\star} / a^{\star}} \, . \label{eq:gapo}
\end{equation}

\begin{figure}
\centering
\includegraphics[width=0.68\textwidth]{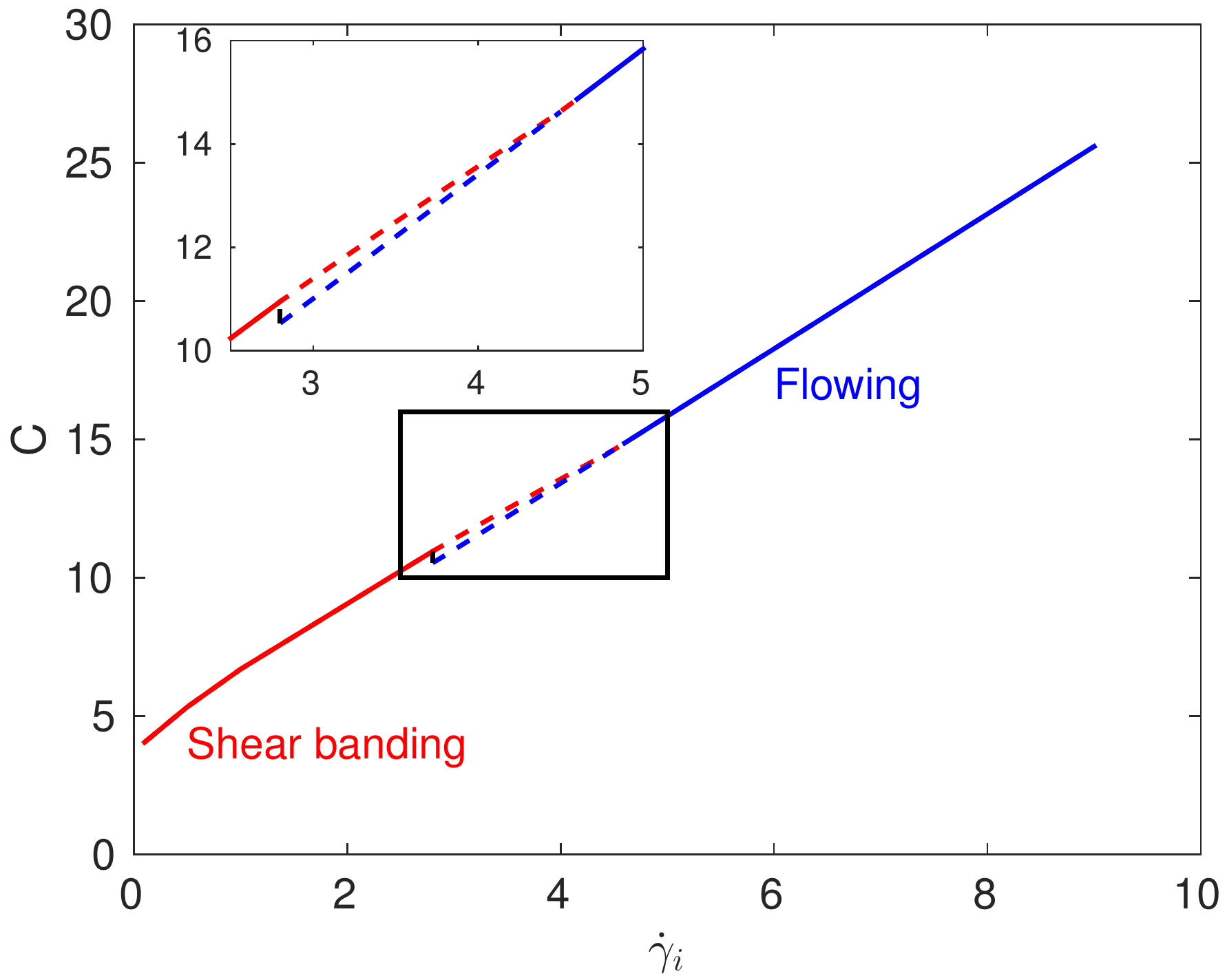}
\caption{Reduced torque $C$ \textit{vs} the global strain rate $\dot{\gamma}_i$. Color dashed-lines stand in the range of $\gap _i$ where either the shear banding (red) or the fully flowing regime (blue) are possible depending on the history of the flow. $Bn=2$, $\Delta K^{\star}=1$, $\tau_1^{\star}=1$, $a^{\star}=1$ and $b^{\star}=10$.\label{fig:torque}}
\end{figure}

\noindent The figure \ref{fig:taudegamma} and eqs. (\ref{eq:gapc}) and (\ref{eq:gapo}) show that $\gap_c \geq \gap_0$ . The figures \ref{fig:localstress}a-b shows the local stress-strain in several cases under controlled global strain rate $\dot{\gamma}_i = v_i / d$ for $b^{\star}=10$. To control the strain rate in practice, the non-dimensional parameters $a^{\star}$, $b^{\star}$ and $Bn$ are re-calculated with the equations (\ref{eq:ab} and \ref{eq:Bn}) using the chosen reference values of the parameters and the aimed values of $v_i$. Then, the non-dimensionalized velocity is multiplied by the value for the inner cylinder in order to compare the flows obtained at different imposed strain rates $\gap_i$. As expected, in shear banded flows, the minimum value of the local strain rate is $\gap(r_o)=\gap_c$. Of course, the outer radius of the fluid region is such as $r_o<r_e$. The minimum value of the local stress is $\tau_o=\tau(r_o)=\tau_{yo}$ as long as $r_o<r_e$. In fact, $r_o$ is a growing function of the controlled strain rate $\dot{\gamma}_i$ as shown in the figure \ref{fig:rheo}b. The lower branch of the composite flow curve corresponds to local stress-strain curve if there is no shear-banding, \textit{i. e.} when $r_o=r_e$ (blue dashed-line in the figures \ref{fig:rheo}). Nevertheless, when the controlled strain rate $\dot{\gamma}_i$ is too low, under the vertical black dashed-line in the figures \ref{fig:rheo}, the local strain rate at $r_e$ falls down zero because the local stress grows. The interface between the fluid and the solid-like region appears. The transient flow ends at a shear banded flow, the stress at the interface is $\tau_{yo}$. Thus, there is a range of controlled strain rates corresponding to the range of local strain rate at $r_e$ $ [ \dot{\gamma}_0,\, \leq \dot{\gamma}_c [$ where the steady state depends on the existence of the interface in the initial state, \textit{i. e.} if the strain rate decreases from the flowing state or grows from a shear-banded state. One can notice that the torque jumps up to the shear-banding value at the end of the flowing state (fig. \ref{fig:torque}). Under controlled torque, the flow may freeze in a Couette with a small gap when the radii ratio $\eta>\sqrt{\tau_{min} / \tau_{yo}}$ and $\dot{\gamma}_o<\dot{\gamma}_0$. This result is consistent with the multiple yield stresses observed by Ovarlez \textit{et al.} \cite{ovarlez2009}.

One can notice that if $n_c<1$ and $m\geq 1$,

\begin{equation}
\frac{\partial \tau}{ \partial \gap}_{\gap \mapsto 0^+} \rightarrow + \infty \label{eq:dtauncneq1}
\end{equation}

\noindent and thus there is a  range of strain rate values close to zero where $\tau(\gap)$ is a growing function of $\gap$ even if there is a  range of positive strain rates where the stress $\tau$ decreases when $\gap$ grows. This case would be similar to the flow curve of a semidilute wormlike micelle solution with a yield stress like in the Fig. 1b of \cite{perge2014}. As there is no diffusion term in our set of equations, sharp discontinuities of the strain rate and the structural parameter can appear within the fluid region at a radius $r_i<r<r_o$ when $n_c<1$. Our numerical method does not allow such discontinuous fields in the fluid domain except at the interface between the fluid and solid-like region, \textit{i. e.} at $r=r_o$. Thus, in the following, we limit our parametric study to shear banded flows with a banding interface between the flowing and static regions only (flow curve like in Fig. 1c of \cite{perge2014}), \textit{i. e.} with $n_c=1$ and shear localization with $n_c \leq 1$.

\subsection{Velocity profiles and structure parameter}

\begin{figure}
\centering
\begin{tabular}{cc}
(a) & (b) \\
\includegraphics[width=0.49\textwidth]{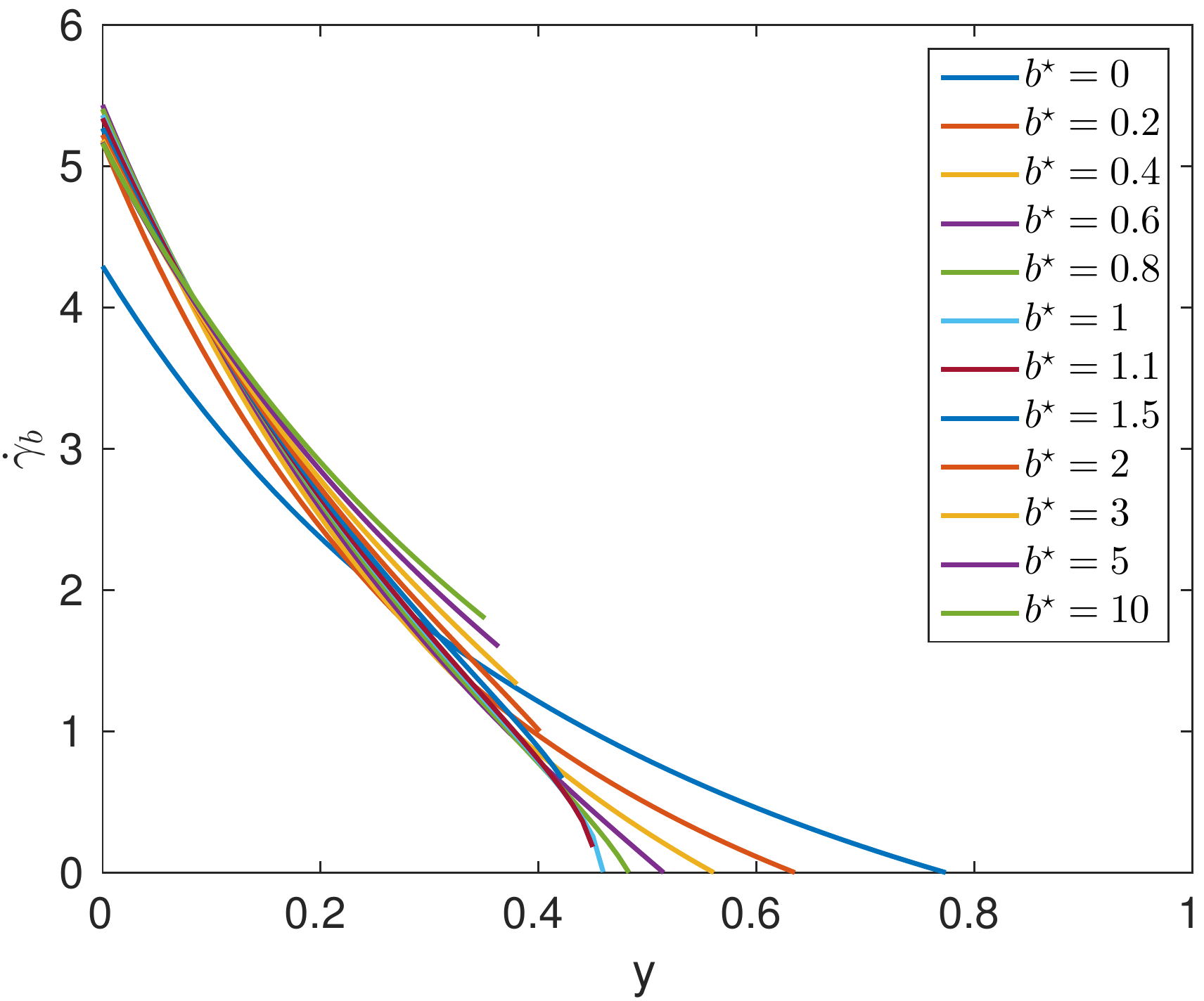} &
\includegraphics[width=0.49\textwidth]{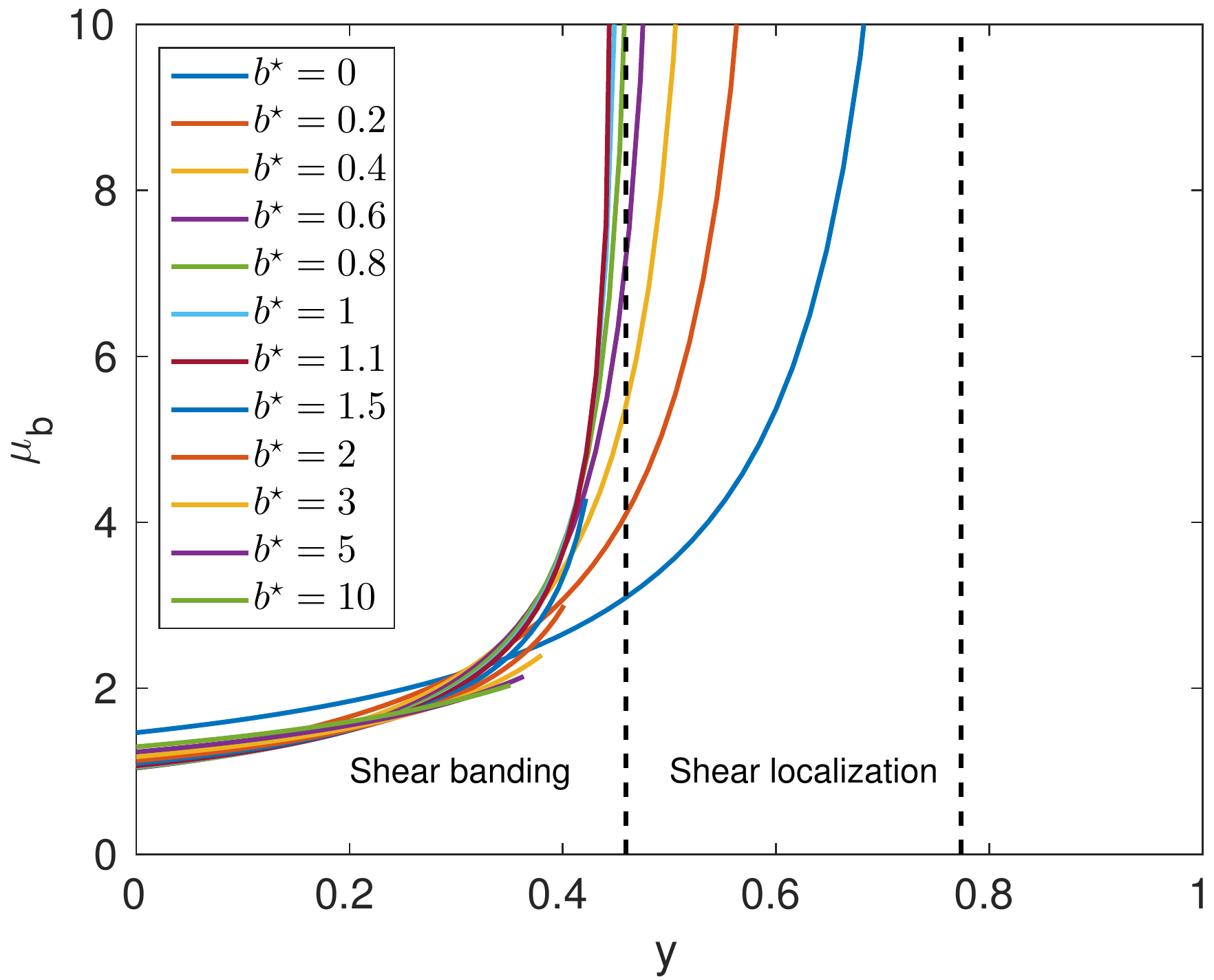} \\
(c) & (d) \\
\includegraphics[width=0.49\textwidth]{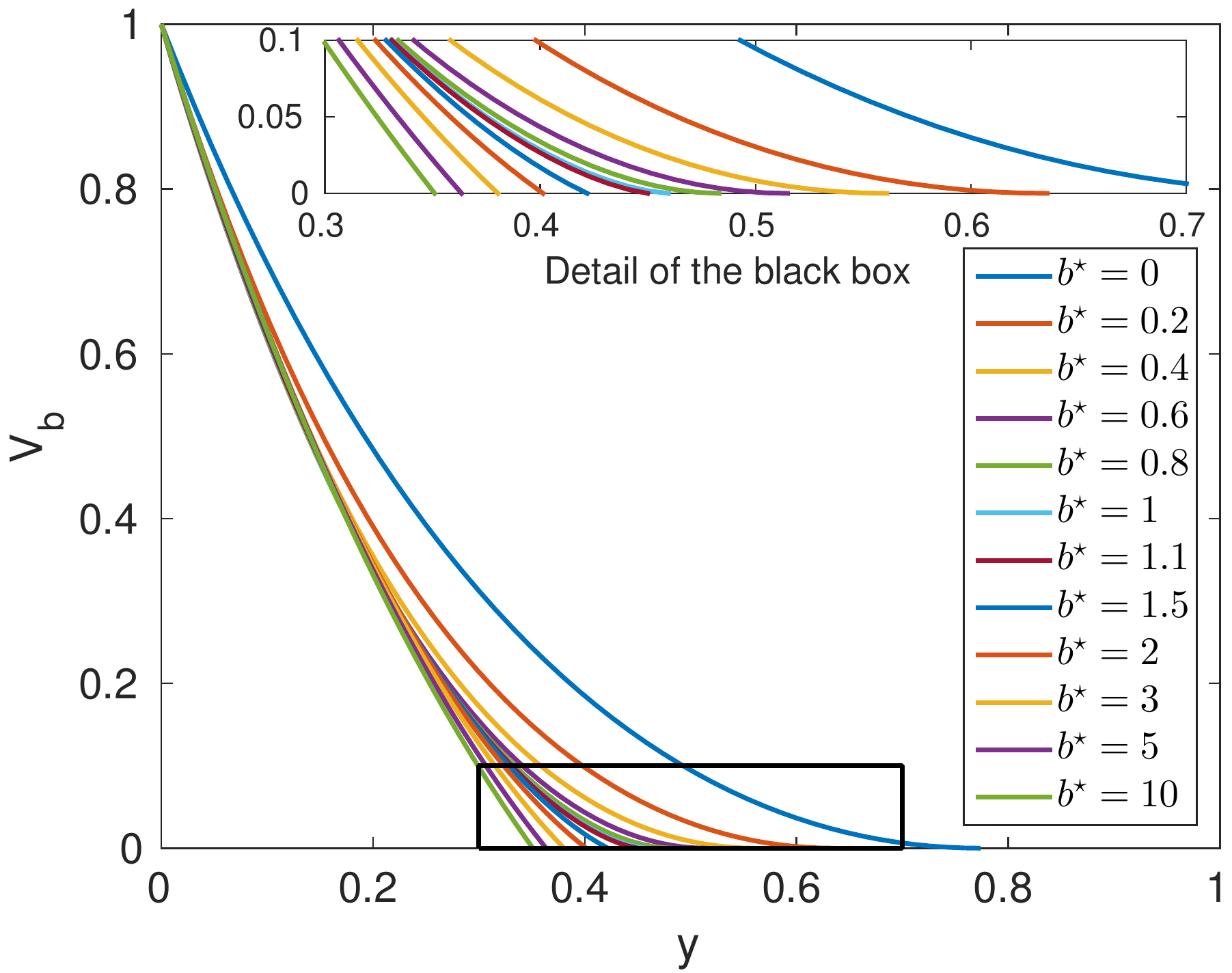} &
\includegraphics[width=0.49\textwidth]{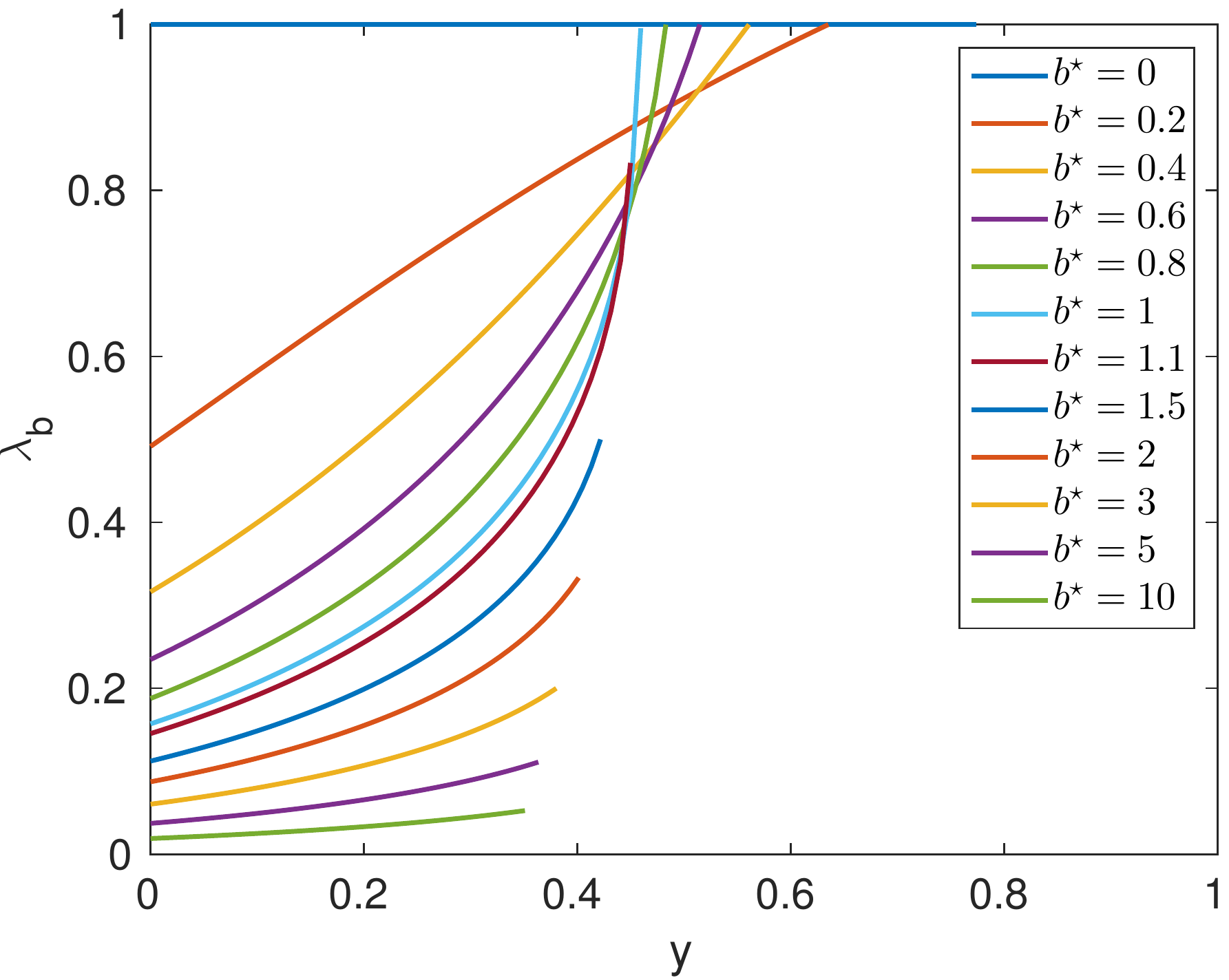}
\end{tabular}
\caption{Strain rate $\dot{\gamma}_b$ (a), viscosity $\mu_b$ (b), azimutal velocity $V_b$ (c) and structural parameter $\lambda_b$ (d) of the base flow \textit{vs} the reduced gap position $y=(r-r_i)/(r_e-r_i)$ with $Bn=2$, $n_c=1$, $\Delta K^{\star}=1$, $\tau_1^{\star}=1$ and $a^{\star}=1$ for a large gap $\eta=0.5$. \label{fig:vlb}}
\end{figure}

The strain rate, the viscosity the velocity and the structure parameter profiles are shown in Fig. \ref{fig:vlb}. We see that whatever the value of $b^{\star}$, a flowing and a static region coexists with the considered value of Bingham number $Bn=2$. However, the discontinuity of the strain rate profile depends on the latter parameter and is related to a discontinuity of the structure parameter. For lower values of $b^{\star}$, the velocity profiles as the structure parameter is continuous as observed for example in carbopol gels, emulsions and foams  \citep{ovarlez2009,ovarlez2013}. In that case, the shear localization is inherent to the existence of the yield stress and no shear banding is observed.
By contrast, for larger values of $b^{\star}$, the shear rate becomes non-zero at the outer boundary of the flowing  region and the structural parameter $\lambda$ does not reach $1$ as it stands in the solid-like region when the breakdown parameter is greater than a critical value $b_c^{\star}=1$ (figs. \ref{fig:vlb}). Discontinuous strain rate profiles between a static and a flowing region have been observed using MRI measurements in cement pastes \citep{jarny2005} and bentonite suspensions \citep{ovarlez2013}. It corresponds to a steady-state shear-banded velocity profile where the shear rate is equal to a critical shear rate in the liquid region and is equal to zero in the solid region.

\begin{figure}
\centering
\begin{tabular}{cc}
(a) & (b) \\
\includegraphics[width=0.49\textwidth]{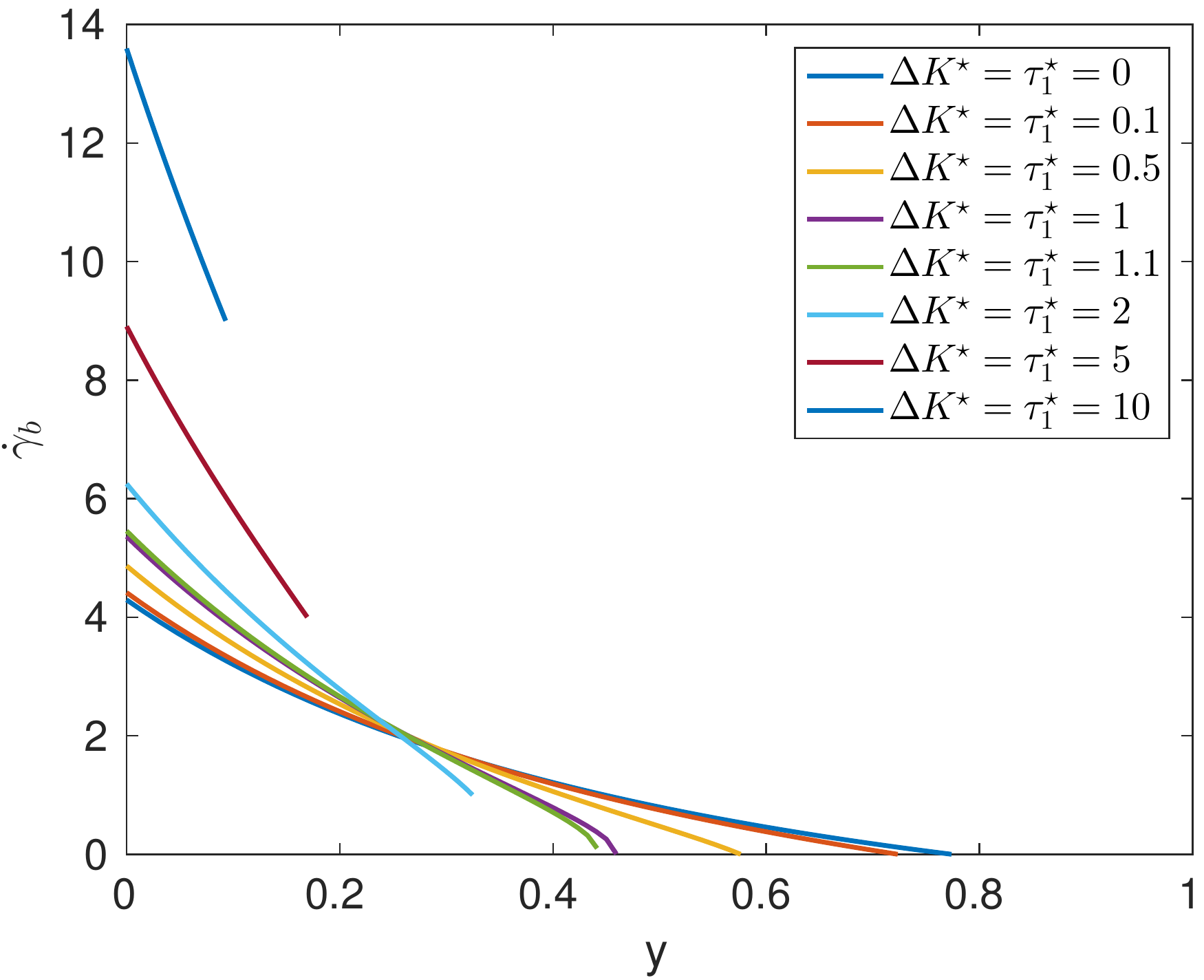} &
\includegraphics[width=0.49\textwidth]{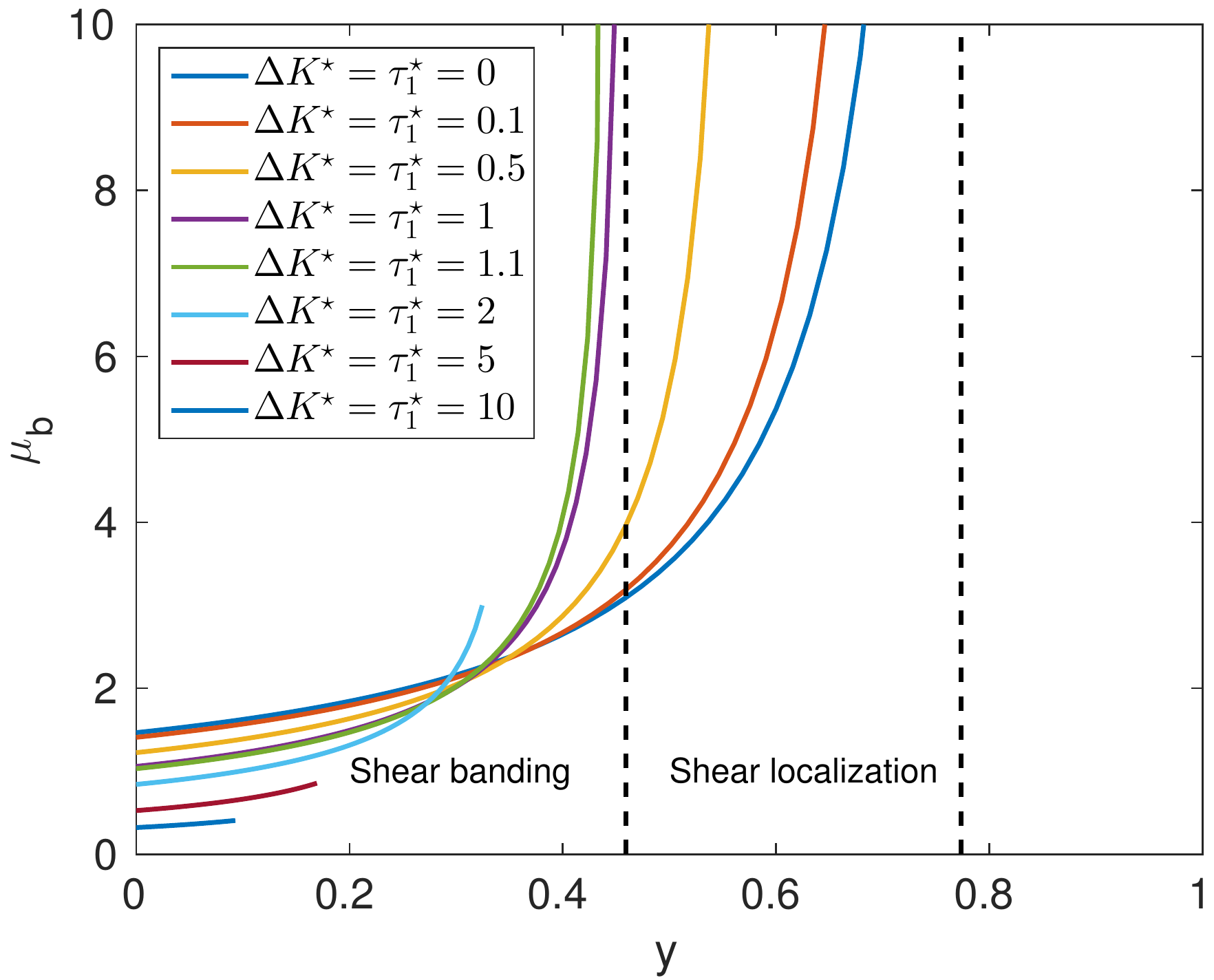} \\
(c) & (d) \\
\includegraphics[width=0.49\textwidth]{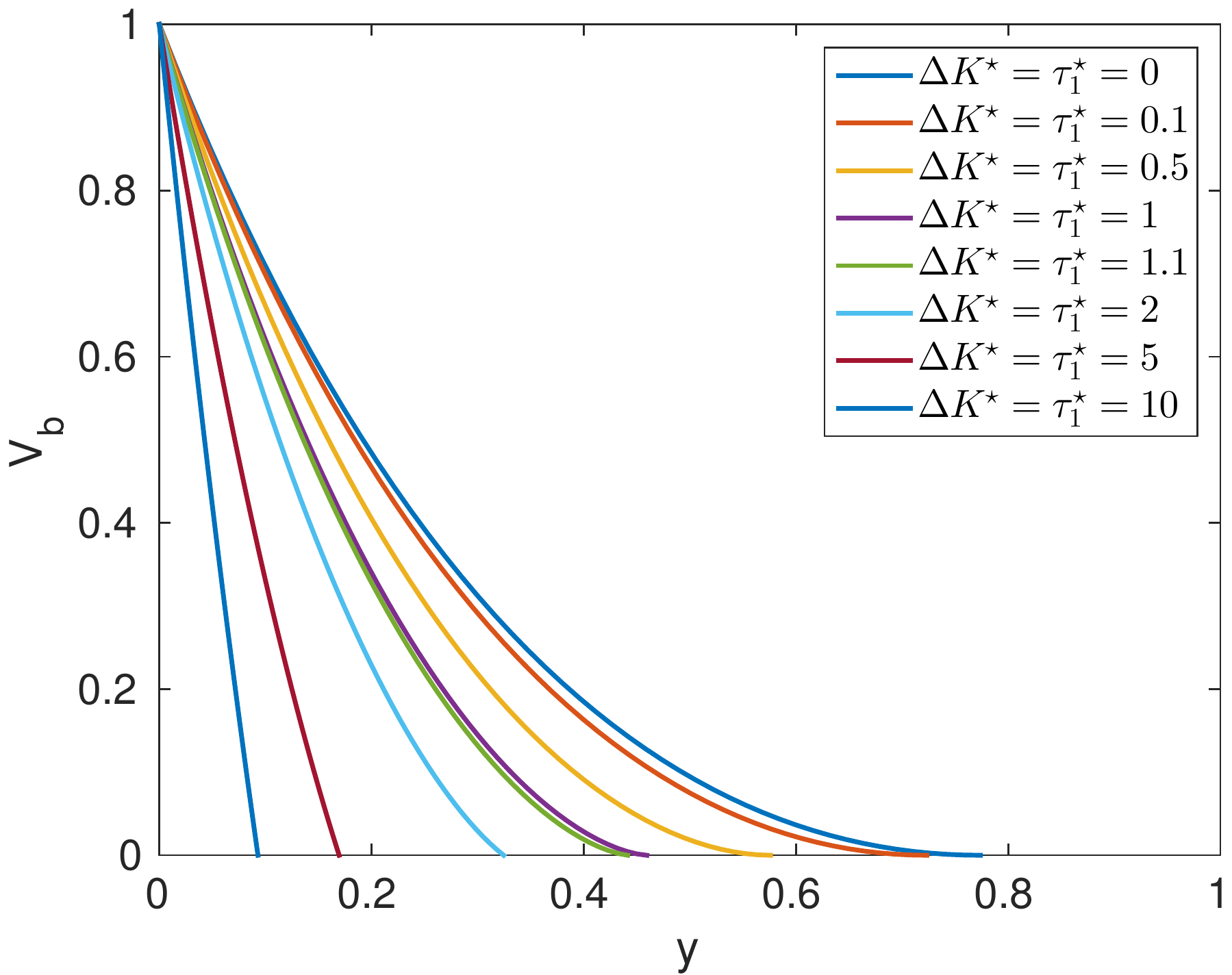} &
\includegraphics[width=0.49\textwidth]{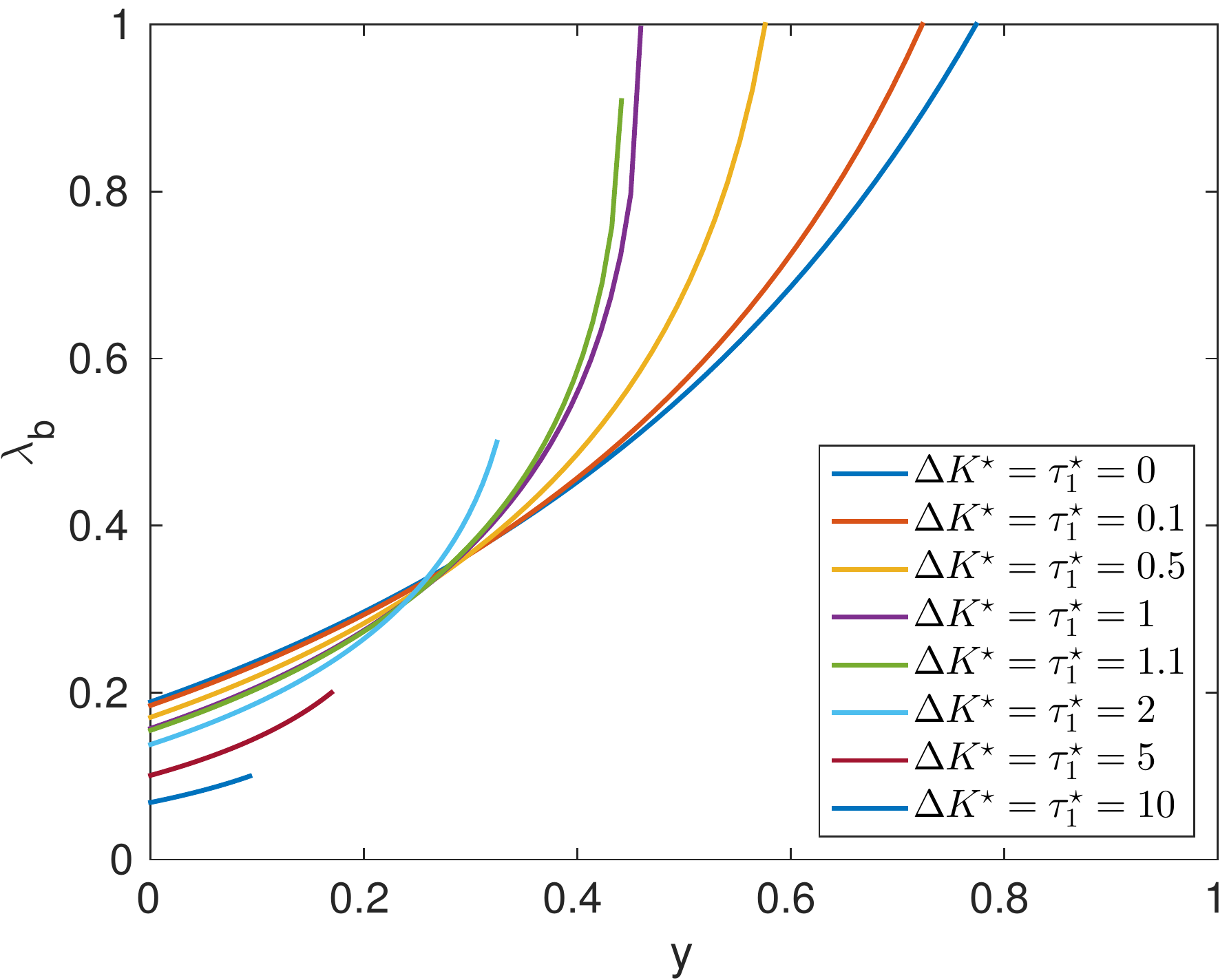}
\end{tabular}
\caption{Strain rate $\dot{\gamma}_b$ (a), viscosity $\mu_b$ (b), azimutal velocity $V_b$ (c) and structural parameter $\lambda_b$ (d) of the base flow \textit{vs} the reduced gap position $y=(r-r_i)/(r_e-r_i)$ with $Bn=2$, $n_c=1$, $a^{\star}=1$ and $b^{\star}=1$ for a large gap $\eta=0.5$. \label{fig:vldktau1}}
\end{figure}

According eq. (\ref{eq:bc}) and figs. \ref{fig:vldktau1}, increasing the parameter $\tau_1^{\star}$ may produce the same effect as increasing $b^{\star}/a^{\star}$. The steady state flow is controlled by the competition between the restructuring and the breakdown effects. The more the structure close to the interface is broken efficiently by the strain rate (fig. \ref{fig:vlb}d and \ref{fig:vldktau1}d), the more the viscosity drops significantly and rapidly. A band in terms of structure takes place next to the unyielded region (fig. \ref{fig:vlb}b and \ref{fig:vldktau1}b). Then, the transition from shear localization to shear banding is feasible.

\begin{figure}
\centering
\begin{tabular}{cc}
(a) & (b) \\
\includegraphics[width=0.49\textwidth]{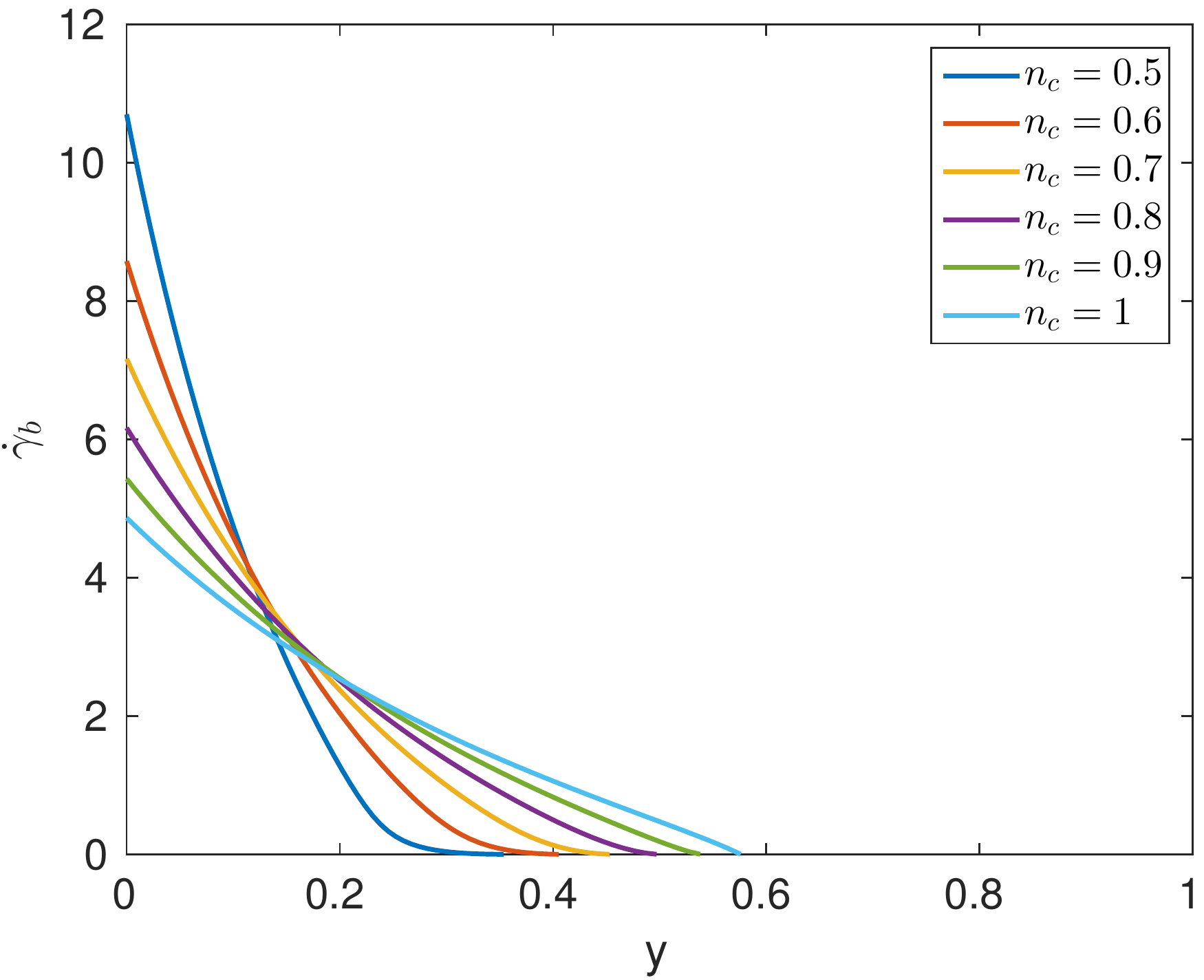} &
\includegraphics[width=0.49\textwidth]{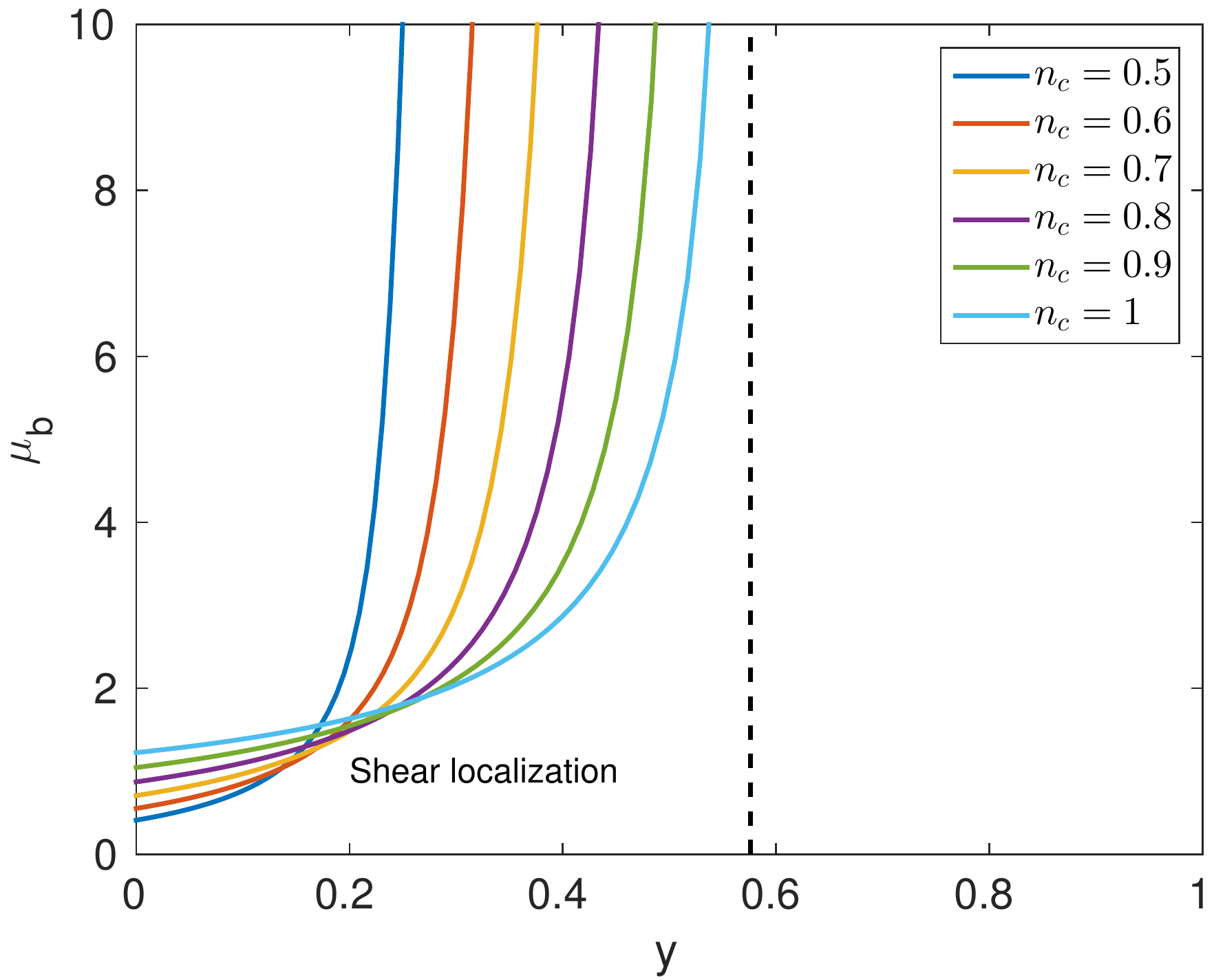} \\
(b) & (c) \\
\includegraphics[width=0.49\textwidth]{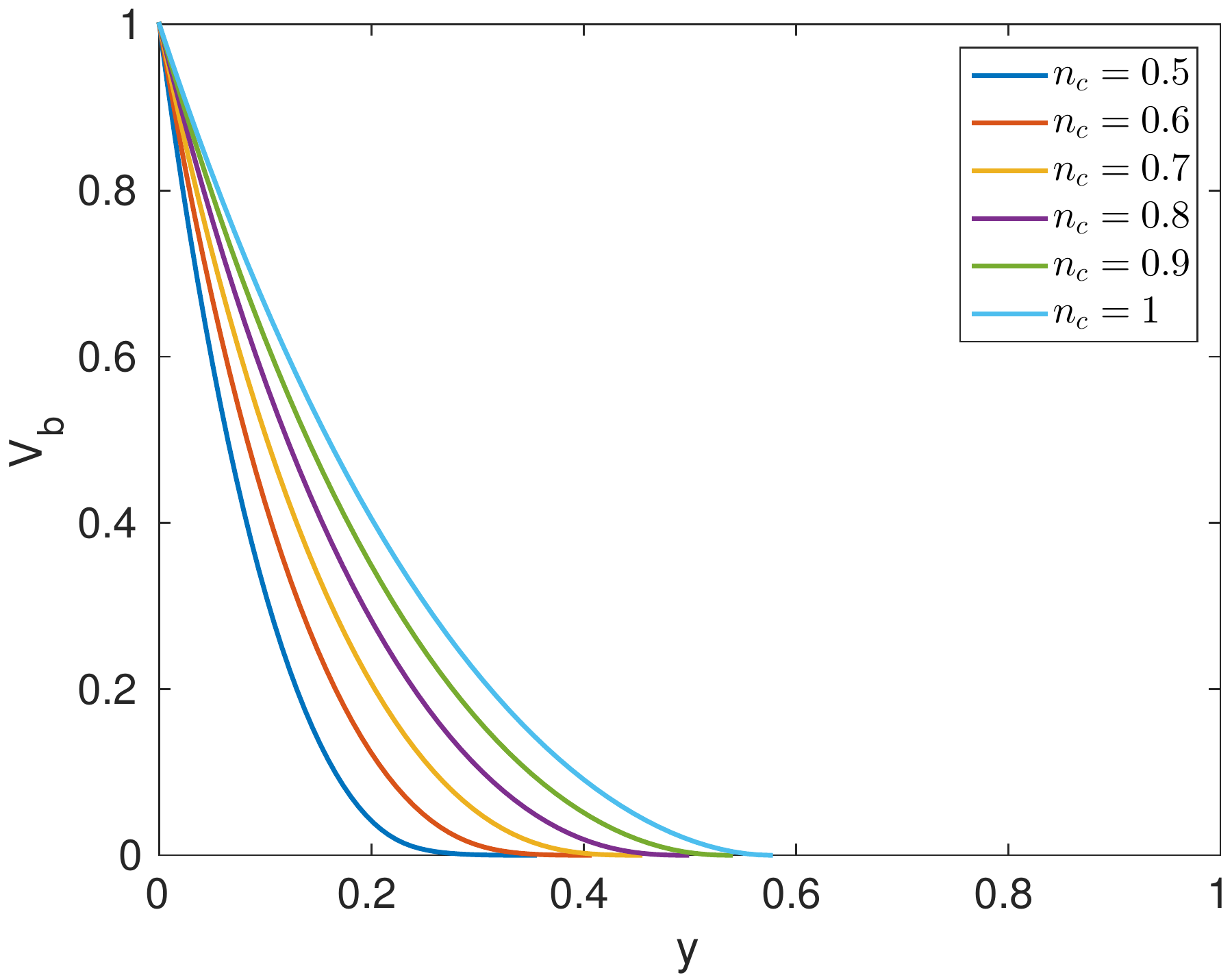} &
\includegraphics[width=0.49\textwidth]{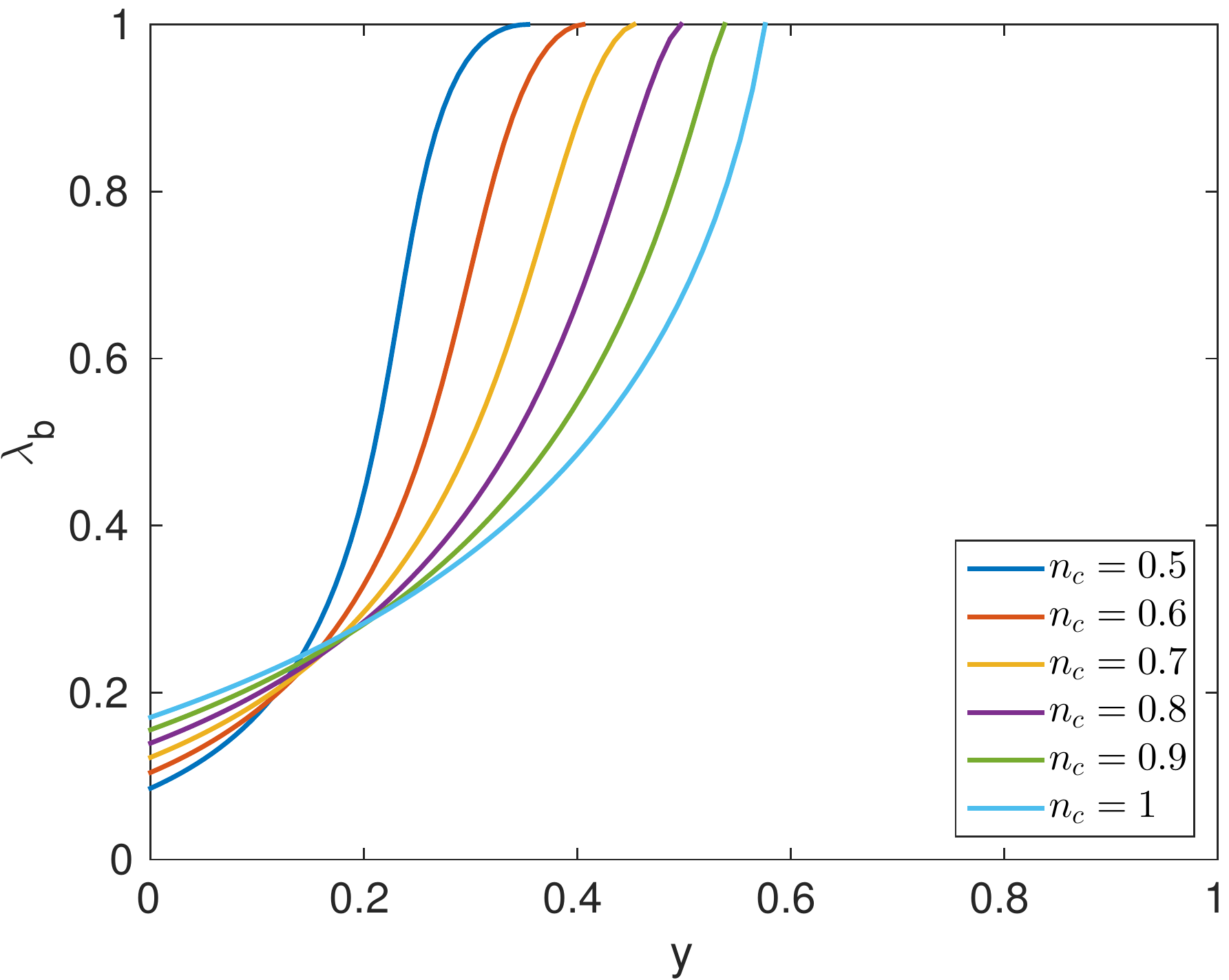}
\end{tabular}
\caption{Strain rate $\dot{\gamma}_b$ (a), viscosity $\mu_b$ (b), azimutal velocity $V_b$ (c) and structural parameter $\lambda_b$ (d) of the base flow \textit{vs} the reduced gap position $y=(r-r_i)/(r_e-r_i)$ with $Bn=2$, $\Delta K^{\star}=0.5$, $\tau_1^{\star}=0.5$, $a^{\star}=1$ and $b^{\star}=1$ for a large gap $\eta=0.5$. \label{fig:vlnc}}
\end{figure}

At last, we explore the effect of the shear-thinning index $n_c$ . When $n_c<1$, the simple shear-banded flows as observed in figs. \ref{fig:vlb} or \ref{fig:vldktau1} are not observed  because the constitutive relation of the material is always a growing function for $\gap$ sufficiently close to zero (eq. \ref{eq:dtauncneq1}),  allowing small values for $\gap$ in the flow . In that case the flow is always shear-localized. This contrast with the previous cases discussed above ($n_c=1$) where small values of $\gap$ fall in the unstable branch of the flow curve and then lead to shear banded flows. In order to keep only one flowing zone in the explored range of $n_c$, we set $\Delta K^{\star}= \tau_1^{\star}=0.5$. Staying on the continuous solution for the structural parameter $\lambda$, the inner strain rate increases with the shear-thinning index $n_c$ and the structural parameter $\lambda$ reaches smoothly the fully structured state of the solid-like zone because of the growing viscosity near $y_o$ (figs. \ref{fig:vlnc}). As it would be expected from the velocity profiles obtained with Carreau fluids by Alibenyahia \textit{et al.} \citep{alibenyahia2012}, the flow is confined close to the inner cylinder when the shear-thinning index $n_c$ decreases. It confirms that the shear-thinning behaviour confined the flow in the inner region of the gap where the viscosity is lower. 

\subsection{Interface between the static and the flowing regions}

\begin{figure}
\centering
\begin{tabular}{cc}
(a) & (b) \\
\includegraphics[width=0.49\textwidth]{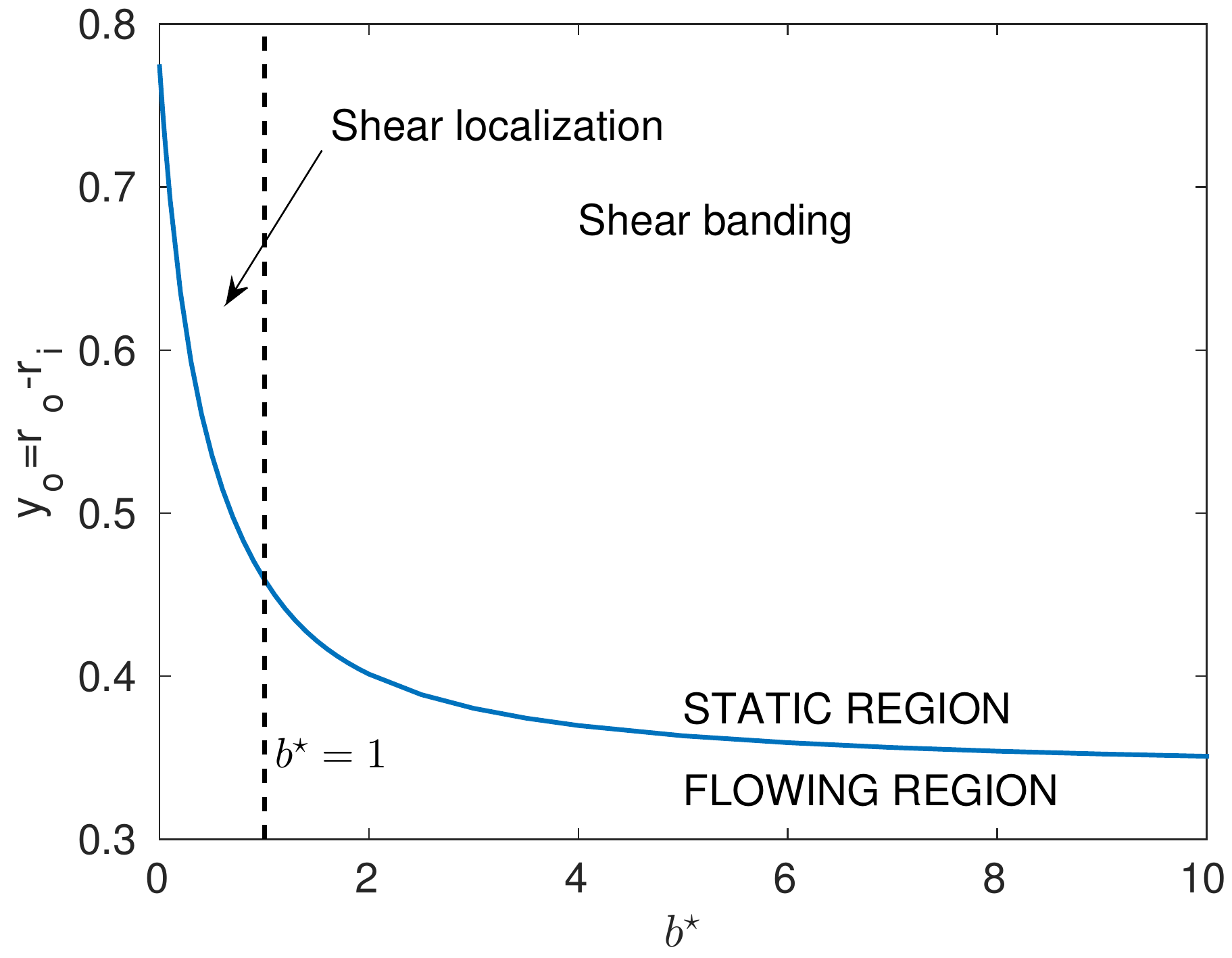} &
\includegraphics[width=0.49\textwidth]{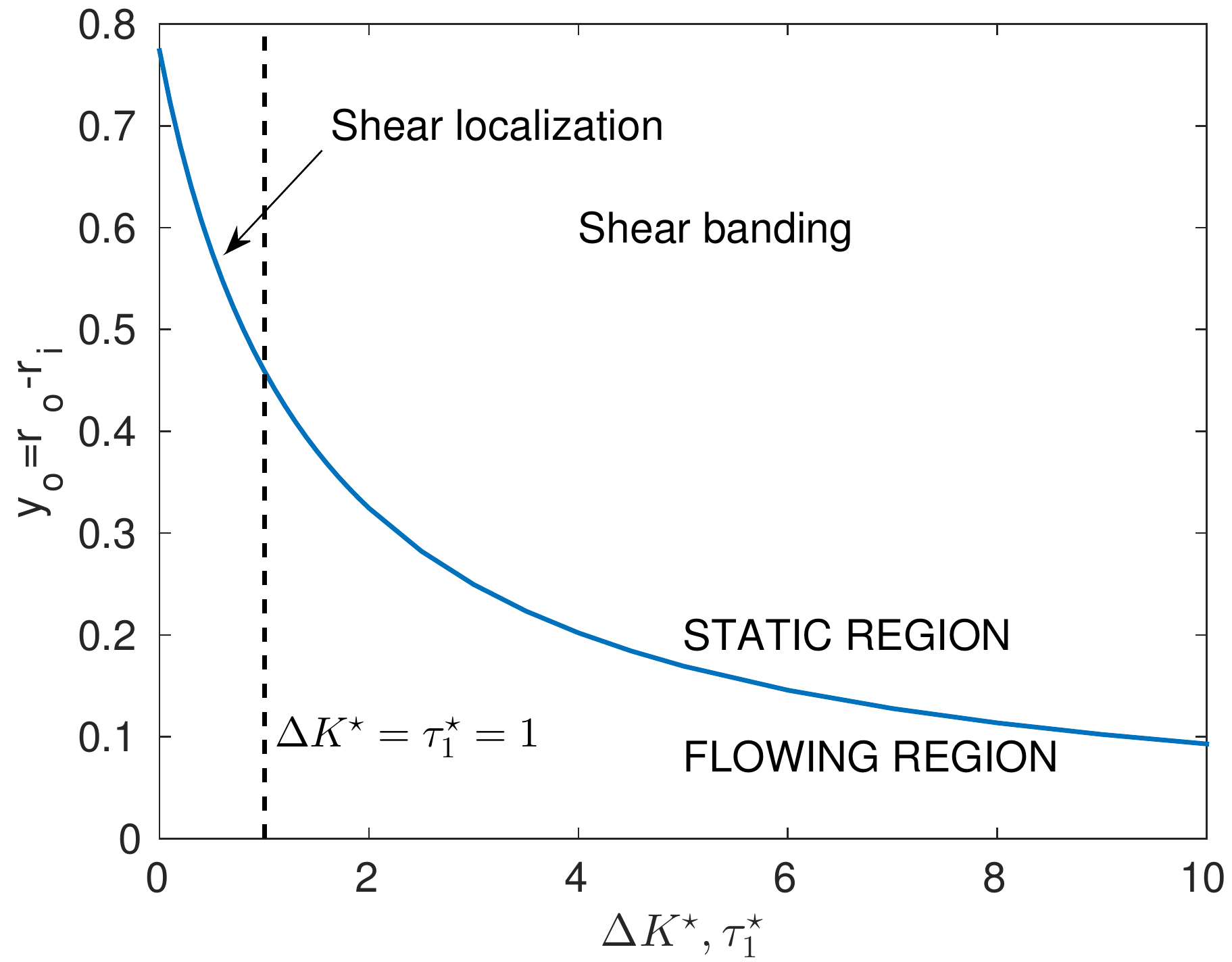}
\end{tabular}\\
(c) \\
\includegraphics[width=0.49\textwidth]{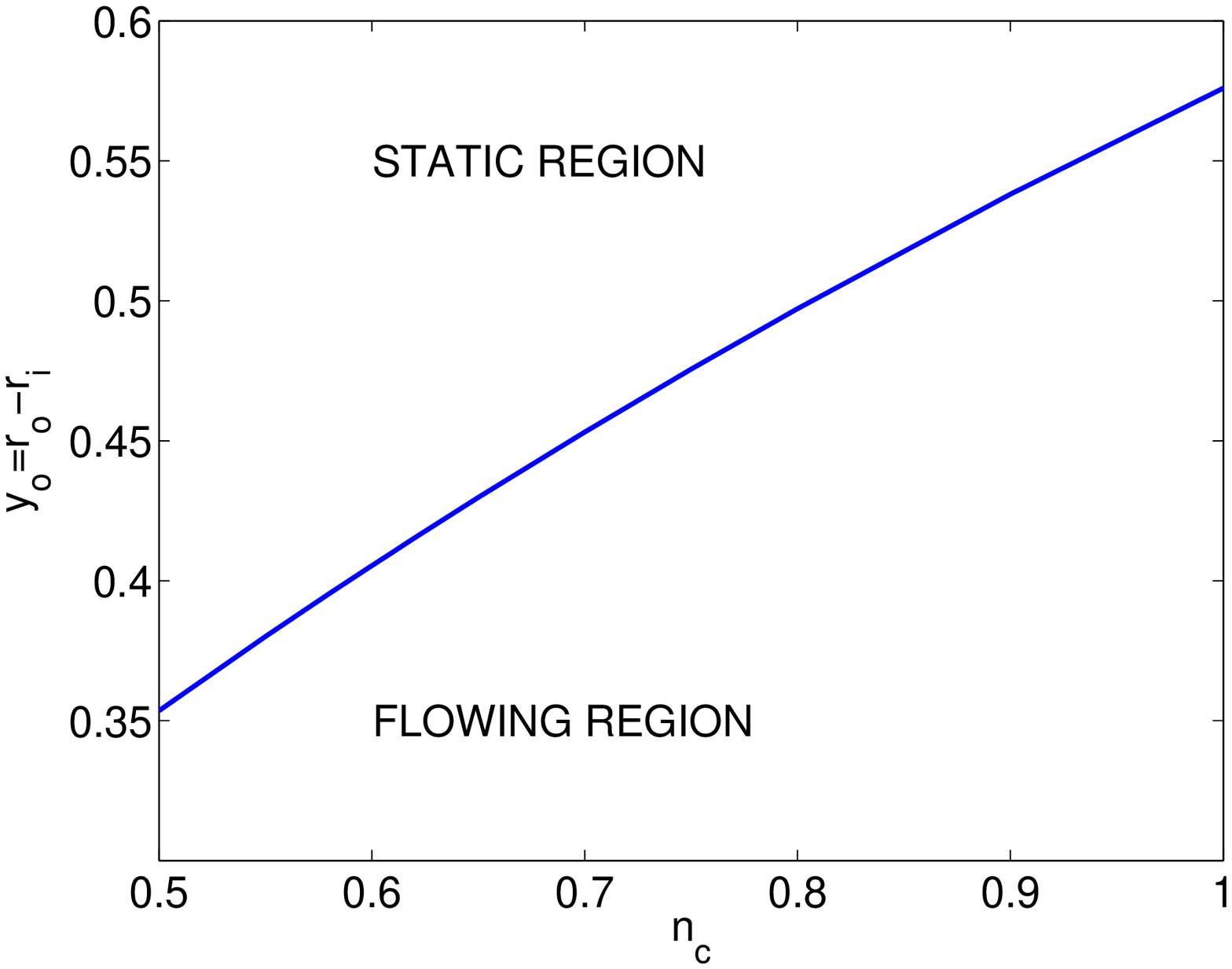}
\caption{ (a) Reduced position of the interface between flowing  and static  regions $y_o=(r_o-r_i)/(r_e-r_i)$ with $Bn=2$, $n_c=1$, $\Delta K^{\star}=1$, $\tau_1^{\star}=1$ and $a^{\star}=1$ for a large gap $\eta=0.5$. Vertical dashed-line stands for the critical value of $b^{\star}=1$ where the strain rate at the interface $\gap_o$ becomes non-zero. (b) Reduced position of the interface between flowing  and static  regions $y_o=(r_o-r_i)/(r_e-r_i)$ with $Bn=2$, $n_c=1$, $a^{\star}=1$ and $b^{\star}=1$  for a large gap $\eta=0.5$. Vertical dashed-line stands for the critical value of $\Delta K^{\star}=1$ (or $\tau_1^{\star}$) where the strain rate at the interface $\gap_o$ becomes non-zero. (c) Reduced position of the interface between flowing  and static  regions $y_o=(r_o-r_i)/(r_e-r_i)$ with $Bn=2$, $\Delta K^{\star}=0.5$, $\tau_1^{\star}=0.5$, $a^{\star}=1$ and $b^{\star}=1$ for a large gap $\eta=0.5$. Only shear-localization cases are considered when $n_c \neq 1$. \label{fig:yob}}
\end{figure}

Now, we focus on the evolution of the width of the flowing region $y_o=r_o-r_i$ depending on the thixotropic parameters and the shear-thinning index $n_c$. $y_o$ can be obtained on figs. \ref{fig:vlb}-d, \ref{fig:vldktau1}-d and \ref{fig:vlnc}-d by reading the abscissa where $\lambda$ reaches $1$. On figs. \ref{fig:yob}-a, b and c, the curve of $y_o$ separates the inner flowing region from the outer static region. The transition between the shear localization and the shear banding is smooth because, at $r=r_o$, the strain rate $\gap_c$ increases continuously from zero at the critical conditions and the stress at the interface is still the yield stress of the fully structured material. Nevertheless, it is observed on fig. \ref{fig:yob}-a that the variation of $y_o$ is much more important in the regime of the continuous shear localization than when the shear banding occurs. When the breakdown parameter $b^{\star}$ increases, the size of the flowing  region, characterized by the reduced position $y_o=r_o-r_i$ of the interface between the liquid and the solid-like regions, decreases to a minimum size corresponding to the one of the fully unstructured equivalent Bingham fluid when $b^{\star} \rightarrow + \infty$ (fig. \ref{fig:yob}-a). Quite similar remark can be done when $\Delta K^{\star}$ and $\tau_1^{\star}$ increase. Indeed, increasing $b^{\star}$, $\Delta K^{\star}$ or $\tau_1^{\star}$ makes the shear-thinning behaviour stronger in the steady flowing region. Thus, it is not surprising that when the shear-thinning index $n_c$ decreases, the width of the flowing zone also decreases (fig. \ref{fig:yob}-c).  Nevertheless, the model with one structural parameter allows to predict the shear banding which can not be described by the Bingham law or the Hershel-Bulkley law. When the thixotropic parameters are increasing above some critical values given by eq. (\ref{eq:bc}), the shear-banding appears smoothly from the shear localization where the strain rate is continuous between the flowing and static regions.

Now, we will study the linear stability of the base flow, whatever we have shear-banding or not.

\section{Linear stability analysis}

\subsection{Equations setup}

To perform a linear analysis of stability, the fluid velocity, the structural parameter and the pressure are decomposed such as:
\begin{eqnarray}
\mathbf{v} &=& \mathbf{v}_b + \tilde{\mathbf{v}}(r) \, exp(\sigma t  + i n \theta + i k z) \label{vp} \\
\lambda &=& \lambda_b + \tilde{\lambda}(r) \, exp(\sigma t + i n \theta + i k z) \label{eq:lp} \\
p &=& P_b + \tilde{p}(r) \, exp(\sigma t + i n \theta + i k z) \label{eq:pp}
\end{eqnarray}

\noindent $\tilde{\mathbf{v}}$, $\tilde{\lambda}$ and $\tilde{p}$ describe the perturbation of the base flow considering the azimuthal mode $n$ and the axial wave number $k$. Injecting eqs. (\ref{vp} -- \ref{eq:pp}) in the general setup of equations (\ref{eq:ns}, \ref{eq:lambda}, \ref{eq:div}) and after withdrawing the nonlinear terms, the linear setup of equations for the perturbation of the base flow is:

\begin{eqnarray}
\sigma \tilde{\mathbf{v}} &=& - \dbar{\nabla} \mathbf{v}_b \cdot \tilde{\mathbf{v}} - \dbar{\nabla} \tilde{\mathbf{v}} \cdot \mathbf{v}_b + \frac{1}{Re} \mathbf{div} \left( \left. \frac{\partial \dbar{\tau}}{\partial \gap_{ij}} \right|_b \gap_{ij}(\tilde{\mathbf{v}}) + \left. \frac{\partial \dbar{\tau}}{\partial \lambda} \right|_b \tilde{\lambda} \right) - \bm{\nabla} \tilde{p} \label{eq:nspert} \\
\sigma \tilde{\lambda} &=& - \mathbf{v}_b \cdot \bm{\nabla} \tilde{\lambda} - \tilde{\mathbf{v}} \cdot \bm{\nabla} \lambda_b -(a+b \gap_b^m) \tilde{\lambda} - m b \lambda_b \gap_b^{m-1} \left. \frac{\partial \gap}{\partial \gap_{ij}} \right|_b  \gap_{ij}(\tilde{\mathbf{v}}) \label{eq:lpert} \\
0 &=& div(\tilde{\mathbf{v}}) \label{eq:divpert}
\end{eqnarray}

\noindent The indices $ij$ stand for $r$, $\theta$ or $z$ and Einstein's convention for summation is used. Thus, the generalized eigenvalues problem given by the latter setup of equations (\ref{eq:nspert}--\ref{eq:divpert}) can be straightfully written in matrix form:
\begin{equation}
\sigma \begin{bmatrix} I_v & 0 & 0 \\ 0 & I_{\lambda} & 0 \\ 0 & 0 & 0  \end{bmatrix}
\begin{bmatrix} V \\ \Lambda \\ P \end{bmatrix} = \begin{bmatrix} L_{vv} & L_{v \lambda} & -G \\
L_{\lambda v} & L_{\lambda \lambda} & 0 \\ D & 0 & 0 \end{bmatrix} \begin{bmatrix} V \\ \Lambda \\ P \end{bmatrix} \label{eq:linsyst}
\end{equation}

\noindent $V$ is the vertical matrix of the values of the components of velocity $\tilde{\mathbf{v}}$ at each inner point of the gap. $\Lambda$ is the vertical matrix of the values of $\tilde{\lambda}$ at each point of the mesh, including the inner and outer radii. $P$ is the the vertical matrix of the values of $\tilde{p}$ taken in the middle points of two successive nodes of the velocity and structural parameter mesh.

The linear problem (\ref{eq:linsyst}) admits a number of infinite eigenvalues which is two times the number of degree of freedom of the pressure.  The infinite eigenvalues have to be eliminate because they correspond to non-zero divergence velocity fields. Nevertheless, it is not harmful because Matlab's algorithm generates true infinite eigenvalues avoiding any ambiguity.

\subsection{Convergence test and validation \label{ssec:valid}}

In order to test the convergence of our numerical scheme and to validate our method, the critical Reynolds number $Re_c$ and the critical axial wave number $k_c$ are determined using different number of nodes $M$ in the gap. The results are given in the tables \ref{tab:newt} and \ref{tab:bing}. For Newtonian fluids, many works allows us to validate our results. In a recent work \cite{pourjafar2015}, a similar numerical method gives $Re_c=131.66$ and $k_c=3.130$ in Newtonian fluids with $\eta=0.9$. Their values are in very good agreement with ours. More, for Bingham fluids, Alibenyahia et \textit{al.} \citep{alibenyahia2012} found $Re_c=127.74943$ and $k_c=3.183706$ with a spectral method at $Bn=1$ and $\eta=0.5$. Once again, our results in the table \ref{tab:bing} agree this values within an error below $0.1\%$.

According, the tables \ref{tab:newt} and \ref{tab:bing} and the results of \citep{alibenyahia2012} and \cite{pourjafar2015}, we can estimate the relative error for the critical values of the Reynolds number $Re_c$ and the axial wave number $k_c$ below $0.1\%$ when $M \geq 50$. Thus, we use $M=50$ in the following.

\begin{table}
\begin{tabular}{c|cccccc}
$M$ & 20 & 30 & 40 & 50 & 60 & 100\\
\hline
$Re_c$ & 132.492 & 131.989 & 131.822 & 131.746 & 131.705 & 131.647 \\
\hline
$k_c$ & 3.1270 & 3.1280 & 3.1283 & 3.1285 & 3.1286 & 3.1287
\end{tabular}
\caption{Critical Reynolds number $Re_c$ and critical axial wave number $k_c$ for a Newtonian fluid \textit{vs} the number of nodes $M$ in the gap with a radii ratio $\eta=0.9$. \label{tab:newt}}
\end{table}

\begin{table}
\begin{tabular}{c|cccccc}
$M$ & 20 & 30 & 40 & 50 & 60 & 100\\
\hline
$Re_c$ & 128.472 & 128.057 & 127.919 & 127.857 & 127.823 & 127.776 \\
\hline
$k_c$ & 3.1695 & 3.1776 & 3.1803 & 3.1816 & 3.1822 & 3.1832
\end{tabular}
\caption{Critical Reynolds number $Re_c$ and critical axial wave number $k_c$ for a Bingham fluid \textit{vs} the number of nodes $M$ in the gap with a radii ratio $\eta=0.5$. \label{tab:bing}}
\end{table}

\subsection{Stability analysis of Couette flow of thixotropic yield stress fluids }

To determine the critical eigenmode of the linear setup of equations (\ref{eq:linsyst}), the algorithm seeks the minimum of the critical value of the Reynolds number $Re_c$ depending on the wave number $k$ for a given azimutal mode $n$. The critical Reynolds number is reached when the real part of the eigenvalue $\sigma$ is zero. The minimal value of $Re_c$ is reached at the critical wave number $k_c$. We have verified that the critical perturbation is always axisymetric, \textit{i. e.} $n=0$, by computing the critical Reynolds number for the azimuthal modes $n$ from $0$ to $3$ within the range of our parameters for the thixotropic yielded fluids. The results is that the Taylor vortices, steady and axisymmetric, corresponds always to the most unstable eigenmode of (\ref{eq:linsyst}). One can observed that the equation (\ref{eq:lpert}) does not generate oscillating or three-dimensional modes and only the values of the critical wave number $k_c$ and the critical Reynolds number $Re_c$ are modified compared to the Newtonien or shear-thinning cases. One can notice this result because it means that the unsteady effects of the thixotropy in cylindrical Couette flow which might occur above the threshold of the primary instability are nonlinear.  Nevertheless, shear-banded flows may occur with thixotropic yielded fluids. This is a real difference with simple yielded fluids.

\begin{figure}
\begin{tabular}{cc}
(a) & (b) \\
\includegraphics[width=0.49\textwidth]{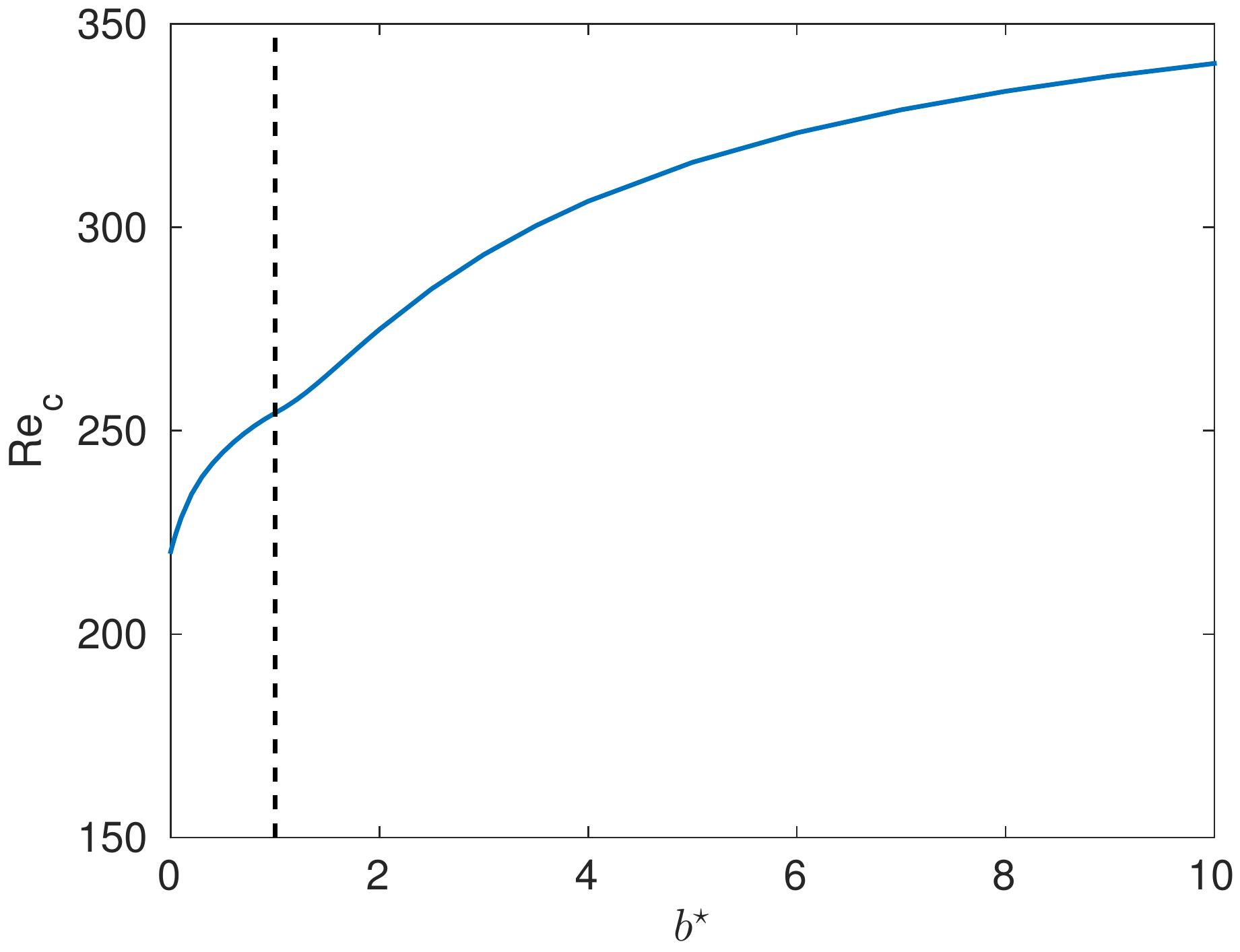} &
\includegraphics[width=0.49\textwidth]{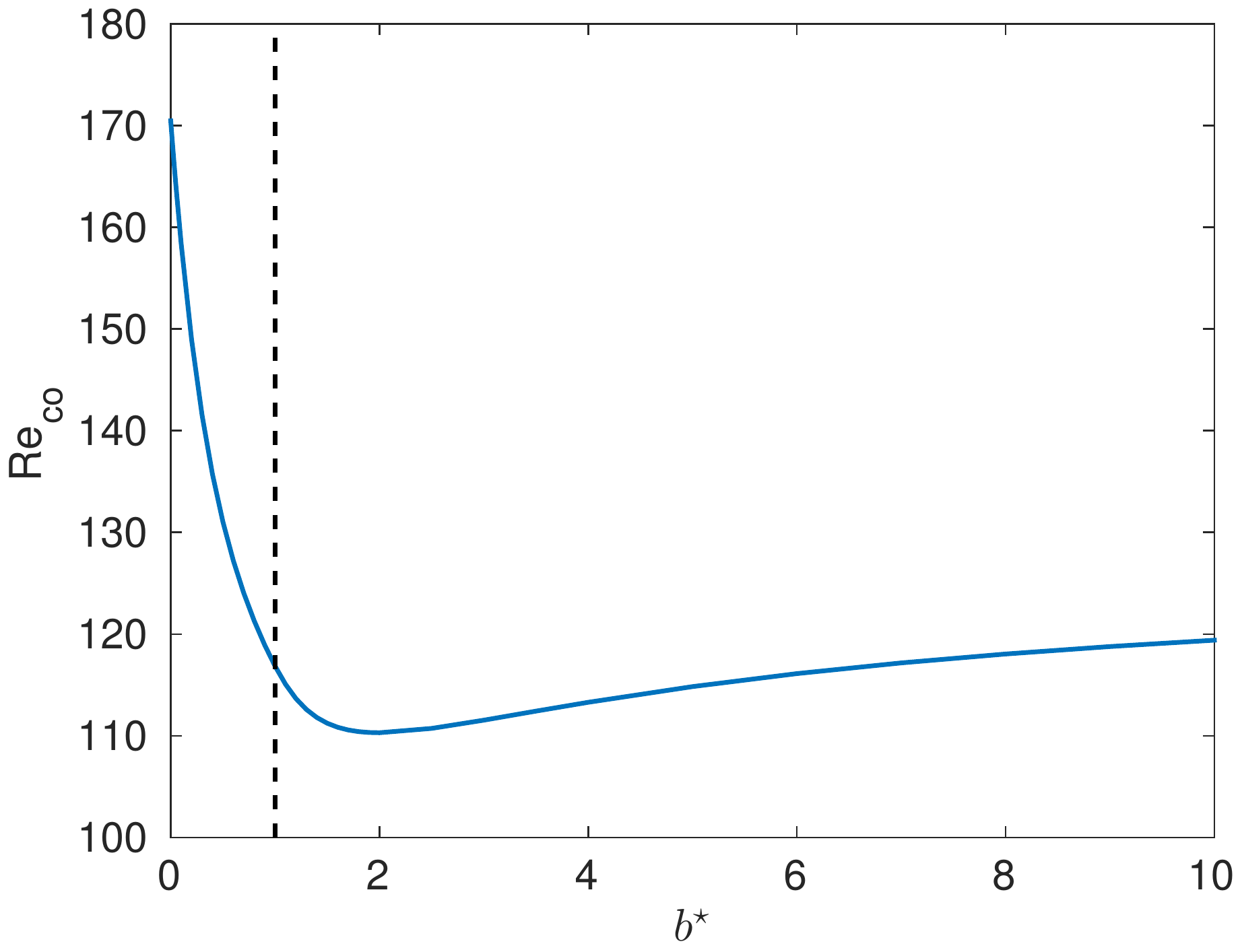} \\
(c) & (d) \\
\includegraphics[width=0.49\textwidth]{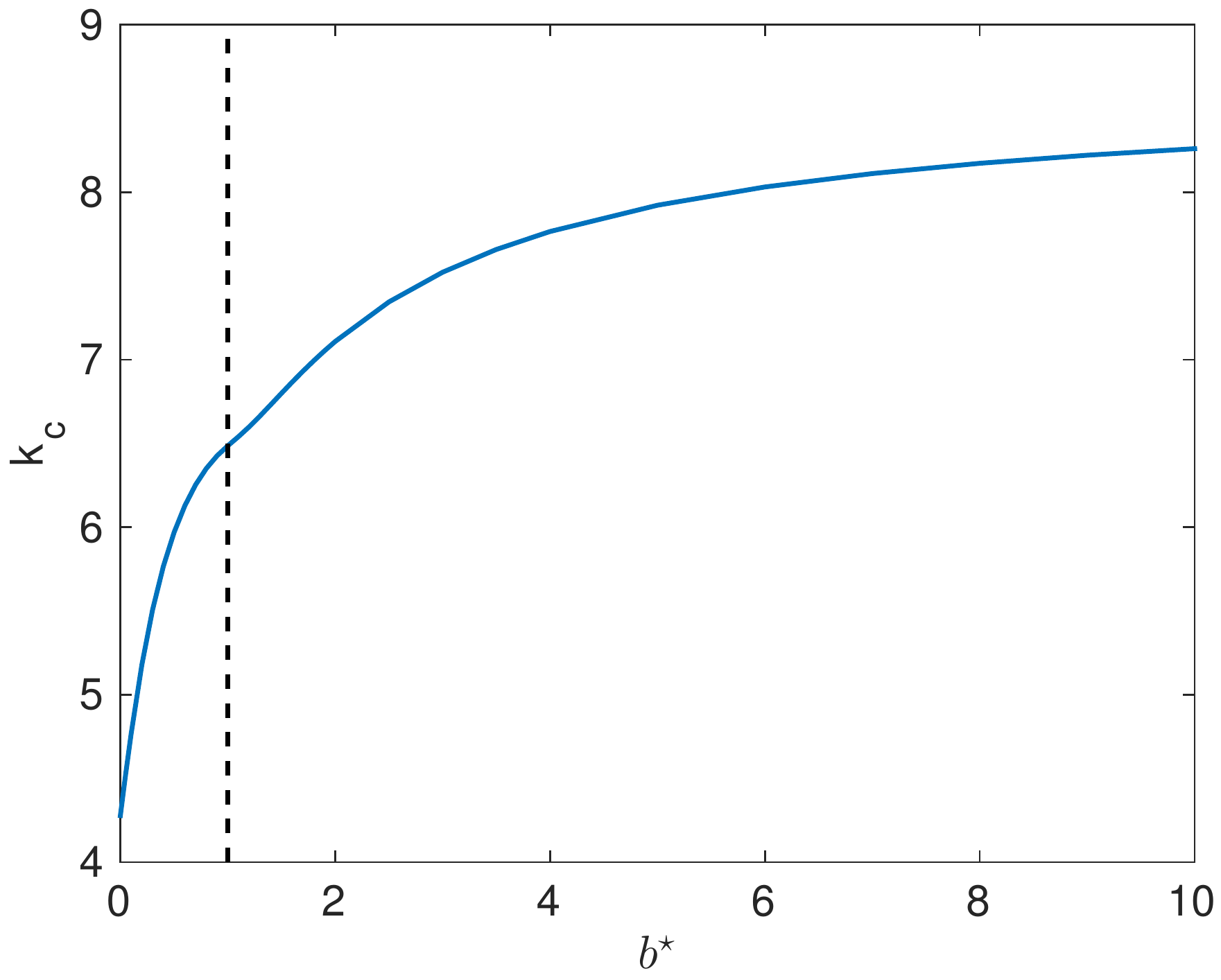} &
\includegraphics[width=0.49\textwidth]{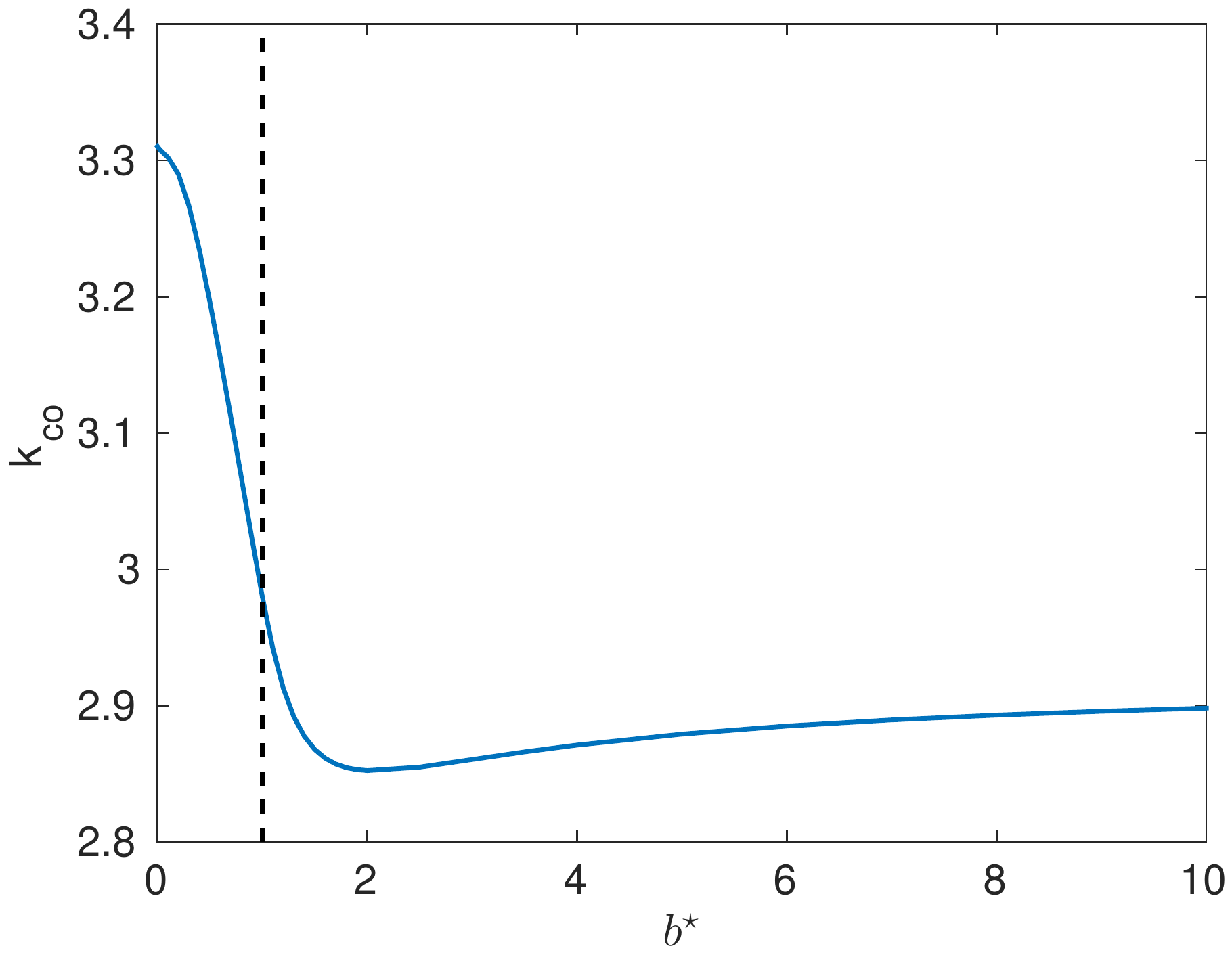}
\end{tabular}
\caption{Critical Reynolds number $Re_c$ (a), $Re_{co}=y_o Re_c$ (b), critical axial wave number $k_c$ (c), $k_{co}=y_o k_c$ (d) \textit{vs} $b^{\star}$. Large gap $\eta=0.5$, $Bn=2$, $n_c=1$, $\Delta K^{\star}=\tau_1^{\star}=1$ and $a^{\star}=1$. Vertical dashed-line stands for the critical value of $b^{\star}=1$ where the strain rate at the interface $\gap _o$ becomes non-zero. \label{fig:Recb}}
\end{figure}

\begin{figure}
\begin{tabular}{cccc}
(a) & (b) & (c) & (d) \\
\includegraphics[width=0.24\textwidth]{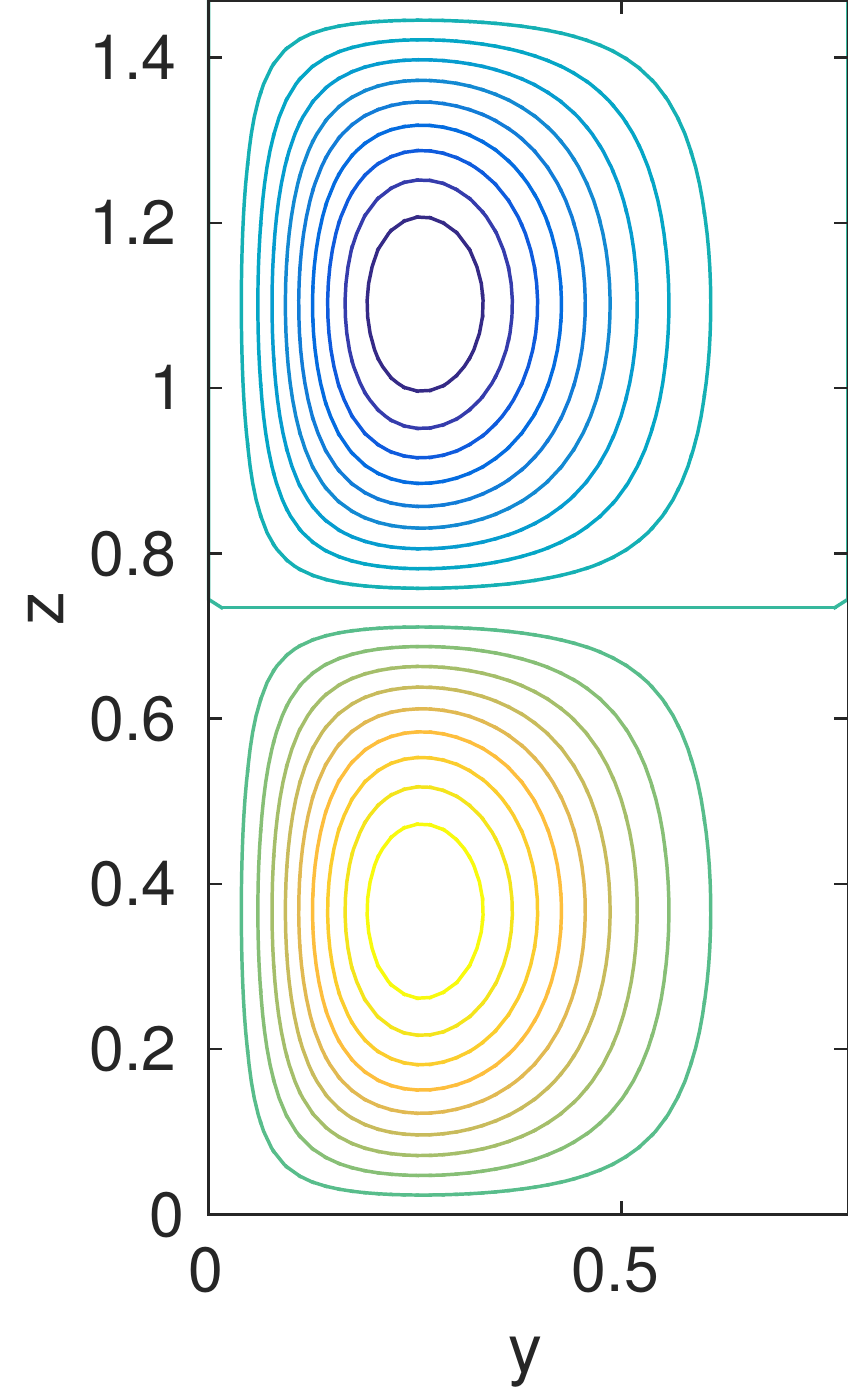} &
\includegraphics[width=0.24\textwidth]{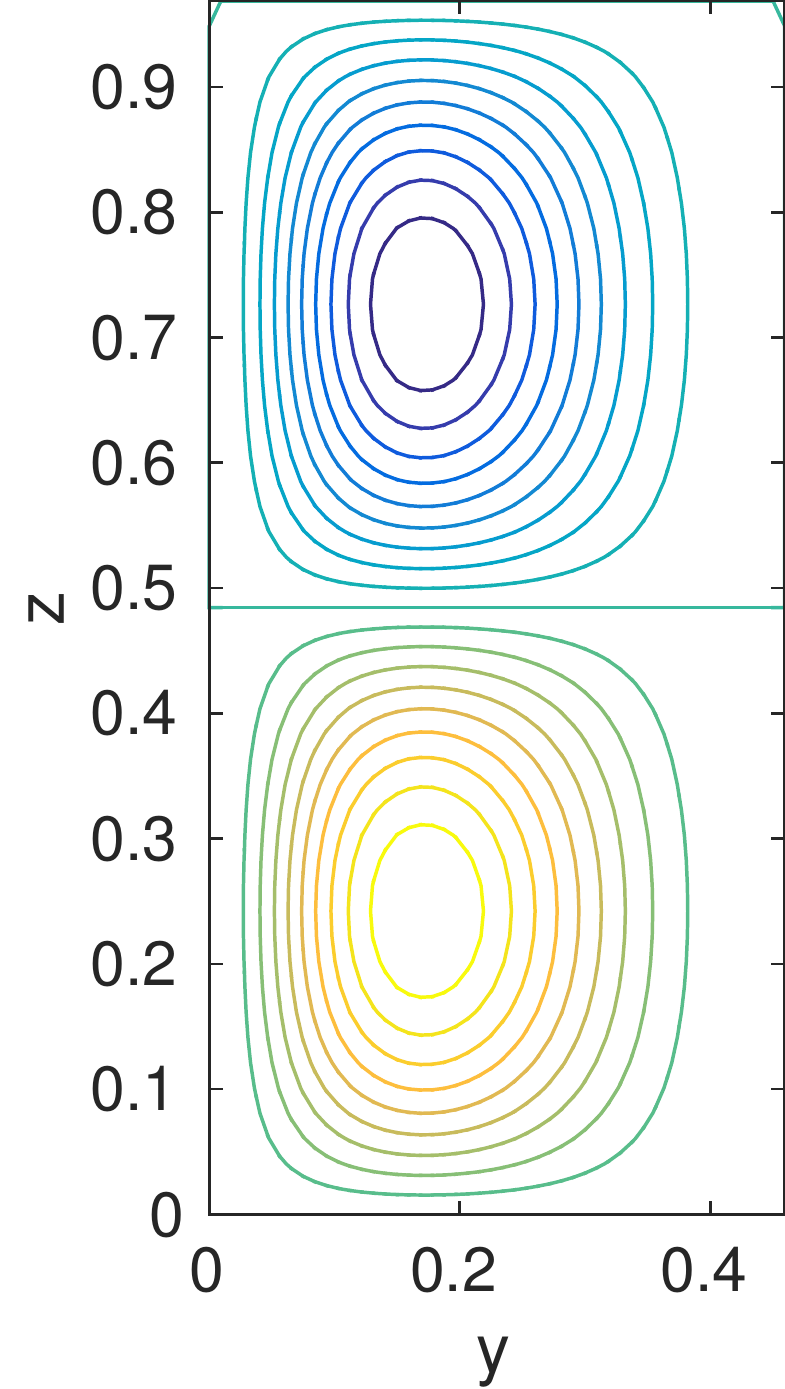} &
\includegraphics[width=0.24\textwidth]{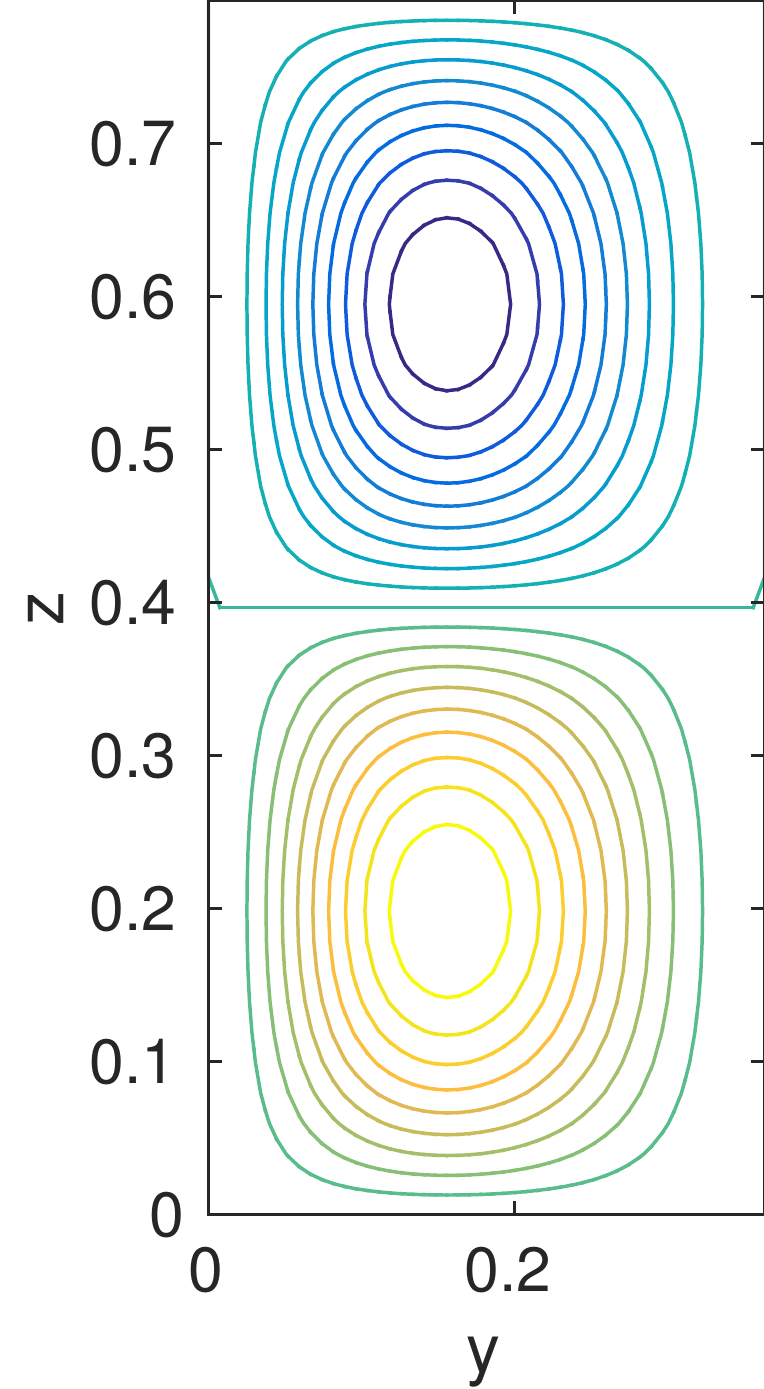} &
\includegraphics[width=0.24\textwidth]{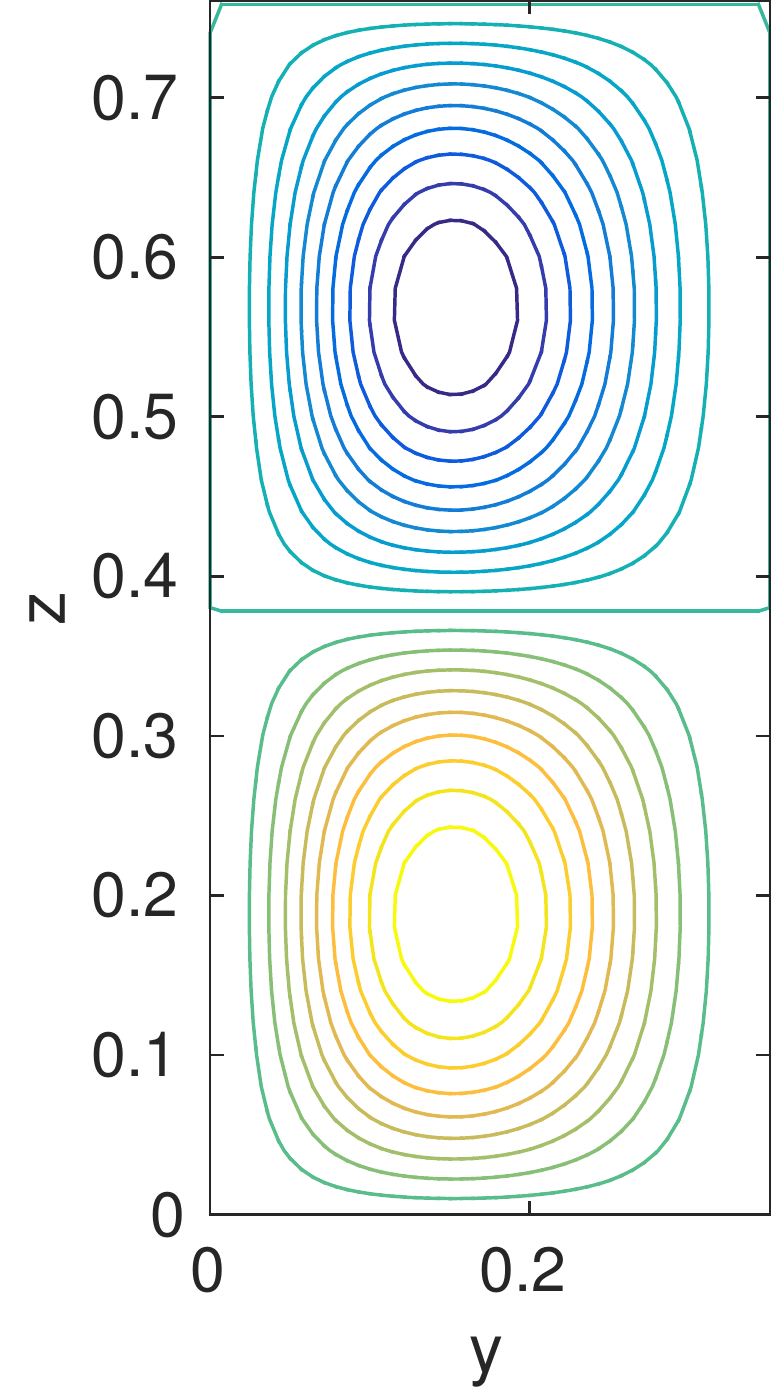} \\
 & (e) & (f) & (g) \\
&
\includegraphics[width=0.24\textwidth]{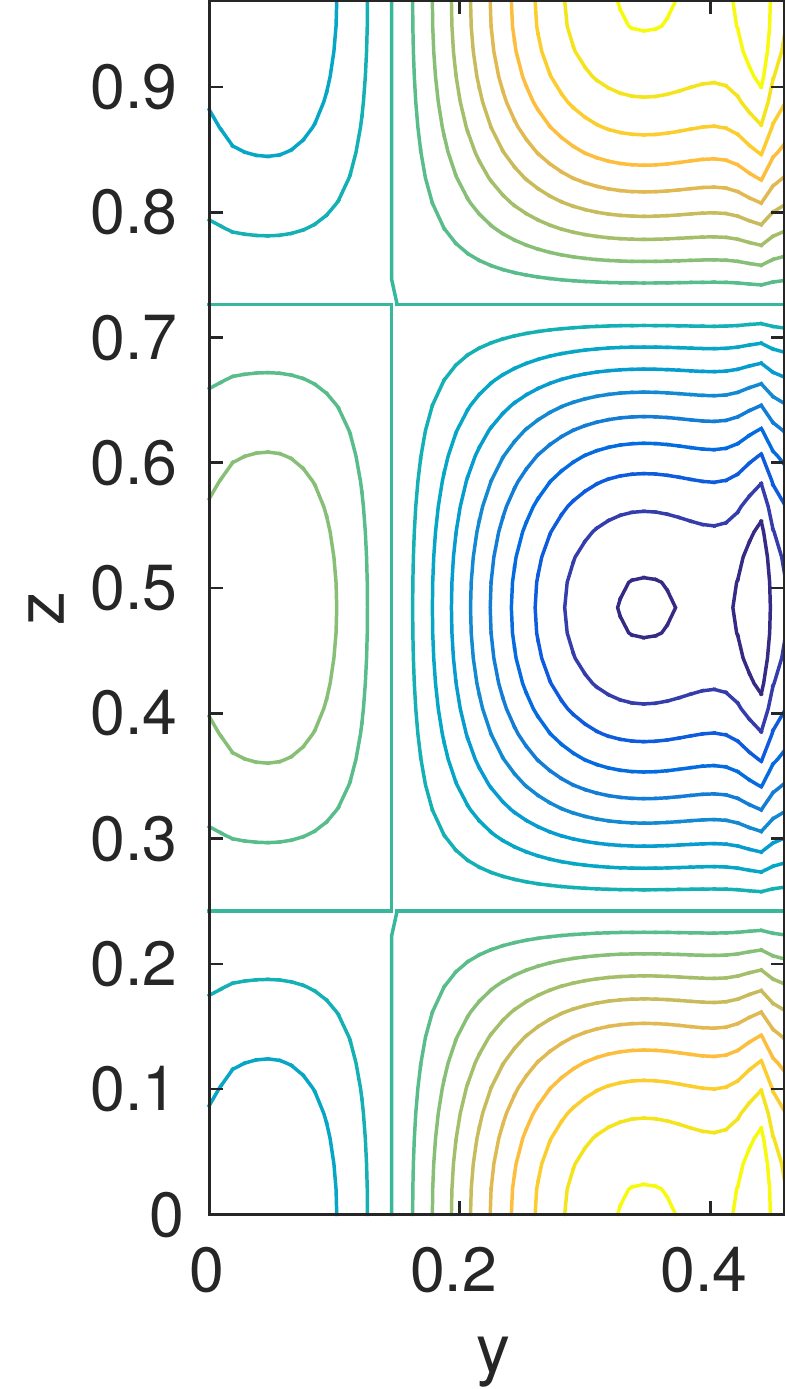} &
\includegraphics[width=0.24\textwidth]{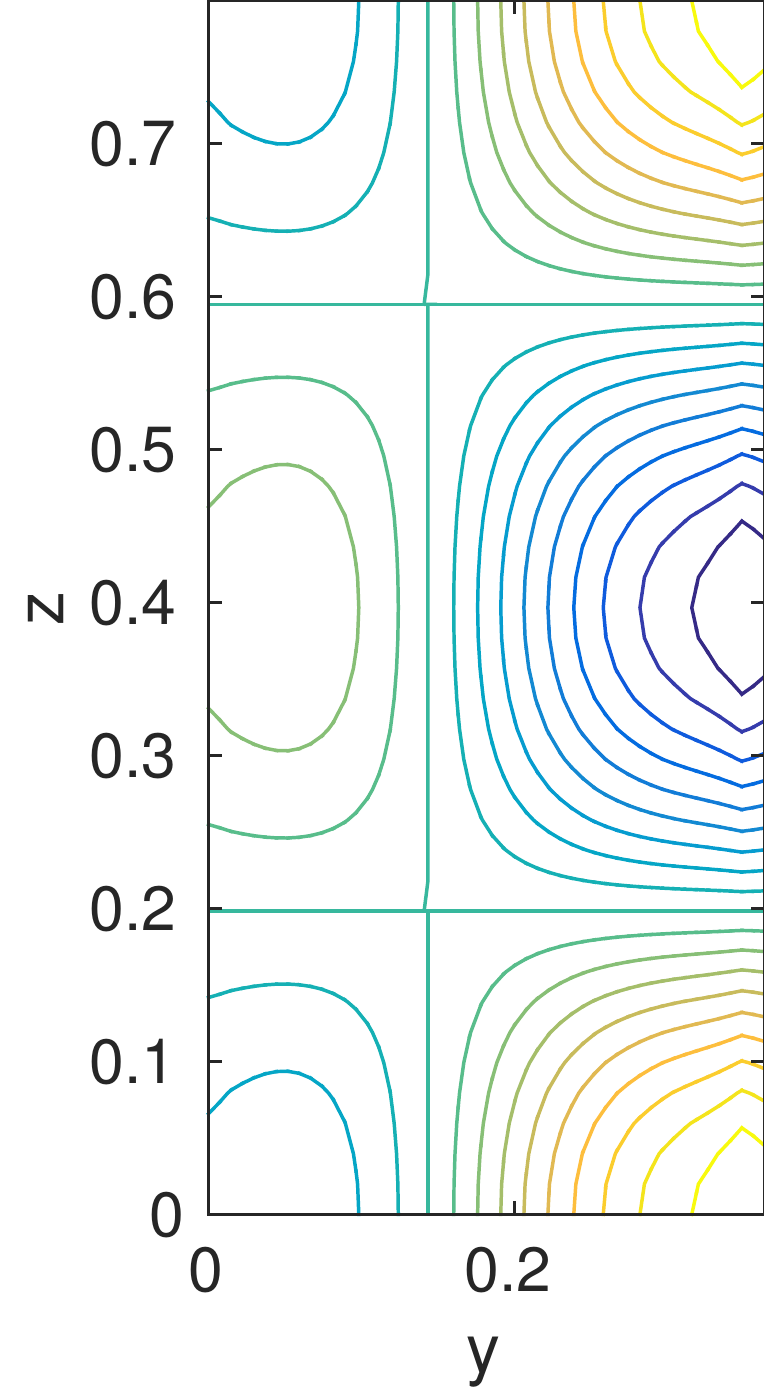} &
\includegraphics[width=0.24\textwidth]{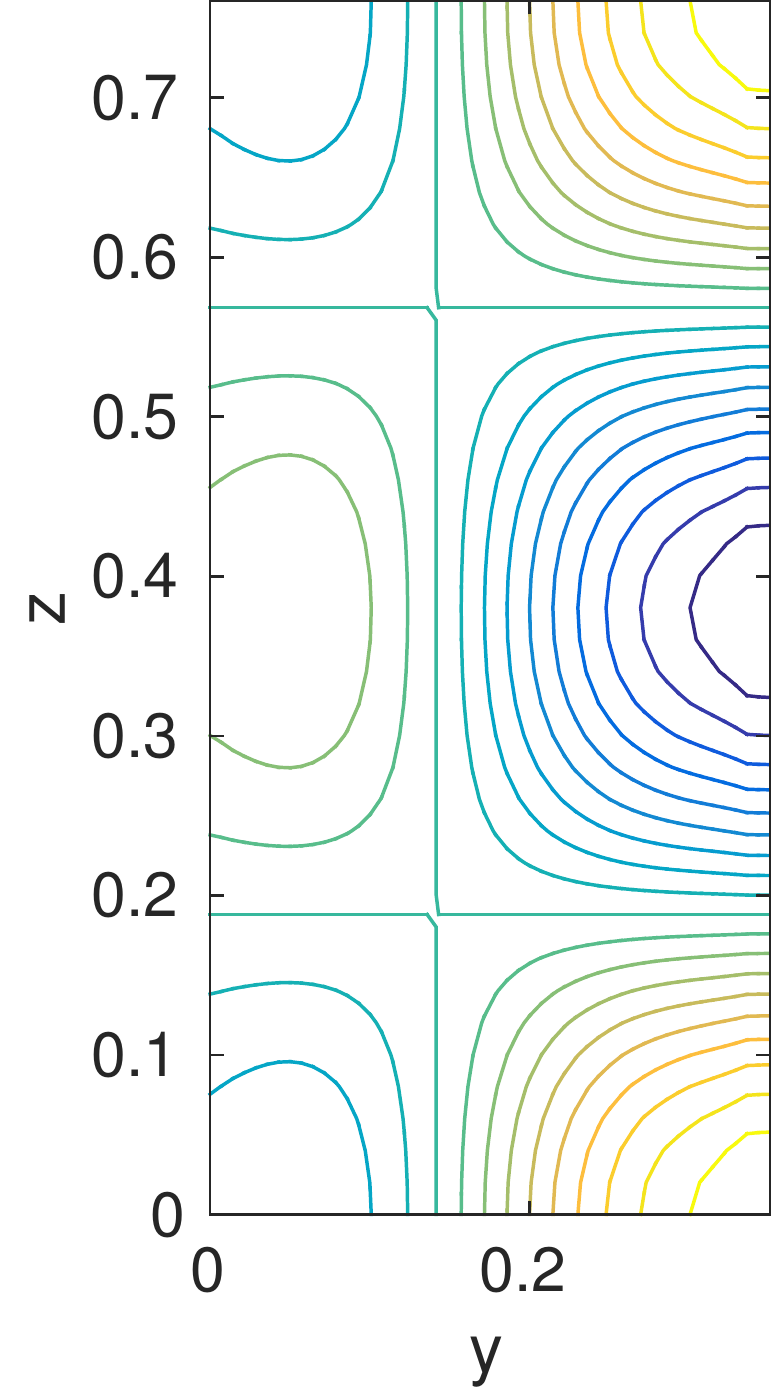}
\end{tabular}
\caption{First line (a-d): streamlines of the critical velocity field of the perturbation. Second line (e-g): contour plot of the perturbation of the structural parameter $\lambda$. Contour lines step is $5\%$ of the normalized amplitude of the perturbation. Blue color stands for negative values and red color for positive values (counter-clockwise and clockwise spin for Taylor vortices). Large gap $\eta=0.5$, $Bn=2$, $n_c=1$, $\Delta K^{\star}=\tau_1^{\star}=1$, $a^{\star}=1$ and respectivelly $b^{\star}=0$ (a), $1$ (b, e), $5$ (c, f) and $10$ (d, g). \label{fig:contourb}}
\end{figure}

The ratio $b^{\star}/a^{\star}$ denotes the resistance against the strain rate $\gap$ of the structure described by $\lambda$. The higher $b^{\star}/a^{\star}$  is, the easier the inner structure of the fluid is broken down by the shear. As shown previously in fig. \ref{fig:yob}, the flowing region  decreases because the yield stress collapses with the breakdown of the structure. If the fluid would be a viscous Newtonian fluid, the critical Reynolds number would increases because of the gap becomes small. In the fig. \ref{fig:Recb}-a, the variation of $Re_c$ with $b^{\star}$ suggests that our choice for the reference viscosity $\mu_{ref}$ is representative of an equivalent Newtonian fluid and thus we retrieve the stabilizing effect of the reduction of the gap width. Moreover, as the viscosity of the fluid decrease when the inner structure is broken, the fluid is stronger and stronger shear-thinning. Thus, it is not surprising that the grow of $b^{\star}/a^{\star}$ ends by stabilizing the flow (fig. \ref{fig:Recb}-a) as it is observed experimentally for large gap by  Escudier \textit{et al.} \citep{escudier1995} and shown by Alibenhahia \textit{et al.} \citep{alibenyahia2012} when the shear-thinning index $n_c < 0.6$ for $\eta=0.5$. Nevertheless, for yielded fluids, one can argue that the effective gap corresponds to the fluid zone. The Reynolds number $Re_o = y_o Re$ is calculated with the gap width of the yielded region $y_o$. In this case, the collapse of the width $y_o$ of the fluid zone (fig. \ref{fig:yob}) because of the breakdown of the inner micro-structure is stronger than the stabilizing effect of shear-thinning and the critical Reynolds number $Re_{co}$ decreases until $y_o$ is close to its minimum value corresponding to a fully destructured fluid (fig. \ref{fig:Recb}-b). The critical wave number $k_c$ mainly follows the evolution of the fluid gap width and thus it increases with $b^{\star}$ (fig. \ref{fig:Recb}-c). Of course, by recalculating the wave number considering the fluid gap as the effective gap, show that the wave length $L_z=2\pi/k_{co}$ first increase because the high viscosity near the end of the fluid area decrease when $b^{\star}$ increases. Thus, the vortices are squeezed toward the inner wall as shown in the fig. \ref{fig:contourb}-a. When the shear-banding appears, the stratification of the viscosity is weaken and the Taylor vortices takes all the place in the fluid zone (fig. \ref{fig:contourb}-c,d). Thus, the wave number $k_{co}$ tends to an asymptotic value wich correspond to the rolls of an equivalent fully destructured fluid ($\Delta
K^{\star}=\tau_1^{\star}=0$). To compute the critical Reynolds number of an equivalent non-thixotropic fluid when the gap width correspond to $y_o$ for $b^{\star}=10$, we have to set $\eta=0.7403$ and $Bn=0.7018$. The critical Reynolds number is $Re_c=126.9870$ and the critical wave number $k_c=2.9177$. This values have to be compared to $Re_{co}=119.4021$ and $k_{co}=2.8981$ found when $b^{\star}=10$. As expected, for high values of $b^{\star}$, it fits with an equivalent non-thixotropic (simple) fluid flowing in a smaller gap.

It is worthy to notice that the perturbation of the structural parameter $\lambda$ corresponds to the convection of the structure by the Taylor vortices. In the figs. \ref{fig:contourb}-(e-g), the negative zone of the perturbation corresponds to the convection from the inner cylinder where the structural parameter is low toward the outer cylinder. Thus, in our parameter range, the linear stability is driven by the Navier-Stokes equation and the perturbation of the structural parameter is a passive mode.

\begin{figure}
\begin{tabular}{cc}
(a) & (b) \\
\includegraphics[width=0.49\textwidth]{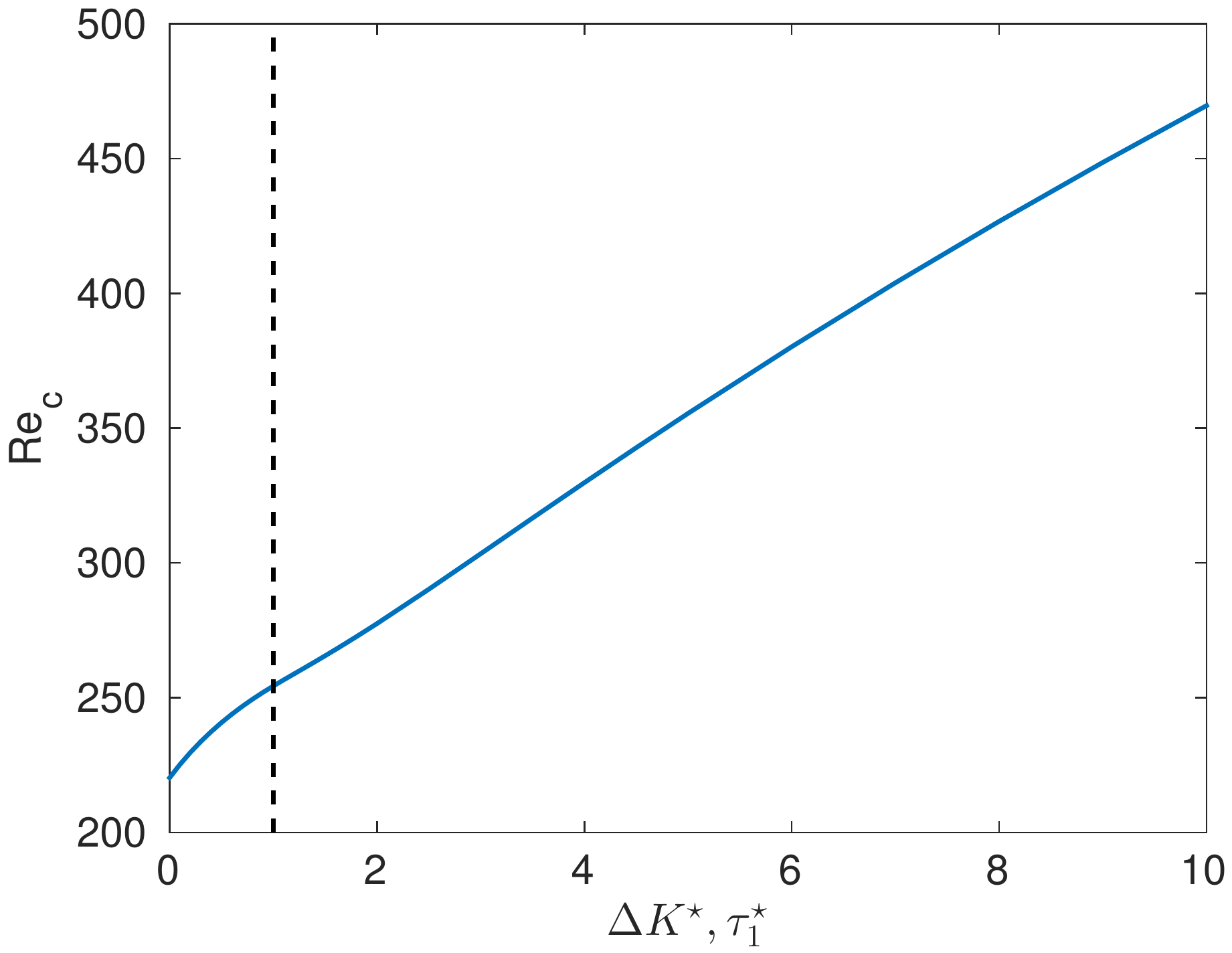} &
\includegraphics[width=0.49\textwidth]{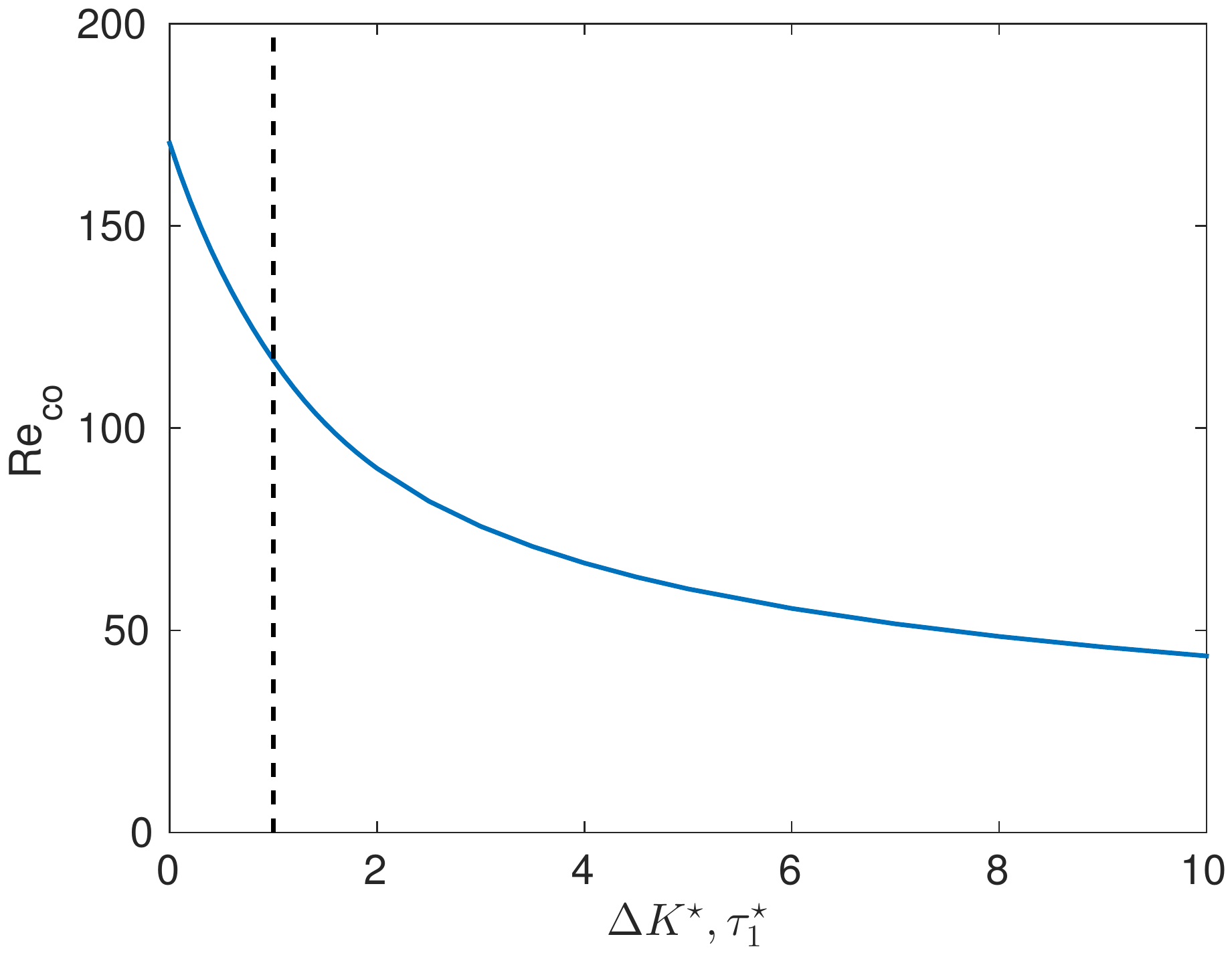} \\
(c) & (d) \\
\includegraphics[width=0.49\textwidth]{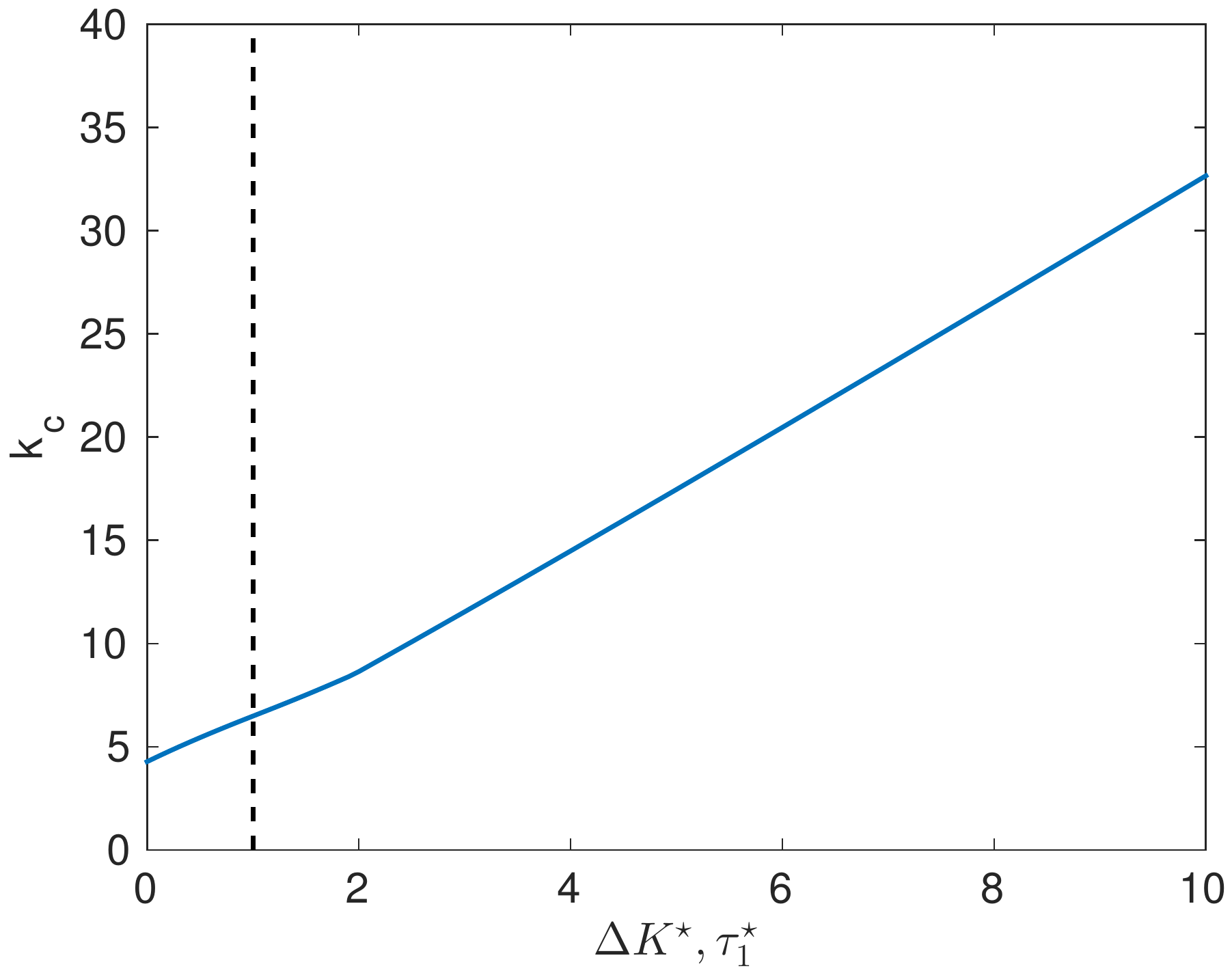} &
\includegraphics[width=0.49\textwidth]{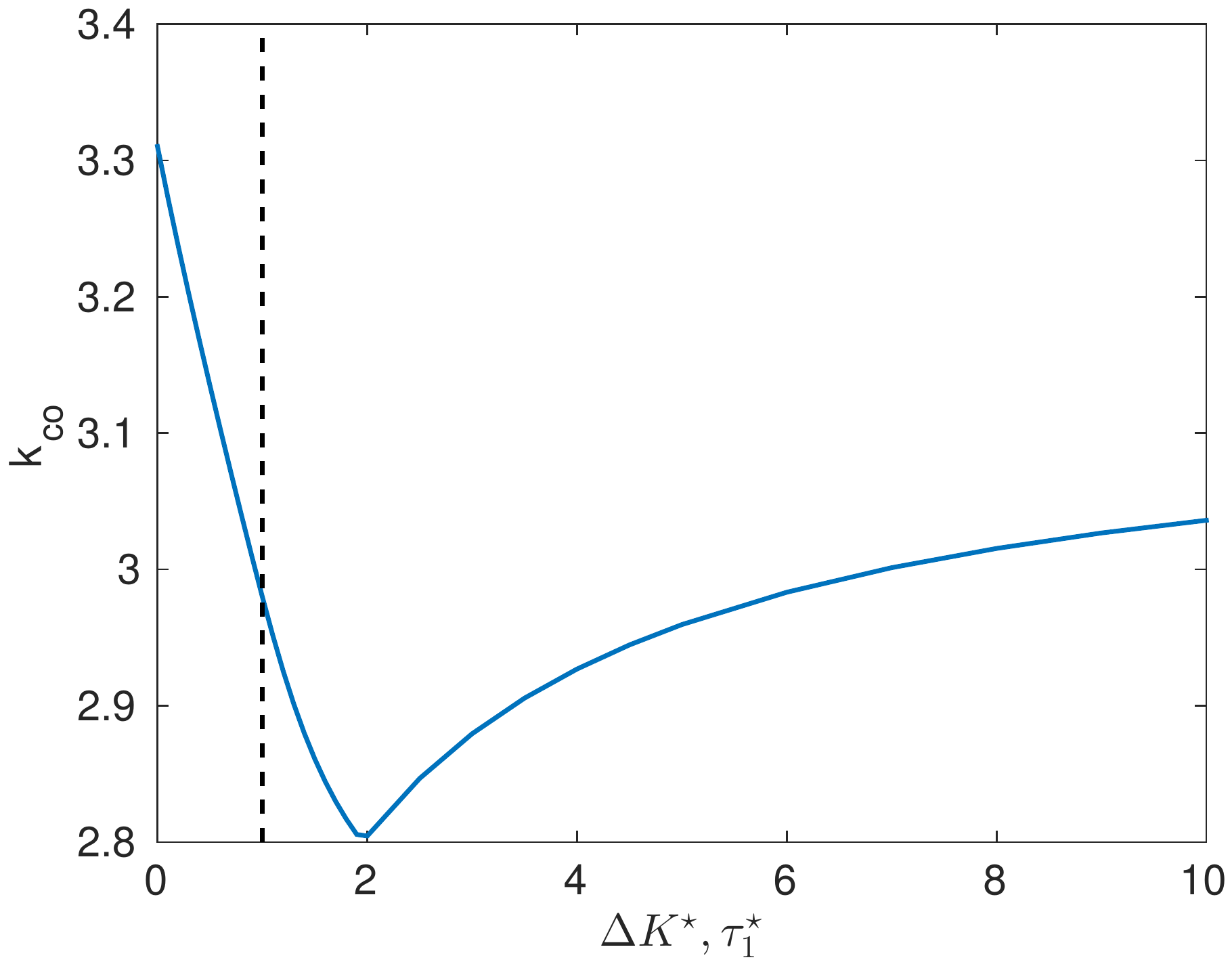}
\end{tabular}
\caption{Critical Reynolds number $Re_c$ (a), $Re_{co}=y_o Re_c$ (b) and critical axial wave number $k_c$ (c), $k_{co}=y_o k_c$ (d). Large gap $\eta=0.5$, $Bn=2$, $n_c=1$, $a^{\star}=1$ and $b^{\star}=1$. Vertical dashed-line stands for the critical value of $\Delta K^{\star}=1$ where the strain rate at the interface $\gap_o$ becomes non-zero. \label{fig:Rectau1}}
\end{figure}


The parameters $\Delta K^{\star}$ and $\tau_1^{\star}$ stabilise the flow according the figure \ref{fig:Rectau1}-a. Nevertheless, as previously, the material gap size is not the most relevant to define the critical Reynolds number. From this point of view, the Reynolds number $Re_{co}$ decreases and the flow is destabilized (fig. \ref{fig:Rectau1}-b). Indeed, the decrease of the viscosity with $\lambda$ is greater when $\Delta K^{\star}$ and $\tau_1^{\star}$ increase. The collapse of the dead zone due to high viscosity near the transition between the fluid and solid-like zones after the onset of the shear-banding is responsible of the increasing of $k_{co}$ in fig. \ref{fig:Rectau1}-d.

\begin{figure}
\begin{tabular}{cc}
(a) & (b) \\
\includegraphics[width=0.49\textwidth]{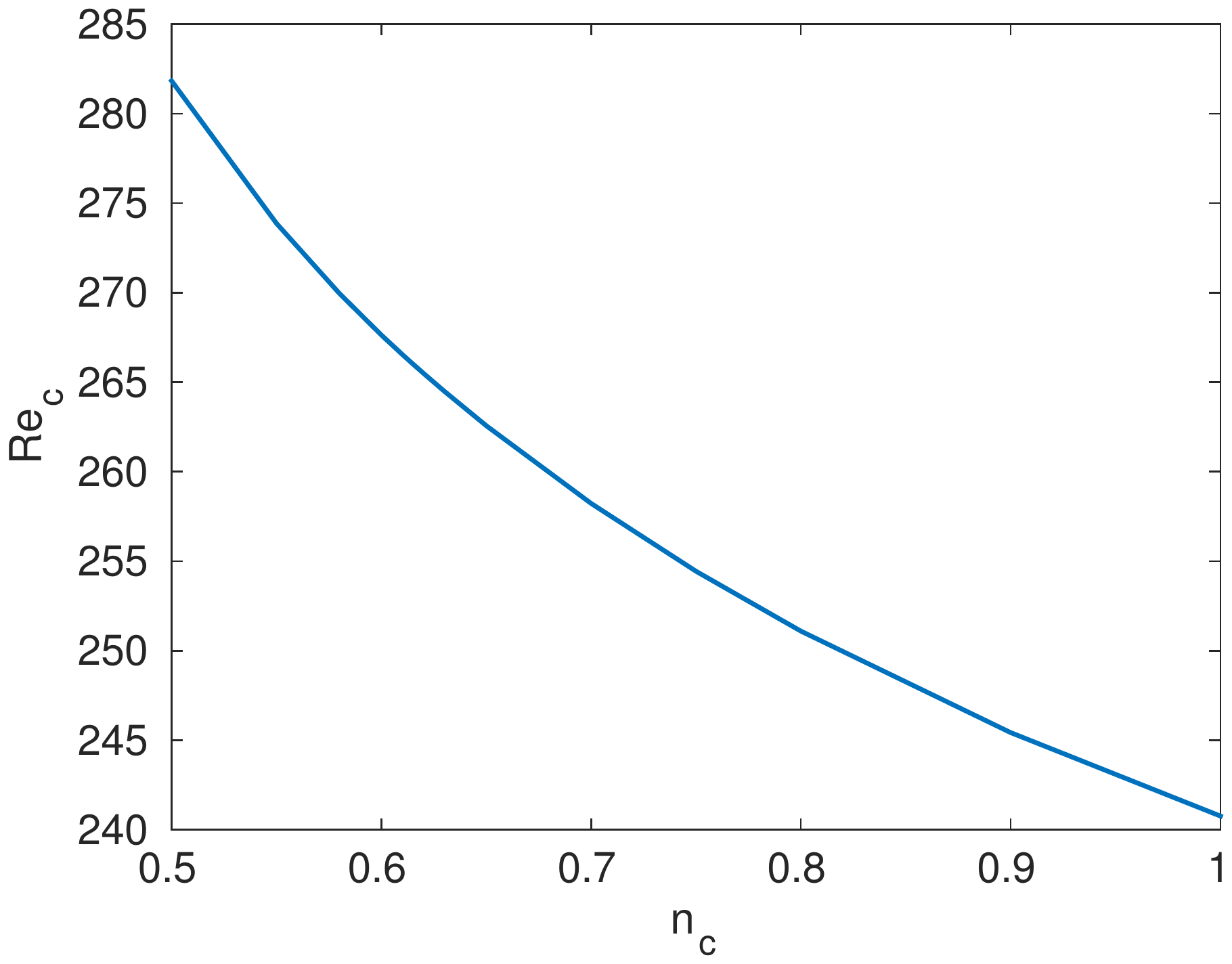} &
\includegraphics[width=0.49\textwidth]{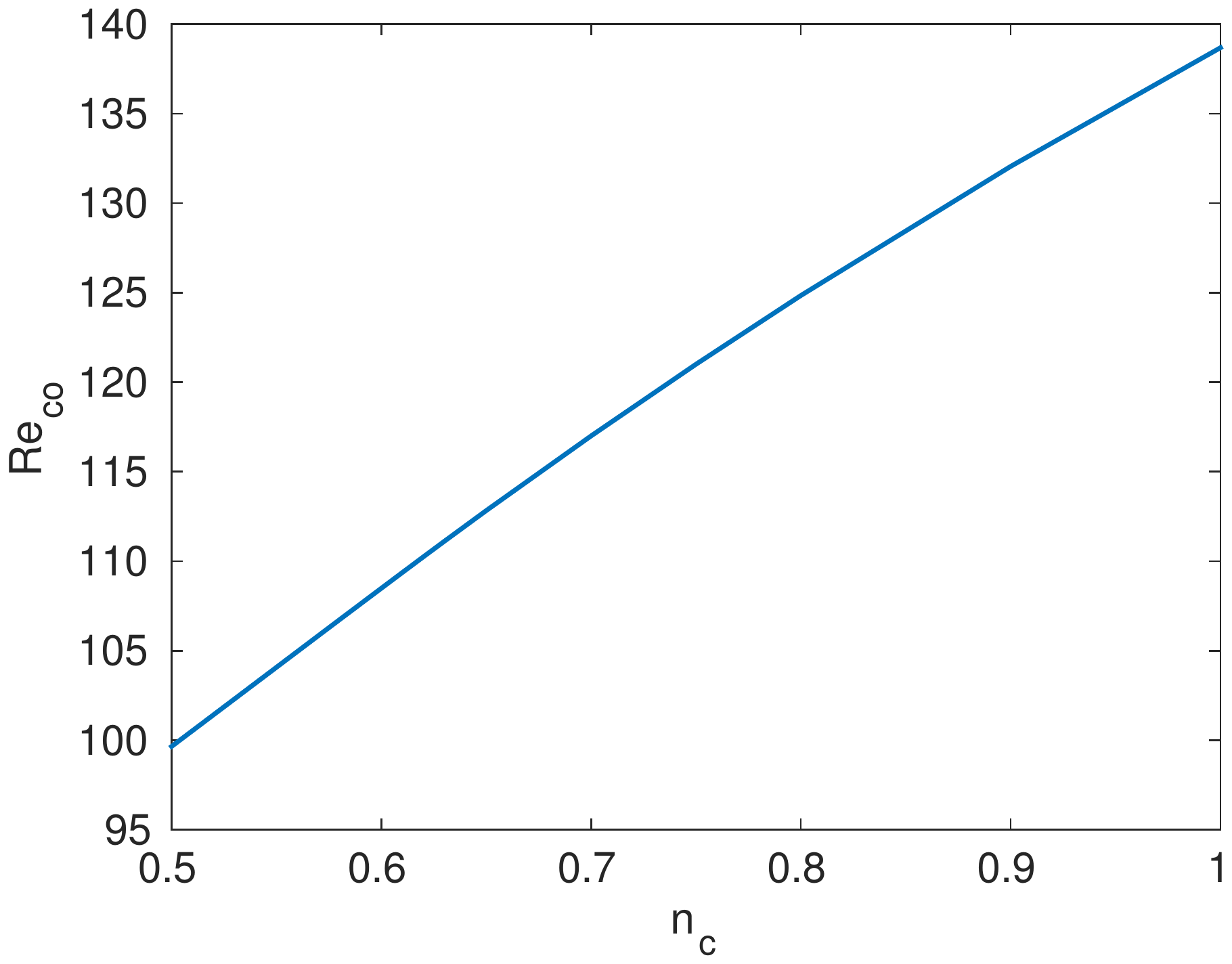} \\
(c) & (d) \\
\includegraphics[width=0.49\textwidth]{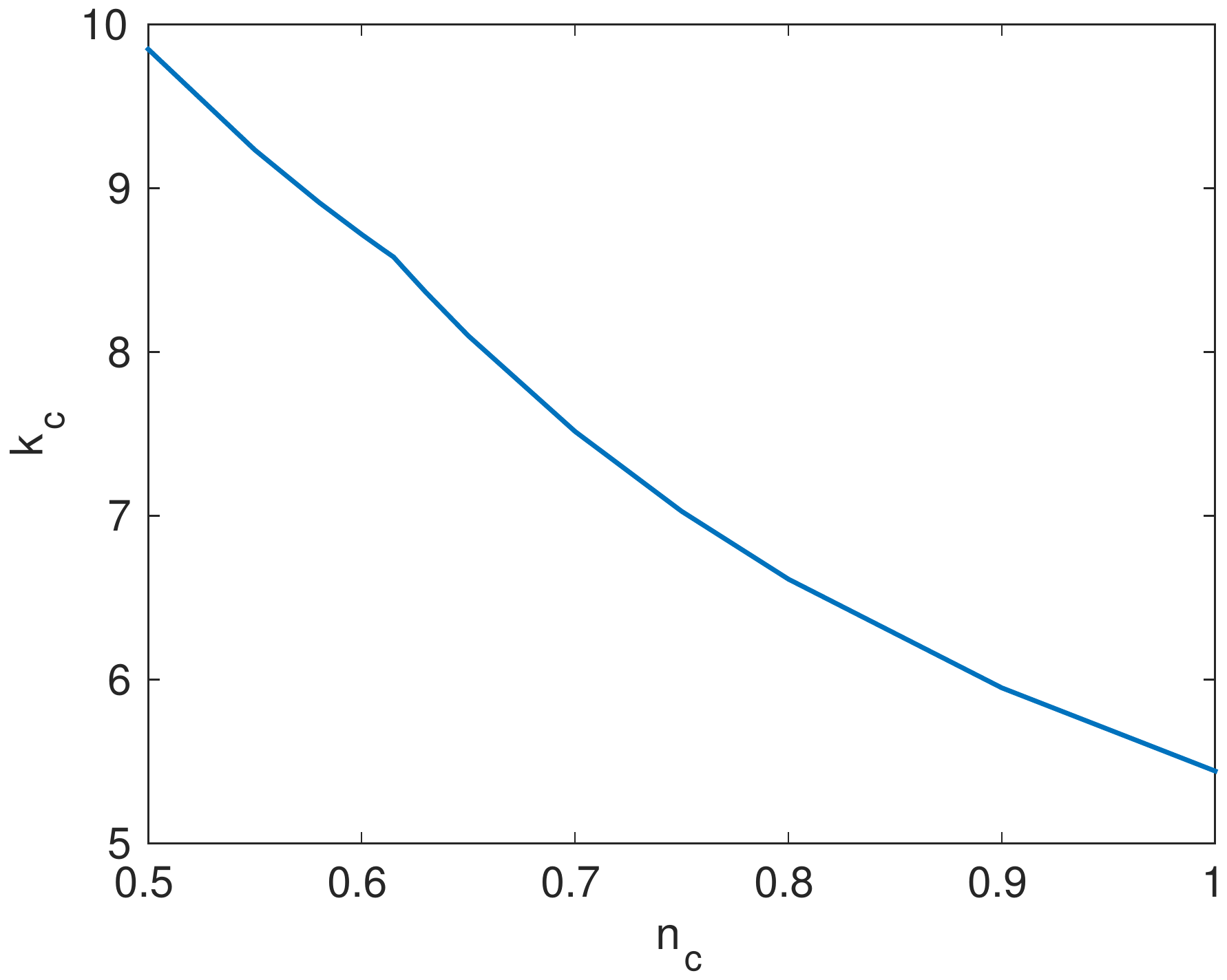} &
\includegraphics[width=0.49\textwidth]{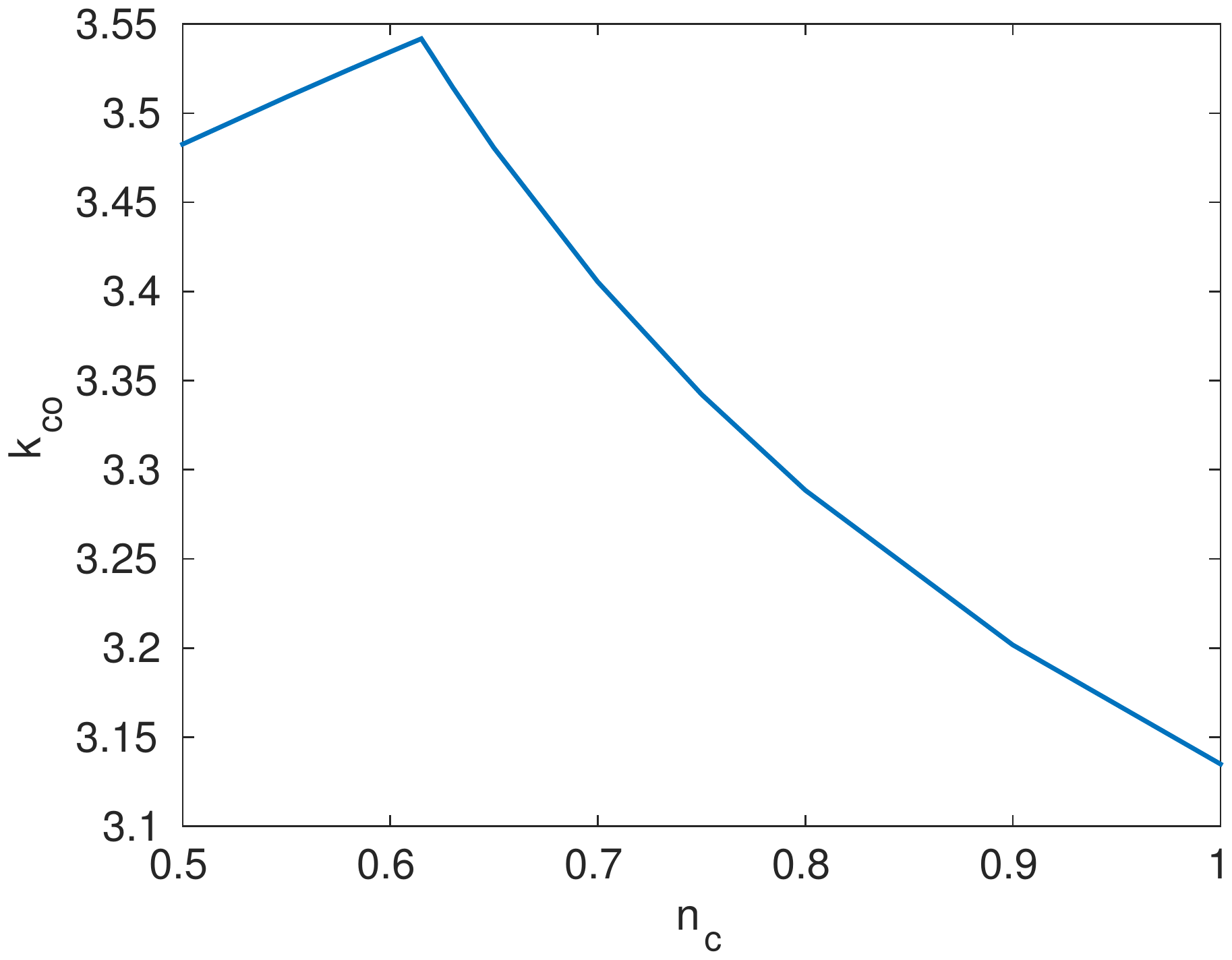}
\end{tabular}
\caption{Critical Reynolds number $Re_c$ (a), $Re_{co}=y_o Re_c$ (b)and critical axial wave number $k_c$ (c), $k_{co}=y_o k_c$ (d).  Large gap $\eta=0.5$, $Bn=2$, $\Delta K^{\star}=\tau_1^{\star}=0.5$, $a^{\star}=1$ and $b^{\star}=1$. \label{fig:Recnc}}
\end{figure}


\begin{figure}
\begin{tabular}{cccc}
(a) & (b) & (c) & (d) \\
\includegraphics[width=0.24\textwidth]{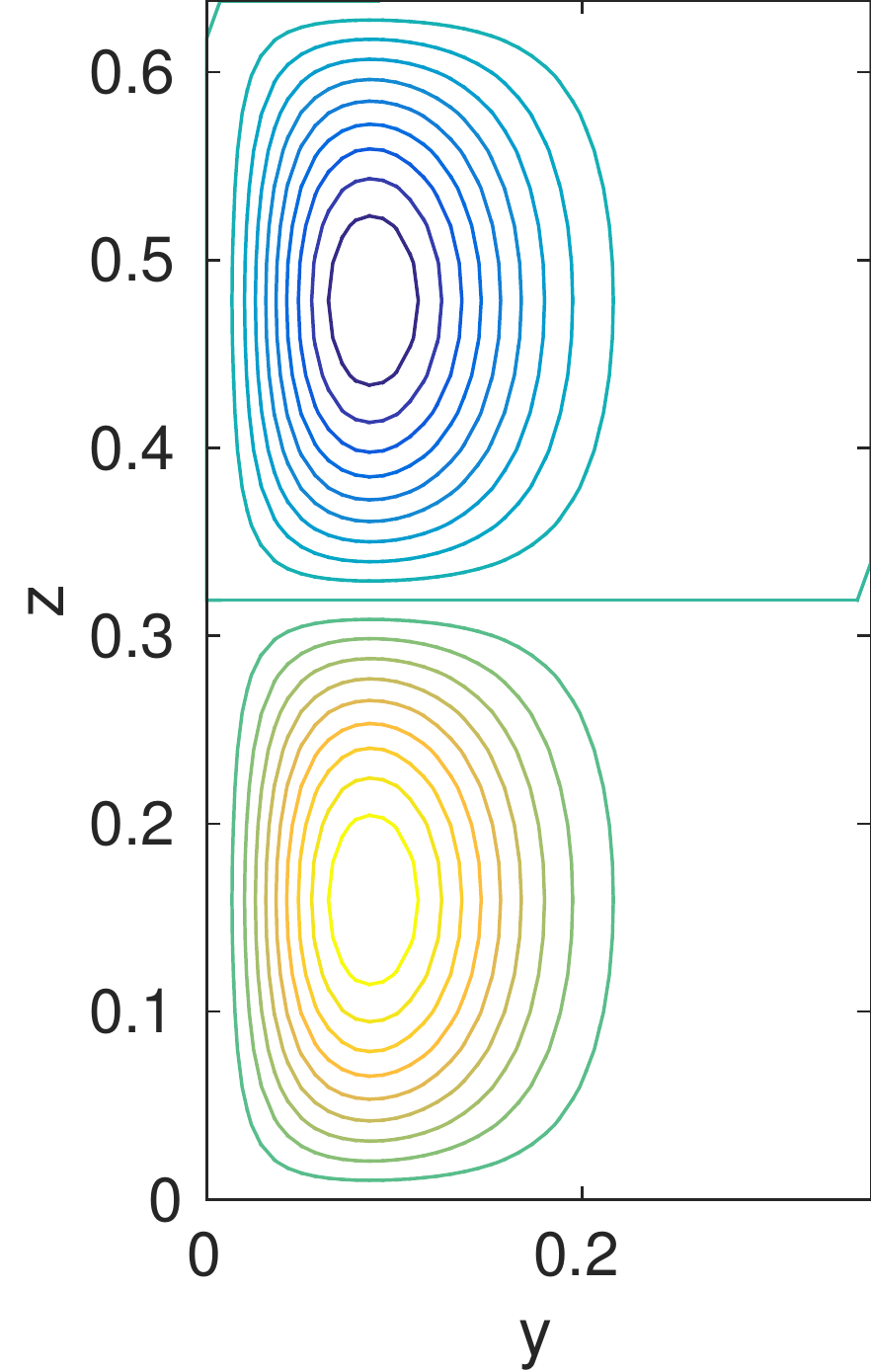} &
\includegraphics[width=0.24\textwidth]{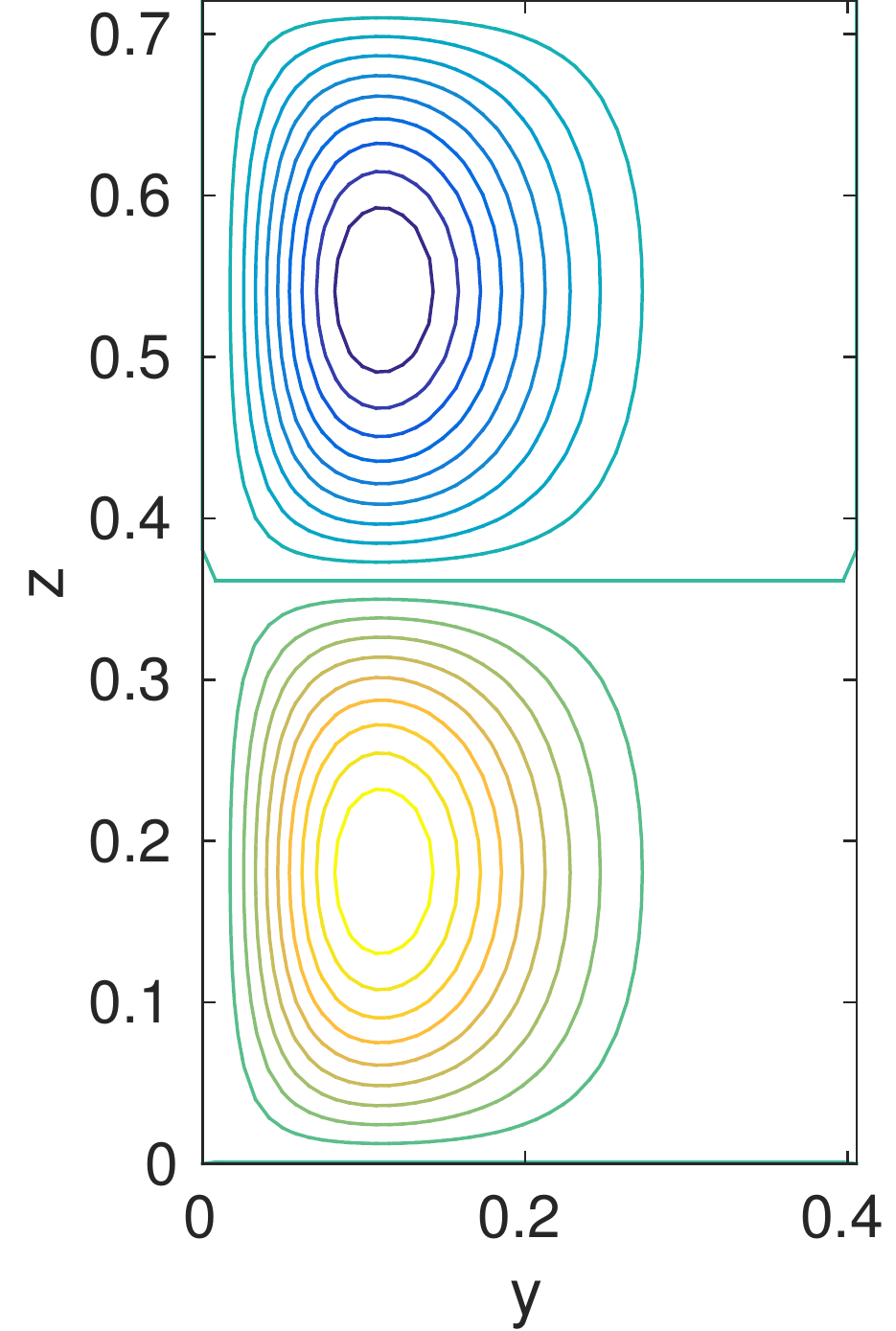} &
\includegraphics[width=0.24\textwidth]{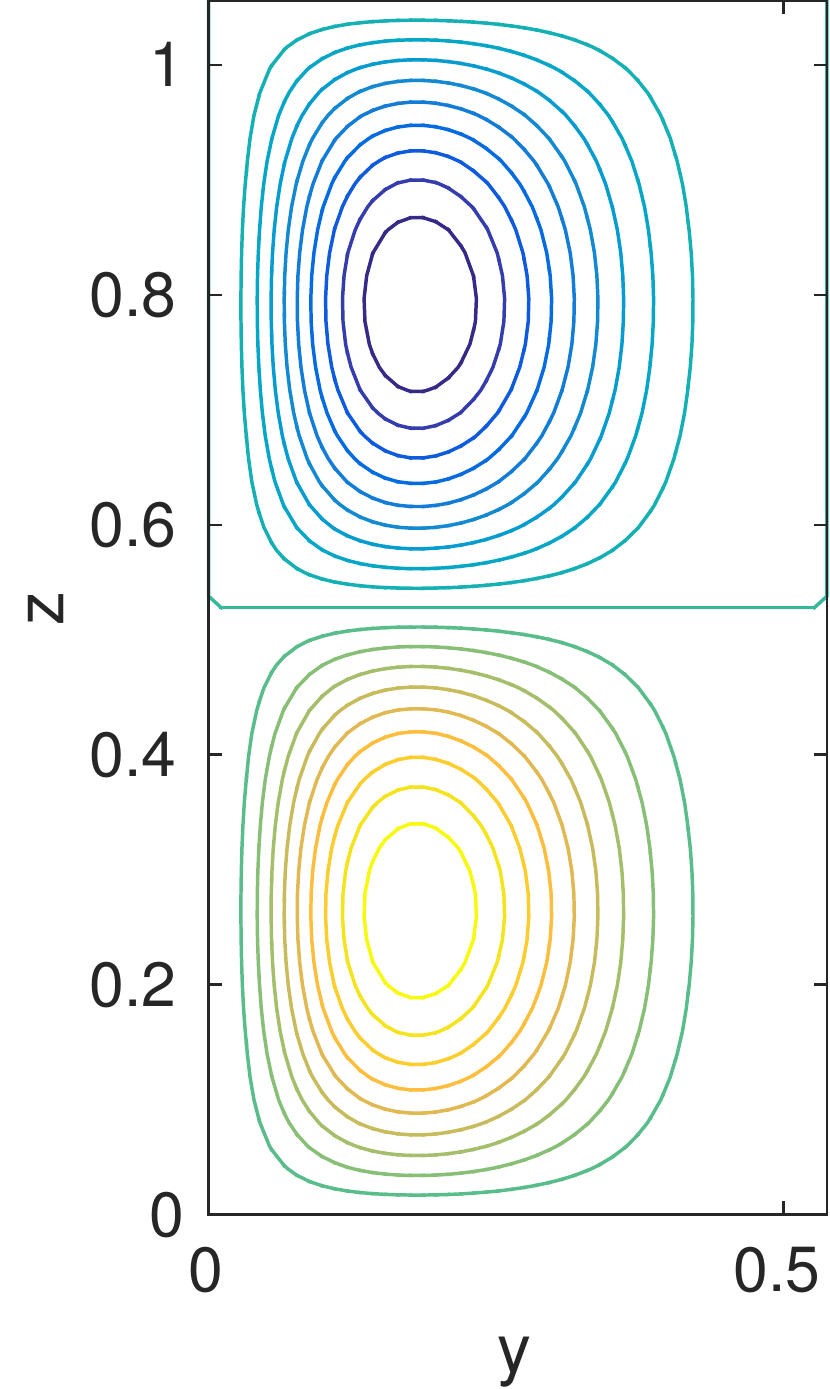} &
\includegraphics[width=0.24\textwidth]{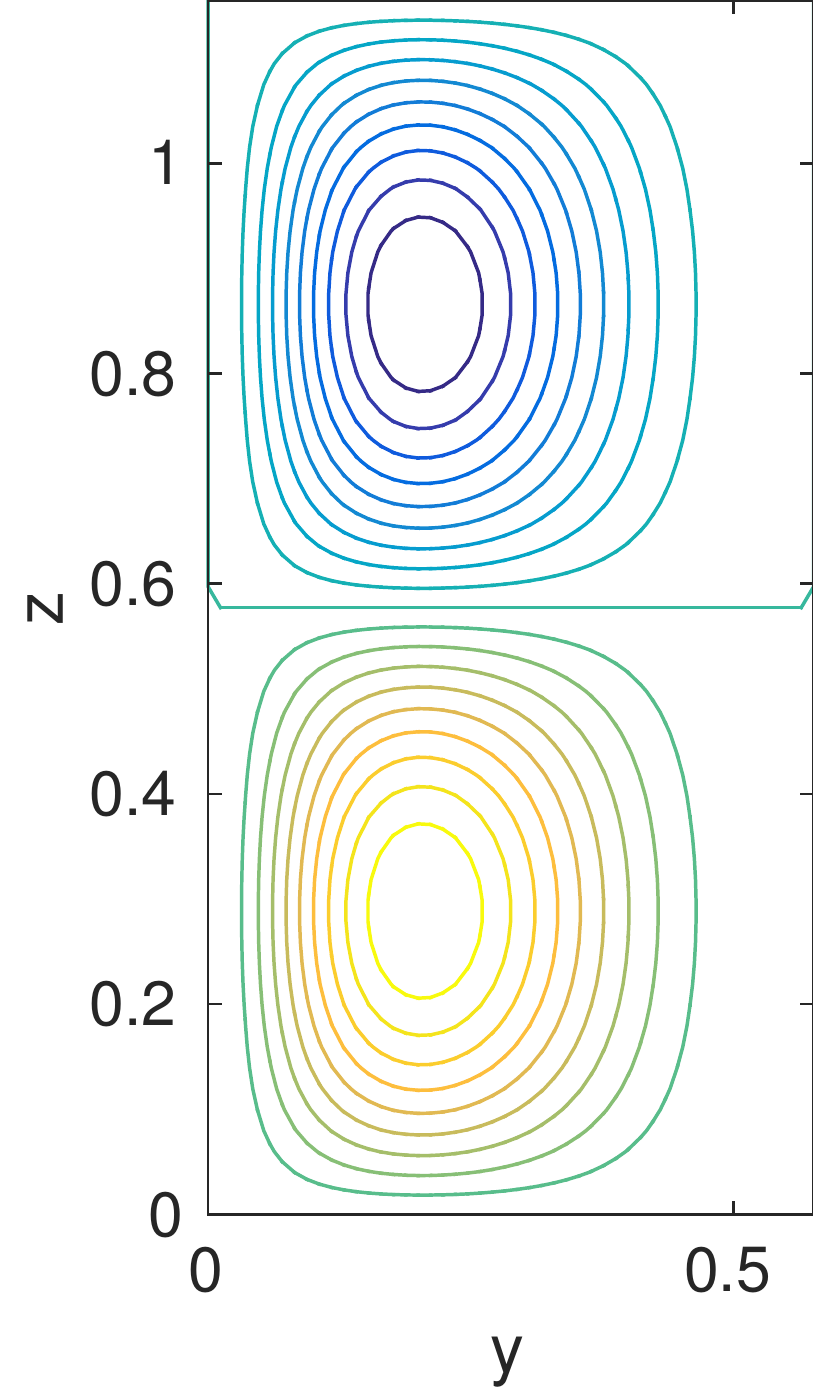} \\
(e) & (f) & (g) & (h) \\
\includegraphics[width=0.24\textwidth]{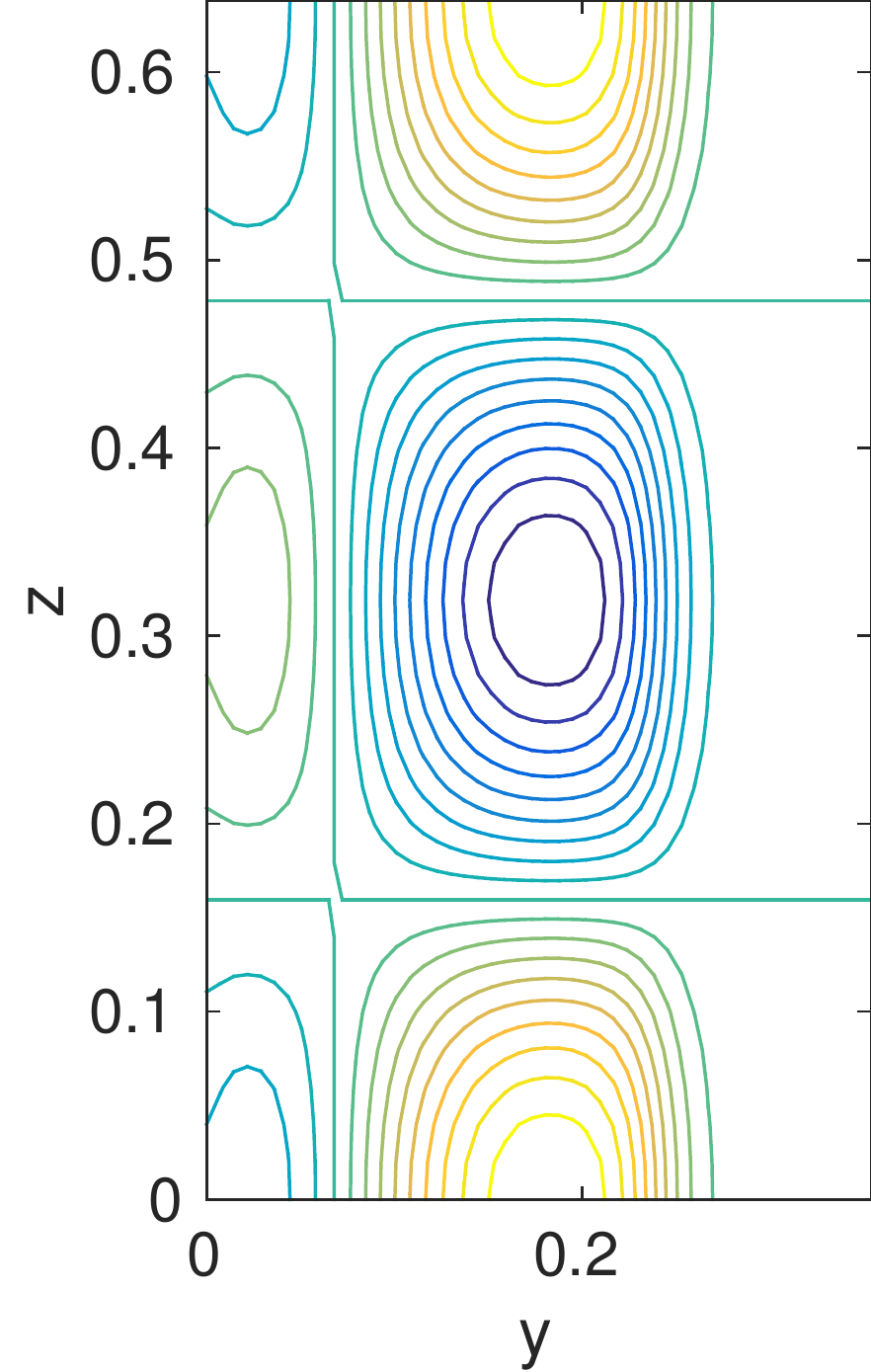} &
\includegraphics[width=0.24\textwidth]{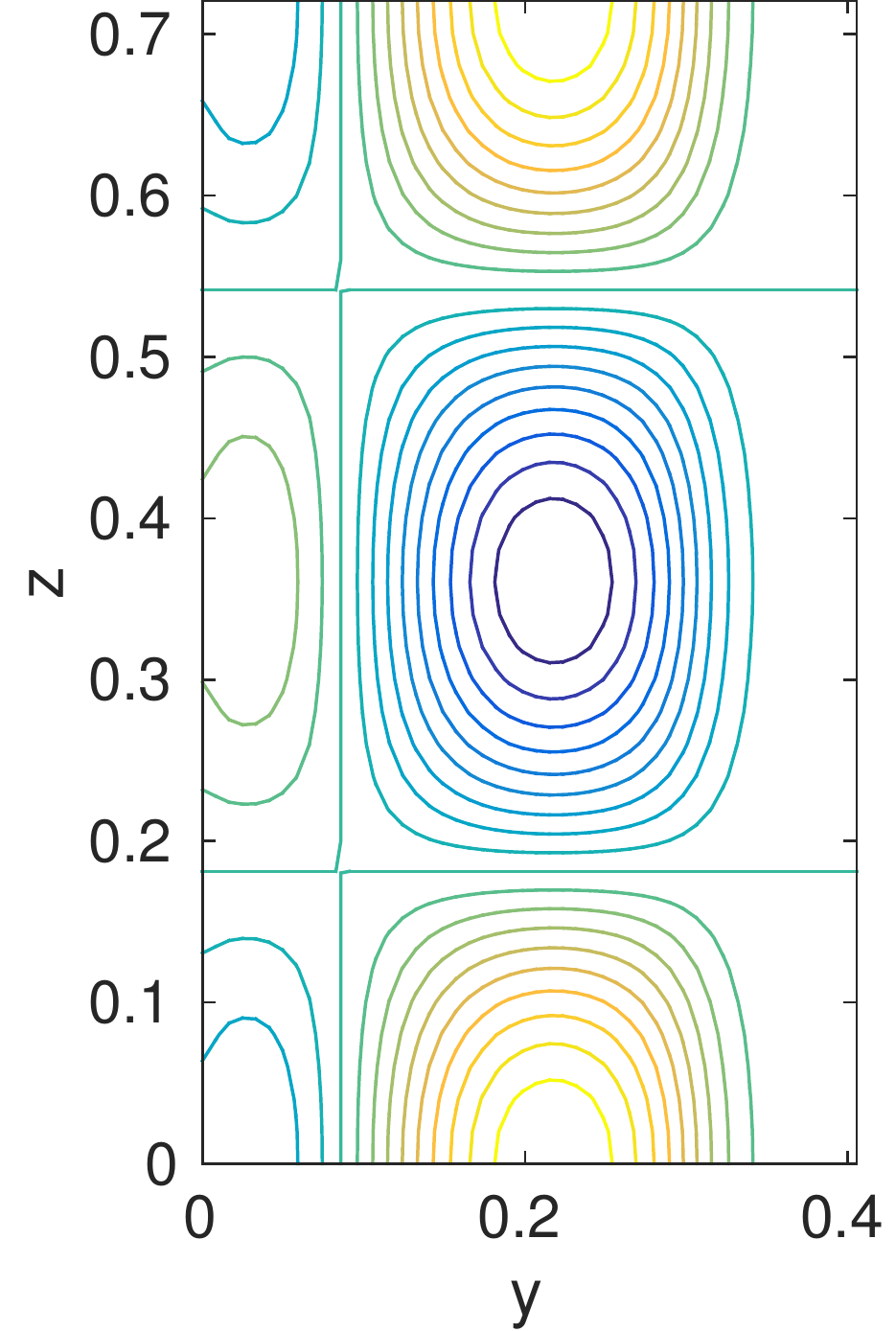} &
\includegraphics[width=0.24\textwidth]{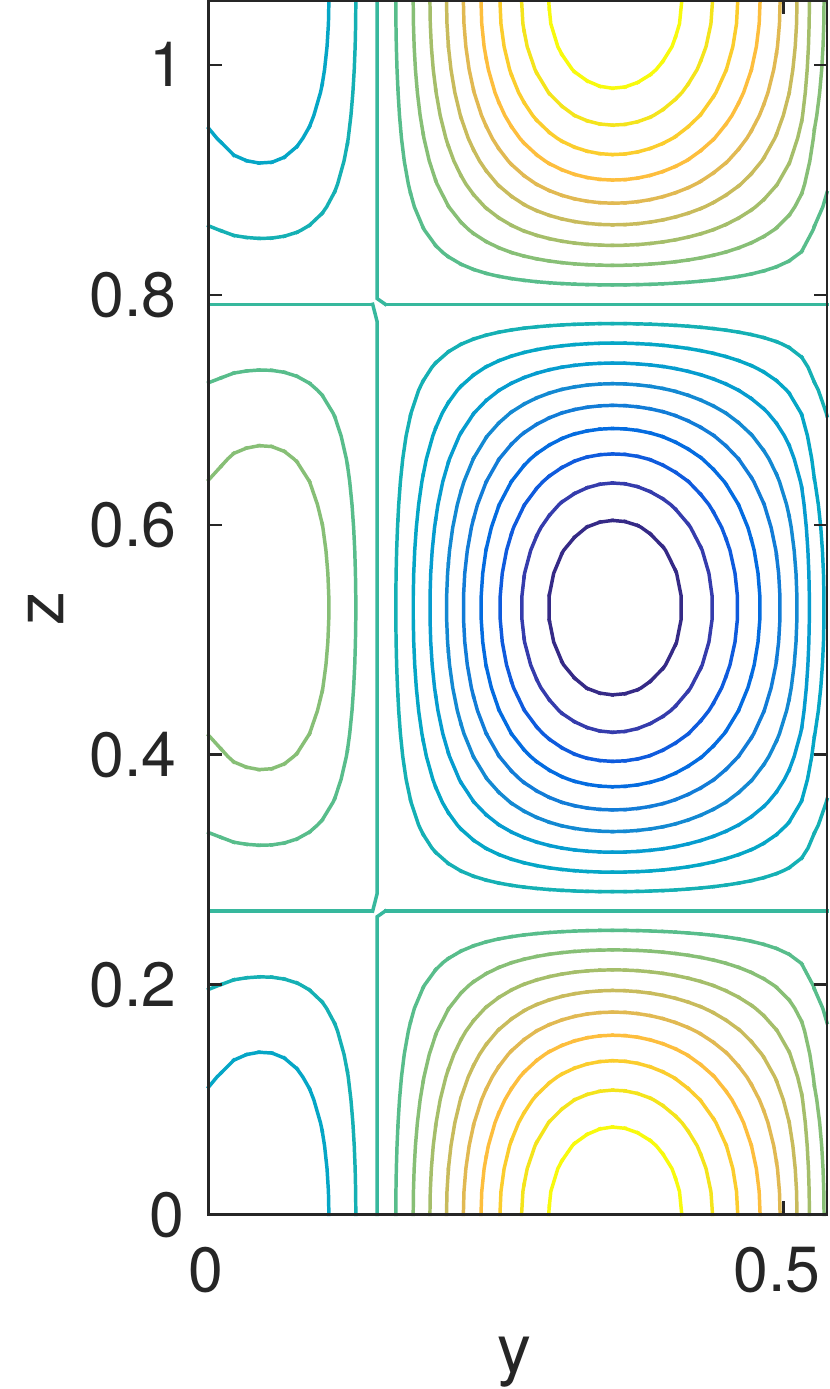} &
\includegraphics[width=0.24\textwidth]{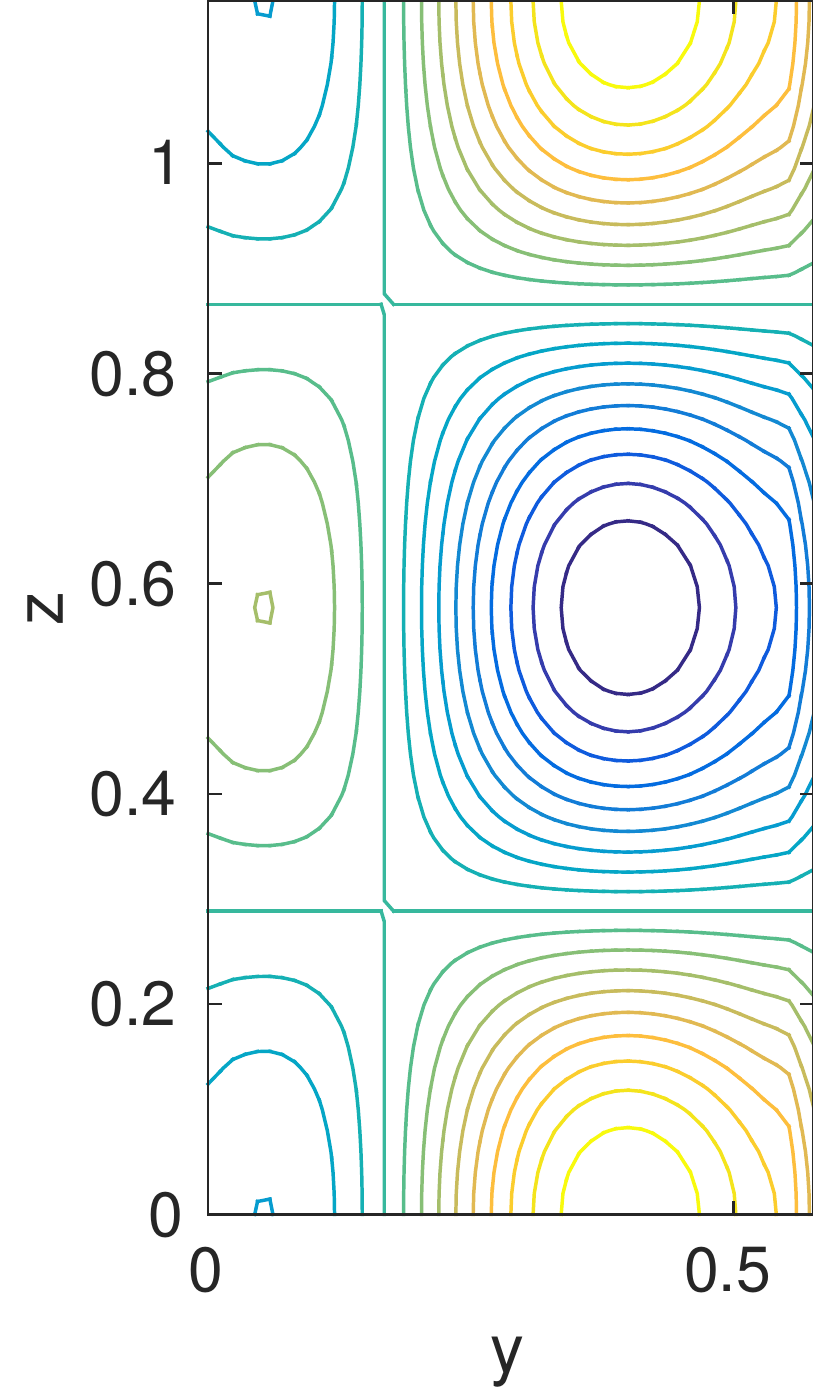}
\end{tabular}
\caption{First line (a-d): Streamlines of the critical perturbation. Second line (e-h): Contour plot of the perturbation of the structural parameter $\lambda$. Contour lines step is $5 \%$ of the normalized amplitude of the perturbation. Blue color stands for negative values and red colors for positive values. Large gap $\eta=0.5$, $Bn=2$, $n_c=0.5$ (a, e), $0.6$ (b, f), $0.9$ (c, g) and $1$ (d, h), $\Delta K^{\star}=\tau_1^{\star}=0.5$, $a^{\star}=1$ and $b^{\star}=1$. \label{fig:contournc}}
\end{figure}


Finally, the effect of $n_c$ seems to be either destabilizing or stabilizing depending if we track $Re_c$ (fig. \ref{fig:Recnc}-a) or  $Re_{co}$ (fig. \ref{fig:Recnc}-b). The figures \ref{fig:contournc} show that the Taylor-vortices are shifted toward the inner cylinder as the shear-thinning behavior increases. The effect is similar as increasing the shear-thinning behavior with the parameters $b^{\star}$, $\Delta K^{\star}$ and $\tau_1^{\star}$) for the shear-localized flows. In shear-banded flows (figs \ref{fig:contourb}c-d), the width of the dead zone collapses and thus, the Taylor-vortices expand in the whole fluid region.

To conclude on the effect of the thixotropy on the linear stability of the Couette flow, first the critical mode corresponds to the axisymmetric Taylor-vortices which are also found for simple yield stress fluids, such as Bingham fluids \cite{landry2006}. The critical eigenvalue is real as in simple fluids. The results found by Landry \textit{et al.} \cite{landry2006} with Bingham fluids or Alibenyahia \textit{et al.} \citep{alibenyahia2012} with shear-thinning fluids are retrieved. The critical perturbation is driven by the inertial term, \textit{i. e.} by the centrifugal force. The yield stress and the shear thinning behavior confine the rolls toward the inner cylinder in the wide-gap case. In a Bingham fluid, Chen \textit{et al.} \cite{chen2015} show that the optimal perturbation is also shifted toward the inner cylinder in the wide-gap case. Thus, even during the transient growth preceding the onset of the instability, only the inner zone of the gap is perturbed. The perturbation of the structural parameter is driven by the convection of the material because of the radial and axial velocity of the Taylor vortices. The stabilizing or destabilizing effect depends on the reference viscosity used for the definition of the Reynolds number. Nevertheless, the key point is that increasing the shear-thinning behaviour reduces the width of the inner region where the viscosity is low. As it would be the case if the material gap size would be reduced, it stabilizes the flow. The shear-thinning is not only driven by $n_c$ but also by the thixotropic parameters, \textit{i. e.} the ratio $b^{\star}/a^{\star}$, $\Delta K^{\star}$ and $\tau_1^{\star}$. Although the thixotropy does not produced a qualitative modification of the linear stability of the flow in case of shear localization, it can make appear a second kind of base flow. In the shear localization case, the strain rate drops smoothly to zero at the interface between the fluid and solid-like regions. Thus, there is a so-called dead zone where the viscosity becomes very strong near the interface. The width of the dead zone collapses in the shear banded flows (Figs. \ref{fig:vlb}b, \ref{fig:vldktau1}b). Thus, the Taylor vortices can occupied the fluid gap as it would be the case with shear-thinning fluids in an equivalent ratio of radii $\eta_{eq}=r_o/r_i$, \textit{i. e.} in a narrow-gap case.

In every cases, the width of the fluid gap $y_o$ is the most relevant reference length to describe the onset of the linear instability, as it would be the case for simple yield stress fluids, such as Bingham's fluids.

\section{Reference viscosity and reference yield stress \label{sec:muref}}

\begin{figure}
\centering
\includegraphics[width=0.49\textwidth]{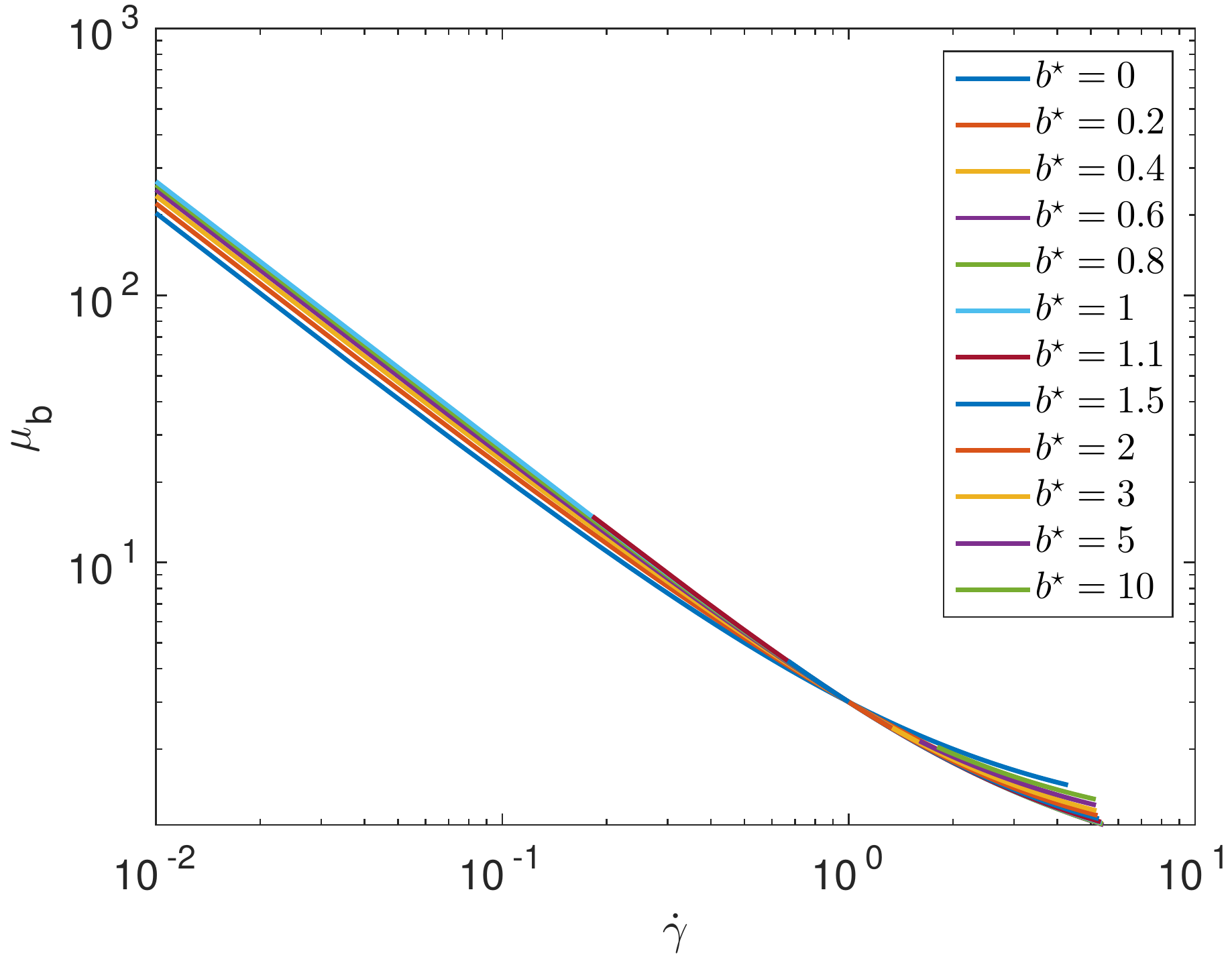}
\caption{Local viscosity $\mu_b$ \textit{vs} strain rate $\gap$ for different $b^{\star}$. Large gap $\eta=0.5$, $Bn=Bn_0=2$, $\Delta K^{\star}=\tau_1^{\star}=1$ and $a^{\star}=1$. \label{fig:mubg}}
\end{figure}

\begin{figure}
\begin{tabular}{cc}
(a) & (b) \\
\includegraphics[width=0.49\textwidth]{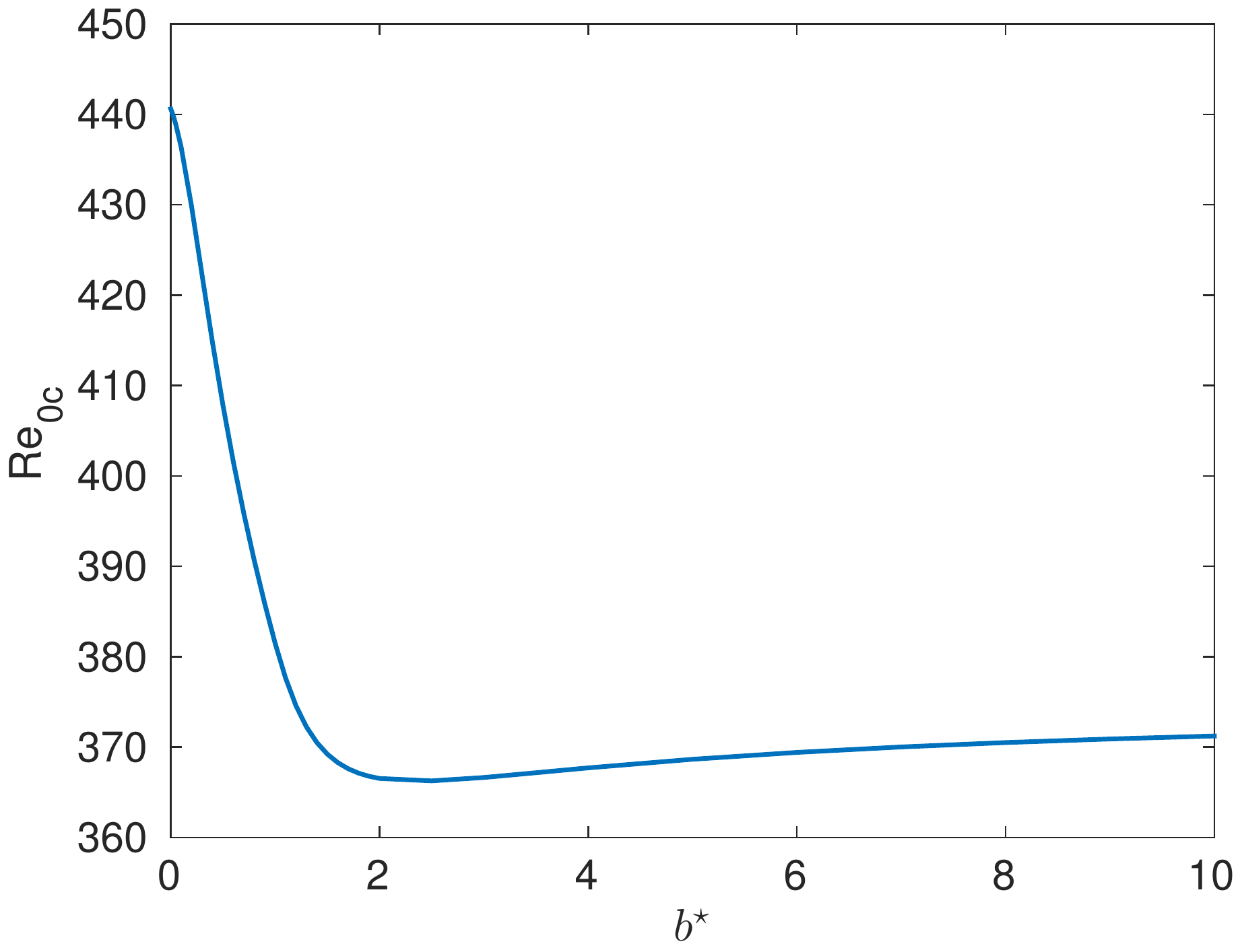} &
\includegraphics[width=0.49\textwidth]{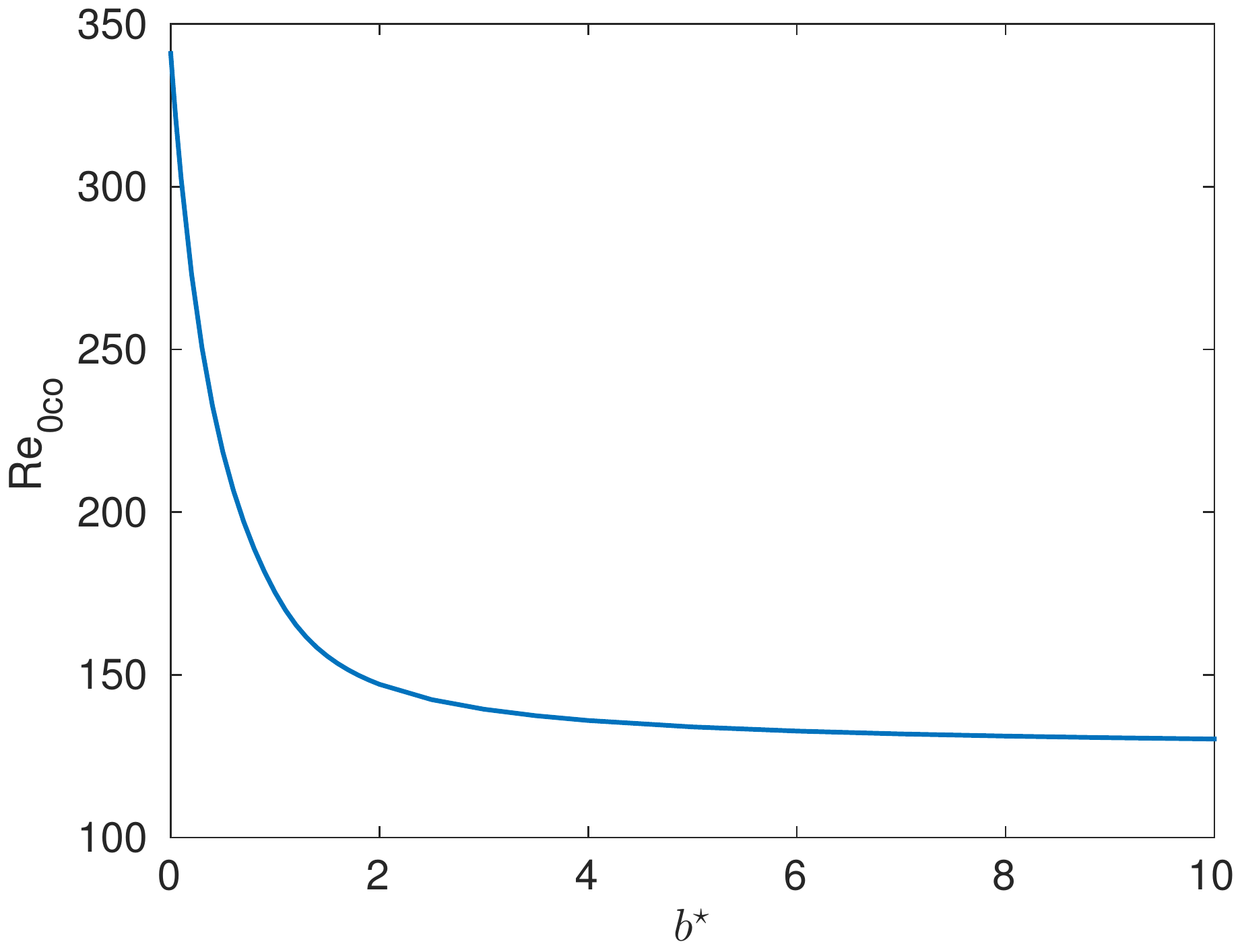}
\end{tabular}
\caption{Critical Reynolds number $Re_{0c}=(1+\Delta K^{\star} \lambda_{ref} ) Re_c$ (a) and $Re_{0co}=y_o Re_{0c}$ \textit{vs} $b^{\star}$. Large gap $\eta=0.5$, $Bn=Bn_0=2$, $\Delta K^{\star}=\tau_1^{\star}=1$ and $a^{\star}=1$. \label{fig:Re0c}}
\end{figure}

The figure \ref{fig:Re0c}-a shows that the reference chosen for the viscosity may dramatically change the conclusion about the effect of the parameters on the critical value of the Reynolds number. Nevertheless, the comparison between figs. \ref{fig:Recb}-b and \ref{fig:Re0c}-b shows that the relevant length is the width of the fluid gap $y_o$. Moreover, the asymptotic behaviour for large $b^{\star}$ can also be retrieved in fig. \ref{fig:Re0c}-b: the critical Reynolds $Re_{co}=130.2569$ close to the one of the equivalent case with Bingham fluid, $Re_{c}=126.9870$.

\begin{figure}
\begin{tabular}{cc}
(a) & (b) \\
\includegraphics[width=0.49\textwidth]{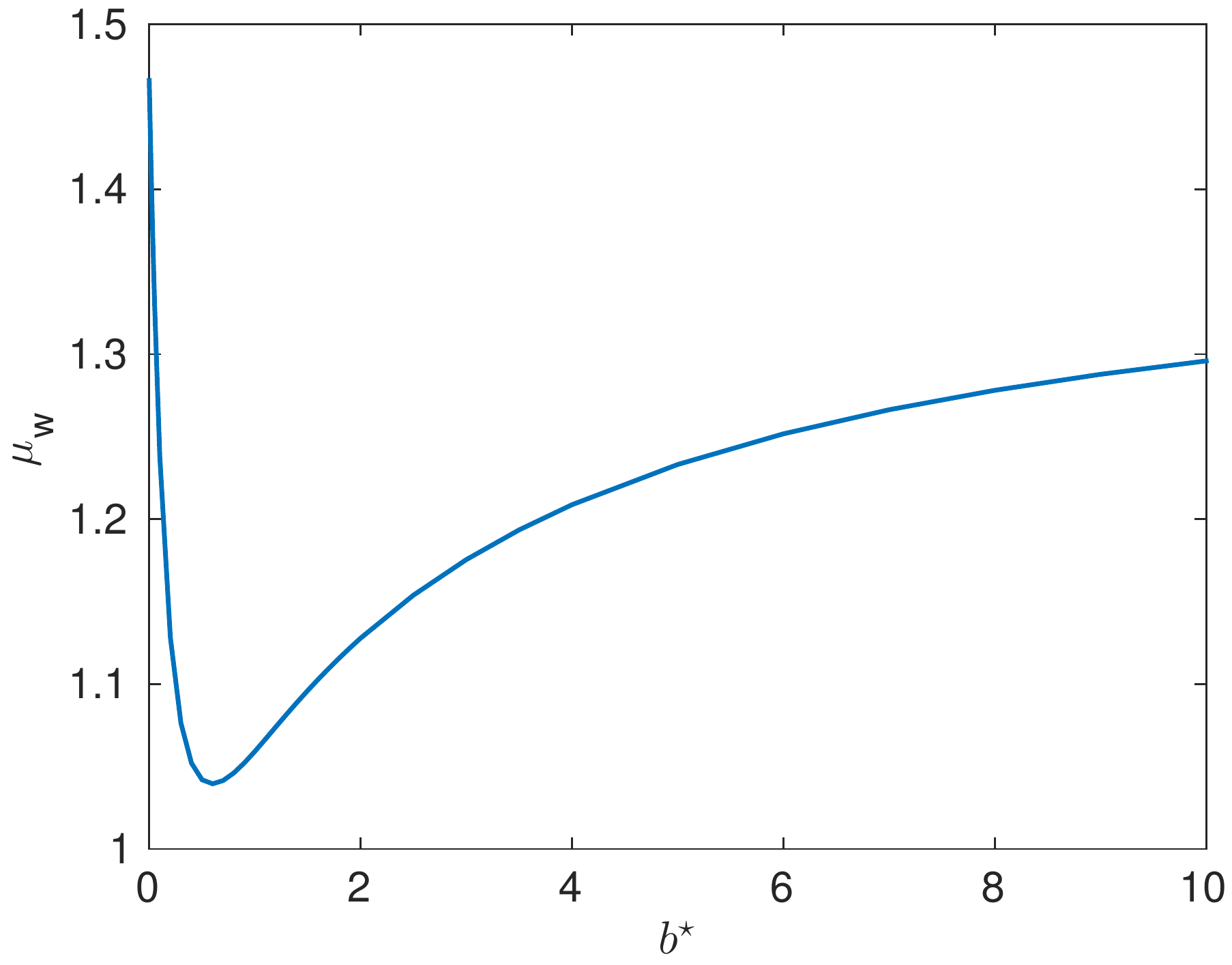} &
\includegraphics[width=0.49\textwidth]{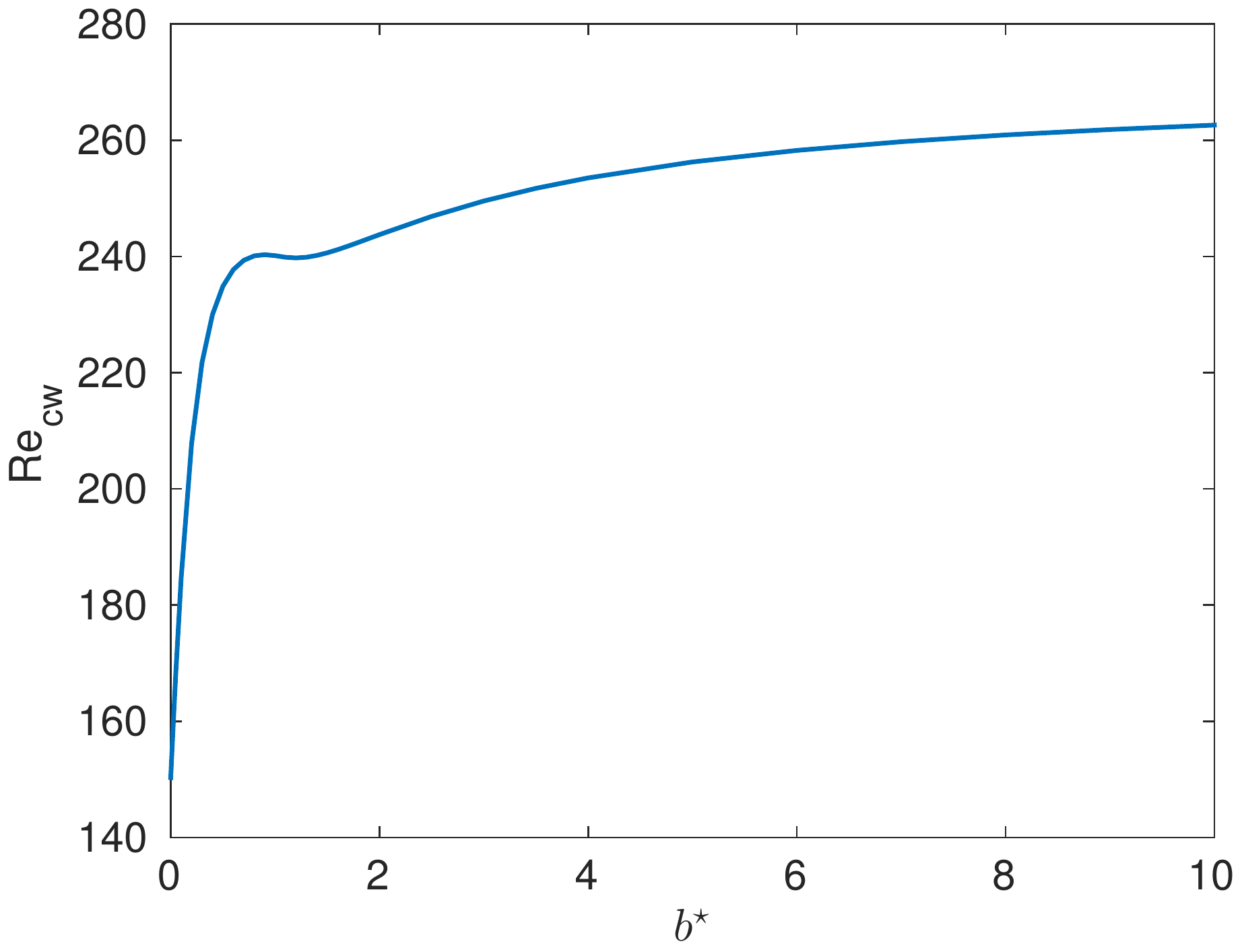}
\end{tabular}
\caption{Inner wall shear-viscosity $\mu_w$ (a) and critical Reynolds number $Re_{cw}=Re_c/\mu_w$ scaled with $\mu_w$ (b) \textit{vs} $b^{\star}$. Large gap $\eta=0.5$, $Bn=Bn_0=2$, $\Delta K^{\star}=\tau_1^{\star}=1$ and $a^{\star}=1$. \label{fig:Rew}}
\end{figure}

The viscosity $\mu_0$ is also a good choice to interpret the results but our reference $\mu_{ref}$ might be more relevant in a practical point of view. Moreover, it reproduces the stabilizing effect of thinning the gap which would be observed with Newtonian fluids. Finally, we defined the wall Reynolds number as

\begin{equation}
Re_w = Re / \mu_w
\end{equation}

\noindent where $\mu_w$ is the shear-viscosity of the fluid on the inner cylinder. This viscosity is relevant because it determines the resistive torque on the rotating cylinder which is measured in rheological experiment. Moreover, the centrifugal instability at the origin of the onset of the Taylor vortices is triggered at the low viscosity region close to the inner cylinder. The fig. \ref{fig:Rew}-a shows that the inner wall shear-viscosity $\mu_w$ decreases for $b^{\star}=0.6$ just before the onset of the shear-banding. For the shear-banded flows, the inner wall shear-viscosity increases to reach a limit value when $b^{\star}$ becomes high, \textit{i. e.} when the structure is broken down even when the strain rate is low. The critical Reynolds number calculated with the wall viscosity is strongly growing before the onset of the shear-banding but it is slightly constant ($\sim 250$) for the shear-banded flow (fig. \ref{fig:Rew}-b). This observation suggests that the inner wall shear-viscosity is more relevant for the onset of the Taylor vortices when the velocity profile of the base flow corresponds to the shear-banding, \textit{i. e.} when the strain rate does not fall down zero and the viscosity values are finite and moderate.

Thus, the thixotropic yielded fluids behave mainly as a viscous fluid when the structure is fragile and the shear-banding appears. In this case, the inner wall viscosity is a good reference to predict the onset of the Taylor vortices.

\section{Conclusion}

In this work we have studied the base flow and the linear stability in a Couette cell of a thixotropic yield stress material modelized by the Houska's model. This model with one structural parameter allows  non-monotonic composite flow curves  depending on the ratio between the building and breakdown parameters $b^{\star}/a^{\star}$ (Fig. \ref{fig:taudegamma}). In case of non-monotonic composite curves, a shear banded flow occurs if only a part of the gap flows. In shear banding, the structural parameter $\lambda$ jumps abruptly from a value $<1$ to $1$ across the interface between the fluid and the solid-like zones. The shear rate $\gap$ is sharply discontinuous across the interface and the width of the high viscosity zone, called the dead zone, collapses with the shear-banding. In shear banded flows, the transient flow shows that the selected stress at the interface is the yield stress of the solid material where $\lambda=1$ if there is no diffusive term for the structural parameter. One can notice that adding a diffusive term would select a lower stress corresponding to the yield stress of the partially structured material. For the following linear stability analysis, we only consider the case where the diffusive parameter is neglectable.  The primary instability of the Couette flow is studied for a large gap ($\eta=0.5$) when the Bingham number $Bn$ is sufficiently high to have a solid-like region in the gap. The thixotropy does not modify the kind of the linear unstable mode which is still steady and axisymmetric in the large range of parameters explored in comparison with simple yield stress fluids \cite{landry2006}. We retrieve that the choice of the viscosity to scaled the Reynolds number may modify the conclusion about the effect of the thixotropy as for shear-thinning fluids. Considering a Reynolds number based on our reference viscosity $\mu_{ref}$ or the inner-wall shear-viscosity $\mu_w$, lead to the conclusion that the thixotropy stabilizes the Couette flow because it increases the stratification of the viscosity over the gap. This stabilizing effect was also found by \citep{alibenyahia2012} with shear-thinning fluids. Nevertheless, re-scaling the Reynolds number with the flowing gap size $y_o$ shows that there is a competition between the reduction of the stratification of the viscosity which destabilizes the flow and the reduction of the effective gap size because of the breakdown of the fluid structure. The shear-banding dramatically reduces the stratification of the viscosity which remains finite and moderate in the fluid region but it squeezes the fluid area near the inner wall. Thus, the $y_o$-scaled Reynolds number $Re_{co}$ is only slightly growing and tends to a constant value for high values of the breakdown parameter $b^{\star}$. 

Finally, The thixotropy allows for shear banding flows but does not induce transition to unsteady and non-axisymmetric flows according to linear perturbations. For shear localized flows, the transition is driven by both the stratification of the viscosity and the size of the flowing region as in Bingham's fluids for instance \cite{landry2006}. When shear banding occurs, the stratification of the viscosity vanishes and the transition is mainly controlled by the effective gap ($y_o$). The shear-banding modifies the linear stability of the flow which becomes similar to the one of a circular Couette setup with a smaller gap and without any dead zone. The unsteady effects of the thixotropy may thus appears at the secondary instability of the Taylor vortices or it could be due to non-linear effects which have to be studied in further works.

\appendix
\section{Details of the linear operators}

In cylindrical coordinates, the perturbation vector of the velocity is written:
\begin{equation}
\tilde{\mathbf{v}}= u \mathbf{e}_r + v \mathbf{e}_{\theta} + w \mathbf{e}_{z}
\end{equation}

\noindent The expressions of the linear operators which appear in eqs. (\ref{eq:nspert}--\ref{eq:divpert}) in the cylindrical coordinates system are:

\begin{equation}
- \dbar{\nabla} \mathbf{v}_b \cdot \tilde{\mathbf{v}} - \dbar{\nabla} \tilde{\mathbf{v}} \cdot \mathbf{v}_b = \frac{V_b}{r} \left( \frac{2}{r} v - i n u  \right) \mathbf{e}_r - \left( \left( \frac{V_b}{r} + \frac{\partial V_b}{\partial r} \right)u + \frac{i n V_b}{r} v  \right) \mathbf{e}_{\theta} - \frac{i n V_b}{r} w \mathbf{e}_{z},
\end{equation}

\noindent for the inertial term and
\begin{equation}
- \mathbf{v}_b \cdot \bm{\nabla} \tilde{\lambda} - \tilde{\mathbf{v}} \cdot \bm{\nabla} \lambda_b =  - \frac{\partial \lambda_b}{\partial r} u - \frac{i n V_b}{r} \tilde{\lambda}
\end{equation}

\noindent for the convective term of the structural parameter. The stress terms are:
\begin{eqnarray}
\left. \frac{\partial \dbar{\tau}}{\partial \gap_{ij}} \right|_b \gap_{ij}(\tilde{\mathbf{v}}) &=& \left[ \frac{1}{2}(\mu_1 - \mu_b) ( \delta_{ir}\delta_{j \theta}+\delta_{i \theta} \delta_{j r} ) (\mathbf{e}_r \otimes \mathbf{e}_{\theta} + \mathbf{e}_{\theta} \otimes \mathbf{e}_r) + \mu_b (\mathbf{e}_i \otimes \mathbf{e}_j) \right] \gap_{ij} (\tilde{\mathbf{v}}) \\
\left. \frac{\partial \dbar{\tau}}{\partial \lambda} \right|_b \tilde{\lambda} &=& \tau_2 \lambda (\mathbf{e}_r \otimes \mathbf{e}_{\theta} + \mathbf{e}_{\theta} \otimes \mathbf{e}_r)
\end{eqnarray}

\noindent with

\begin{eqnarray}
\mu_b &=& \left( \frac{1+\Delta K^{\star} \lambda_b}{1+\Delta K^{\star} \lambda_{ref}} \right) \gap_b^{n_c-1} + \frac{Bn}{\gap_b} \left( \frac{1 + \tau_1^{\star} \lambda_b}{1 + \tau_1^{\star} \lambda_{ref}} \right) \, , \\
\mu_1 &=& n_c  \left( \frac{1+\Delta K^{\star} \lambda_b}{1+\Delta K^{\star} \lambda_{ref}} \right) \gap_b^{n_c-1} \,  ,\\
\tau_2 &=& \left( \frac{\Delta K^{\star}}{1 + \Delta K^{\star} \lambda_{ref}} \gap_b^{n_c-1} + \frac{Bn}{\gap_b} \frac{\tau_1^{\star}}{1 + \tau_1^{\star} \lambda_{ref}} \right) \gap_{r \theta , b}.
\end{eqnarray}

The divergence of the stress tensor in cylindrical coordinates is:

\begin{equation}
\begin{array}{ll}
\mathbf{div} \left( \left. \frac{\partial \dbar{\tau}}{\partial \gap_{ij}} \right|_b \gap_{ij}(\tilde{\mathbf{v}}) \right) = & \left[ 2 \mu_b \frac{\partial^2 u}{\partial r^2} + 2 \left( \frac{\mu_b}{r} + \frac{\partial \mu_b}{\partial r}\right) \frac{\partial u}{\partial r} - \left( n^2 \frac{\mu_1}{r^2} + k^2 \mu_b +\frac{2 \mu_b}{r^2}\right) u \right. \\
& \left. +i n \frac{\mu_1}{r} \frac{\partial v}{\partial r} - i n \left( \frac{2 \mu_b+ \mu_1}{r^2} \right) v + i k \mu_b \frac{\partial w}{\partial r} \right] \mathbf{e}_r + \\
& \left[ i n \frac{\mu_1}{r} \frac{\partial u}{\partial r} + i n \left( \frac{2 \mu_b + \mu_1}{r^2} + \frac{1}{r} \frac{\partial \mu_1}{\partial r} \right) u + \right. \\
& \mu_1 \frac{\partial^2 v}{\partial r^2} + \left( \frac{\mu_1}{r} + \frac{\partial \mu_1}{\partial r} \right) \frac{\partial v}{\partial r} - \left( \frac{2 n^2 \mu_b + \mu_1}{r^2} + k^2 \mu_b + \frac{1}{r} \frac{\partial \mu_1}{\partial r} \right) v \\
& \left. - n k \frac{\mu_b}{r} w \right]  \mathbf{e}_{\theta} + \\
& \left[ i k \mu_b \frac{\partial u}{\partial r} + i k \left(\frac{\mu_b}{r} +\frac{\partial \mu_b}{\partial r} \right) u - n k \frac{\mu_b}{r} v + \right. \\
& \left. \mu_b \frac{\partial^2 w}{\partial r^2} + \left(\frac{\mu_b}{r} +\frac{\partial \mu_b}{\partial r} \right) \frac{\partial w}{\partial r} - \left( n^2 \frac{\mu_b}{r^2}+ 2 k^2 \mu_b \right) w \right] \mathbf{e}_{z}
\end{array}
\end{equation}

\noindent and

\begin{equation}
\mathbf{div} \left(\left. \frac{\partial \dbar{\tau}}{\partial \lambda} \right|_b \tilde{\lambda} \right) = i n \frac{\tau_2}{r} \tilde{\lambda} \mathbf{e}_{r} + \left[ \tau_2 \frac{\partial \tilde{\lambda}}{\partial r} + \left( \frac{2 \tau_2}{r} + \frac{\partial \tau_2}{\partial r}\right) \tilde{\lambda} \right] \mathbf{e}_{\theta} \, .
\end{equation}

Finally, the coupling term with the velocity in eq. (\ref{eq:lpert}) is:

\begin{equation}
\left. \frac{\partial \gap}{\partial \gap_{ij}} \right|_b  \gap_{ij}(\tilde{\mathbf{v}}) = \frac{i n}{r} u + \frac{\partial v}{\partial r}- \frac{v}{r}
\end{equation} 

\bibliographystyle{unsrt}
\bibliography{rheobib}


\end{document}